\documentclass[fleqn,usenatbib]{mnras}

\usepackage{newtxtext,newtxmath}
\usepackage{placeins}   % it should prevent interupting 

\usepackage[T1]{fontenc}

\DeclareRobustCommand{\VAN}[3]{#2}
\let\VANthebibliography\thebibliography
\def\thebibliography{\DeclareRobustCommand{\VAN}[3]{##3}\VANthebibliography}
\usepackage{xcolor}

\usepackage{graphicx}	% Including figure files
\usepackage{amsmath}	% Advanced maths commands
\newcommand{\angstrom}{\text{\normalfont\AA}}
\usepackage[caption=false]{subfig}
\usepackage{float}
\usepackage{lscape}
\usepackage{longtable}
\usepackage{booktabs} 
\usepackage[flushleft]{threeparttable}
\usepackage{cancel} %%% tmp package - to be removed upon submission

\title{Intraday Optical Variability of BL Lacertae during its Highly Active 2020--2024 Phase }

\author[A. Takey et al.]{
Ali Takey,$^{1}$
Ergün Ege,$^{2}$\thanks{E-mail: ergunege@istanbul.edu.tr}
B. M. Mihov,$^{3}$
E. G. Elhosseiny,$^{1}$
Aykut Ozdonmez,$^{4}$
Murat Tekkesinoglu,$^{5}$
\newauthor H\"useyin Er,$^{4}$
Aditi Agarwal,$^{6}$
L. S. Slavcheva-Mihova,$^{3}$
M. Emir Kenger,$^{5,8}$
M. S. Rizk,$^{7}$
I. Zead,$^{1}$
\newauthor F. I. Elnagahy, $^{1}$
W. A. Badawy, $^{1}$ 
S. K. J. Pacif $^{6,9}$ 
\\
$^{1}$ National Research Institute of Astronomy and Geophysics (NRIAG), 11421 Helwan, Cairo, Egypt\\
$^{2}$ Istanbul University, Faculty of Science, Department of Astronomy and Space Sciences, 34116 Beyazit, Istanbul, Türkiye\\
$^{3}$ Institute of Astronomy and NAO, Bulgarian Academy of Sciences, 72 Tsarigradsko Chaussee Blvd., 1784 Sofia, Bulgaria\\
$^{4}$ Atatürk University, Faculty of Science, Department of Astronomy and Space Science, 25240 Erzurum, Türkiye \\
$^{5}$ Atatürk University, Graduate School of Natural and Applied Sciences, Department of Astronomy and Astrophysics, 25240 Erzurum, Türkiye \\
$^{6}$ Pacif Institute of Cosmology and Selfology (PICS), Sagara, Sambalpur 768224, Odisha, INDIA \\
$^{7}$ Astronomy, Space Science and Meteorology Department, Faculty of Science, Cairo University, 12613 Giza, Egypt \\
$^{8}$ Türkiye National Observatories, TUG, 07070 Antalya, Türkiye\\
$^{9}$ Research Center of Astrophysics and Cosmology, Khazar University, Baku, 41 Mehseti Street, AZ1096, Azerbaijan
}

\date{Accepted XXX. Received YYY; in original form ZZZ}

\pubyear{2015}

\begin{document}
\label{firstpage}
\pagerange{\pageref{firstpage}--\pageref{lastpage}}
\maketitle

\begin{abstract}
We present an analysis of the intraday flux and spectral variability of BL Lacertae from 2020 September 30 to 2024 December 7, covering its highest recorded brightness and low, intermediate, and high flux states. Our study involved 62 nights of multi-band ($BVRI$) optical monitoring using four ground-based telescopes located in Egypt, T\"urkiye, and Bulgaria.
We assessed intraday flux variability using the power-enhanced $F$-test and the nested ANOVA test. Significant variability was detected in 88 out of 117 light curves, consistent with previous studies of this object during active epochs. The maximum variability amplitude is 44.5\,per cent in the $B$ band. 
Spectral analysis reveals a bluer-when-brighter trend during intraday flares, supporting a synchrotron origin for the variable emission. For a subset of well-sampled flares, we model their profiles with a double exponential function, deriving rise and decay time-scales, thereby constraining the characteristic times of particle acceleration and cooling processes within the relativistic jet. Assuming a turbulent jet model, we determined limits on the radii and magnetic field strengths of the emitting regions. We detected soft time lags for a multi-band flare and from their analysis, derived the Doppler factor and magnetic field strength of the corresponding emitting region.
Our long-term, high-cadence study confirms that BL Lacertae was in an exceptionally active phase during the 2020–2024 period, with intraday variability being a common phenomenon. The results underscore the dynamic nature of the jet emission region and provide valuable observational constraints for models of blazar variability and jet physics.

\end{abstract}


%%%%%%%%%%%%%%%%%%%%%%%%%%%%%%%%%%%%%%%%%%%%%%%%%%

%%%%%%%%%%%%%%%%% BODY OF PAPER %%%%%%%%%%%%%%%%%%

\section{Introduction}

Blazars are a remarkable class of active galactic nuclei with relativistic jets aligned close to the line of sight, and characterized by rapid, large-amplitude flux variability on diverse time-scales across the entire electromagnetic spectrum, along with strong polarized emission \citep{1995PASP..107..803U}.
The optical emission of blazars is predominantly produced by the jet-driven synchrotron mechanism. Based on their optical emission-line properties, blazars are divided into two main subclasses: flat-spectrum radio quasars (FSRQs) and BL Lac objects. Unlike FSRQs, BL Lac objects exhibit either featureless optical spectra or only weak emission lines \citep[typically equivalent width less than $5\,\angstrom$ for BL Lacs;][]{1997A&A...327...61G}.

The spectral energy distributions (SEDs) of blazars are characterized by two broad components: a low-energy hump, extending from the radio to the soft X-ray band, attributed to synchrotron radiation by relativistic electrons in the magnetized plasma of the jet, and a high-energy hump, spanning the X-ray to $\gamma$-ray regime, generally explained by inverse Compton scattering or, alternatively, by hadronic processes \citep{1998MNRAS.299..433F}. In the inverse Compton process, the seed photons may originate from synchrotron emission within the jet itself \citep[synchrotron self-Compton, SSC;][]{1989ApJ...340..181G} or from external photon fields, such as those from the broad-line region, dusty torus, or accretion disc \citep{1987ApJ...322..650B}.
BL Lacs are further classified according to the frequency of their synchrotron peak \citep[$\nu_{\rm p}$;][]{2010ApJ...716...30A}: low-synchrotron-peaked (LSP; $\nu_{\rm p}\! \leq \!10^{14}$~Hz), intermediate-synchrotron-peaked (ISP; $10^{14}\! \leq \nu_{\rm p} \!\leq 10^{15}$~Hz), and high-synchrotron-peaked (HSP; $\nu_{\rm p}\! \geq \!10^{15}$~Hz).

Flux variability of blazars is categorized by time-scale as intraday variability (IDV, minutes to hours), short-term variability (STV, days to months), and long-term variability (LTV, months to years) \citep{1995ARA&A..33..163W, 2003ApJ...586L..25G, 2004A&A...422..505G}.
Quantifying variability amplitude, duty cycle (DC), interband time lags, and spectral variations offers valuable insights into the location, size, structure, and dynamical properties of the regions responsible for non-thermal photon emission \citep{2003A&A...400..487C}.
The IDV, particularly, offers valuable insights into the compact and dynamic emission regions near the central supermassive black hole, highlighting rapid particle acceleration, and jet dynamics \citep{1989Natur.337..627M}. The detection and characterization of IDV provide important constraints on the size and structure of the emitting region, jet dynamics, and the role of relativistic beaming in shaping the observed emission \citep{2003A&A...397..565P, 2016MNRAS.458.1127G, 2019MNRAS.488.4093A, 2025ApJ...979...56O}. 

BL Lacertae is located at a redshift of 0.069 \citep{Miller1978} and is the archetype of BL Lac objects. It is primarily classified as an LSP \citep{2020ApJ...892..105A} or ISP blazar \citep{2011ApJ...743..171A}. However, its long-term variability indicates episodic transitions to an HSP classification at various epochs \citep{2022MNRAS.513.4645S}.
As a confirmed TeV emitter, it exhibits pronounced variability across the entire electromagnetic spectrum, with observed flux changes spanning time-scales from minutes to years \citep{2022MNRAS.513.4645S}.

Previous studies of BL Lacertae have reported significant intraday flux variability; for example, an increase of 0.6\,mag within 40\,min in the optical $V$ band \citep{1999PASJ...51..253M}, a 0.73\,mag decrease within 53\,min in the optical $B$ band \citep{1999ApJ...522..846X}, and a 0.5\,mag decrease over 7\,h followed by a 0.4\,mag brightening within 1.7\,h \citep{2002A&A...390..407V}. In a recent study, \citet{Agarwal2025} reported variability on 39 of 53 intraday light curves (LCs), a dominant bluer-when-brighter (BWB) behaviour on both short and intraday time-scales. 
Additionally, intraday observations of BL Lacertae have revealed time lags between different bands, including less than 12\,min lag between the $B$ and $I$ bands \citep{2003A&A...397..565P}, 11.6\,min lag between the $e$ (4925.0\,{\angstrom}) and $m$ (8023.2\,{\angstrom}) bands \citep{2006MNRAS.373..209H}, 10\,min lag between the $V$ and $R$ bands \citep{2017MNRAS.469.3588M}, and $\sim$16\,min lag between the $B$ and $V$ bands and $\sim$18\,min lag between the $B$ and $R$ bands \citep{2022ApJ...926...91F}; in this context see also \citet{Agarwal2023}.

Monitoring campaigns have determined typical minimum variability time-scales down to 24.6\,min, and interband lags of around 10–18\,min, implying the complex internal structure of the jet \citep{2024MNRAS.528.6823L}. Previous studies have reported similarly short minimum time-scales, such as 30\,min \citep{2020ApJ...900..137W} and 42.5\,min \citep{2017MNRAS.469.3588M}. \citet{2023MNRAS.522..102R} reported that IDV amplitudes increase toward brighter states.
Furthermore, BL Lacertae exhibits a BWB trend during episodes of IDV, particularly during its flaring state \citep{2003A&A...397..565P, 2017MNRAS.469.3588M, 2021RAA....21..259L, 2024MNRAS.528.6823L, Agarwal2025}.

The observed IDV of flaring type could be used to study the structure of jets at smallest spatial scales under the assumption that the flares are result from the particles energy evolution owing to the acceleration/cooling processes. Regarding BL Lacertae, \citet{Agarwal2023} used empirical model function to decompose flaring LCs into individual flares, while \citet{2023ApJS..268...54X} and \citet{2023Galax..11..108W} considered the equation for the electron distribution function $N(\gamma,t)$ to get the synchrotron flare shape and then modelled LCs with a linear combination of individual flares. Both approaches lead to determination of various parameters of the emitting regions, which generate the flux flares on intraday time-scales, like radius, magnetic field strength, etc.

In this work, we conduct a comprehensive investigation of the optical IDV of BL Lacertae using high-cadence monitoring data obtained between 2020 September and 2024 December. The primary objective of this analysis is to characterize the IDV of the source in its different brightness states throughout this extended active period. The paper is organized as follows. Section~2 describes the observations and the procedures employed to generate the LCs. Section~3 outlines the analysis techniques applied to the dataset. Section~4 presents the results and discusses their implications. Finally, Section~5 provides a summary of the main findings and concluding remarks.

%%%%%%%%%%%%%%%%%%%%%%%%%%%%%%%%%%%%%%%%%%%%%%%%%%%%%%%%%%%%%%%%%%%%%%%%%%%%%%%%%%%%%%%%%%%%%%%%%%%%%%%%%%%%%%%%%%%%%%%%%%%%
%%%%%%%%%%%%%%%%%%%%%%%%%%%  Observations and Data Reduction %%%%%%%%%%%%%%%%%%%%%%%%%%%%%%%%%%%%%%%%%%%%%%%%%%%%%%%%

\section{Observations, Data Reduction, and Photometry}

BL Lacertae was monitored on intraday time-scales using four ground-based optical telescopes located at three observatories. Observations were obtained with the 1.88-m Cassegrain telescope at the Kottamia Astronomical Observatory (KAO) in Egypt, the 1.0-m Ritchey-Chr{\'e}tien (RC) telescope at the T\"{U}B\.{I}TAK National Observatory (TUG) located in Antalya site under the T\"{u}rkiye National Observatories, and 2.0-m RC and 50/70-cm Schmidt telescopes of the Rozhen National Astronomical Observatory (NAO), Bulgaria. The technical specifications of these facilities and their detectors are summarized in Table \ref{tbl:telescopes}. The 1.88-m KAO telescope and the 2.0-m NAO %RC
telescope were equipped with the Kottamia Faint Imaging Spectro-Polarimeter \citep[KFISP;][]{Azzam2022} and the two-channel Focal Reducer Rozhen \citep[FoReRo-2;][]{Jockers2000,2026A&A...708A..30N}, respectively, while the remaining two telescopes operated with CCD cameras mounted directly at their focal planes.

Our monitoring campaign was conducted over 62 nights spanning a four-year period, from 2020 September 30 to 2024 December 7. The duration of individual observing sessions was at least 40 minutes and, in some cases, reached up to nearly eight hours. The resulting LCs contain between 17 and 794 data points per night. In total, the campaign yielded 16\,373 photometric measurements obtained through the Johnson-Cousins $BVRI$ filters, with 3\,228, 575, 8\,471, and 4\,099 data points in the $BVRI$ bands, respectively. The complete observing log is presented in Table~\ref{tab:obs_log}. We note that on three nights the target was observed with two different telescopes, although in different filters. Bias frames were acquired each night, and twilight flat-field frames were obtained on the same night whenever possible; when this was not feasible, flat fields from adjacent nights were used.

%%%%%%%%%%%%%%    Table1: Telescopes used here  %%%%%%%%%%%%%%
\begin{table*}
\centering
\begin{threeparttable}
%\tabcolsep 5.8pt
%\scriptsize
    \caption{Telescopes contributing to this study, listed in order (A–D) by the number of nights on which observations were obtained.}
    \label{tbl:telescopes}
    \begin{tabular}{lccccc}
    \hline
    Code                        & A                  & B                        & C                  & D1                  & D2 \\
    \hline  
    Telescope                   & 1.88-m             & 1.0-m                    & 50/70-cm           & 2.0-m              & 2.0-m \\
    Observatory                 & KAO                & TUG                      & NAO                & NAO                & NAO \\
    Country                     & Egypt              & Türkiye                   & Bulgaria           & Bulgaria           & Bulgaria \\
    CCD Model                   & E2V 42-40 2K CCD   & SI 1100 Cryo, UV, AR, BI & FLI PL16803        & Andor iKon-L       & Andor iKon-L \\
    Chip Size (pixel)           & 2048 $\times$ 2048 & 4096 $\times$ 4096       & 4096 $\times$ 4096 & 2048 $\times$ 2048 & 2048 $\times$ 2048 \\
    Scale (arcsec pixel$^{-1}$) & 0.24               & 0.31                     & 1.079              & 0.174              & 0.497 \\
    Field (arcmin)              & 8.2 $\times$  8.2  & 21.5 $\times$ 21.5       & 73.7 $\times$ 73.7 & 5.9 $\times$ 5.9   & 17.0 $\times$ 17.0 \\
    Gain (e$^-$ ADU$^{-1}$)     & 2.14               & 0.57                     & 1.38               & 1.1                & 1.0/1.1 \\
    Read-out Noise (e$^-$ pixel$^{-1}$ RMS) & 3.92   & 4.11                     & 14.1               & 6.9                & 6.7/6.9 \\
    Typical Seeing (arcsec)     & 1-3            & 1-3                      & 1.5-2.5            & 1.5-2.5            & 1.5-2.5 \\
    \hline
    \end{tabular}
    \begin{tablenotes}
    \item A: $1.88$-m Cassegrain telescope at KAO in Egypt
    \item B: $1.0$-m Ritchey-Chr{\'e}tien telescope at TUG in Türkiye
    \item C: $50/70$-cm Schmidt telescope at NAO in Bulgaria
    \item D1: $2.0$-m RC telescope at NAO in Bulgaria
    \item D2: $2.0$-m RC telescope at NAO used with FoReRo-2 (in each of the FoReRo-2 channels, blue/red, an Andor iKon-L CCD camera is used)
    \end{tablenotes}
\end{threeparttable}
\end{table*}

%%%%%%%%%%%%%%%%%%%%%%%%%%%%%%%%%%%%%%%%%%

Raw data from the 1.88-m KAO and 1.0-m TUG telescopes were reduced using Python-based workflows. Bias subtraction and flat-field correction were performed with Ccdproc \citep{Craig2017}, followed by cosmic-ray removal using the L.A.Cosmic algorithm \citep{Dokkum2001}. Instrumental magnitudes for BL Lacertae and the comparison stars were extracted via aperture photometry, and astrometric calibration of all reduced frames was carried out with Astrometry.net \citep{Lang2010}. These latter steps employed the STDPipe framework \citep{2021ascl.soft12006K} and/or Astropy modules, including Photutils \citep{Bradley2022}. Basic reduction of the raw images obtained with the 2.0-m and 50/70-cm telescopes at NAO was conducted using locally developed {\sc idl} scripts. Aperture photometry for these datasets was performed with the {\sc daophot} package within {\sc idl} \citep{Stetson1987,1995ASPC...77..437L}. 

Because starlight from the BL Lacertae host galaxy contaminates the blazar’s observed flux, we adopted the analysis procedure recommended by \citet{2023MNRAS.522..102R} to remove the host-galaxy contribution. Following this approach, aperture photometry of BL Lacertae was extracted using an 8\,arcsec radius aperture, while the background level was estimated from an annulus with inner and outer radii of 10 and 16\,arcsec, respectively. This strategy was applied uniformly to all images obtained with the four telescopes. We also note that the same source and background apertures were used for all comparison stars to ensure consistent photometric measurements.

For the photometric calibration of the blazar, we adopted comparison stars C and H as both are bright and located close to the target on the detector \citep[][see also WEBT database\footnote{https://www.oato.inaf.it/blazars/webt/2200420-bl-lac/}]{1996A&AS..116..403F}. In addition, star 20, taken from \citet{Gonzalez2001}, was used as a control/check star, with its magnitudes calibrated using the same procedure applied to BL Lacertae. The magnitudes and associated uncertainties for the comparison stars 
C (=21) and H (=16) and for the control star 20 were taken from \citet{Gonzalez2001}, which provides the smallest reported photometric errors across all bands. Finally, nightly LCs were constructed for each filter, and any outliers were removed.

%%%%%%%%%%%%%%    Table2: Obs. Log Table  %%%%%%%%%%%%%%

\begin{table*}
\caption{Log of photometric observations for BL Lacertae. The table lists the date of each observation, the code identifying the telescope used, and the number of data points obtained in each filter for that night; the sequence is repeated for clarity of presentation. 
}
\label{tab:obs_log}
\begin{tabular}{cccccccccccc}
\hline\hline
Observation date & Telescope & \multicolumn{4}{c}{Number of data points} & Observation date & Telescope & \multicolumn{4}{c}{Number of data points} \\
\cmidrule(lr){3-6} \cmidrule(lr){9-12}
          &   code        & $B$ & $V$ & $R$ & $I$         &     &     code      & $B$ & $V$ & $R$ & $I$         \\

\hline
    2020 Sep 30 & B & 0 & 0 & 387 & 0 & 2023 Aug 17 & D2& 179 & 0 & 0 & 724 \\ 
    2020 Oct 12 & B & 0 & 0 & 272 & 0 & 2023 Aug 17 & B &  0 & 100 & 101 & 0 \\ 
    2021 May 16 & B & 0 & 0 & 112 & 0 & 2023 Aug 20 & A & 53 & 0 & 0 & 56 \\ 
    2021 Jun 14 & B & 0 & 0 & 87 & 0 & 2023 Aug 23 & A & 33 & 0 & 0 & 29 \\ 
    2021 Jul 11 & B & 0 & 0 & 359 & 0 & 2023 Sep 03 & A & 42 & 49 & 49 & 39 \\ 
    2021 Jul 17 & B & 0 & 0 & 438 & 0 & 2023 Sep 15 & B & 91 & 93 & 91 & 90 \\ 
    2021 Jul 18 & B & 0 & 0 & 373 & 0 & 2023 Sep 23 & A & 85 & 0 & 0 & 101 \\ 
    2021 Jul 30 & B & 0 & 0 & 212 & 0 & 2023 Sep 24 & A & 81 & 0 & 0 & 83 \\ 
    2021 Aug 16 & B & 97 & 95 & 98 & 94 & 2023 Sep 25 & A & 63 & 0 & 0 & 60 \\ 
    2021 Aug 17 & B & 0 & 0 & 285 & 0 & 2023 Sep 27 & A & 81 & 0 & 0 & 81 \\ 
    2021 Sep 05 & B & 0 & 0 & 295 & 0 & 2023 Oct 06 & A & 135 & 0 & 0 & 142 \\ 
    2021 Sep 28 & B & 0 & 0 & 547 & 0 & 2023 Nov 01 & A & 42 & 0 & 0 & 45 \\ 
    2022 Sep 03 & B & 69 & 69 & 67 & 67 & 2023 Nov 06 & A & 122 & 0 & 0 & 114 \\ 
    2022 Oct 18 & A & 0 & 0 & 78 & 0 & 2023 Nov 22 & A & 54 & 0 & 0 & 59 \\ 
    2022 Oct 19 & A & 26 & 30 & 23 & 0 & 2023 Dec 02 & A & 0 & 0 & 176 & 0 \\ 
    2022 Nov 10 & A & 27 & 29 & 29 & 0 & 2023 Dec 09 & A & 76 & 0 & 0 & 76 \\ 
    2022 Nov 12 & A & 19 & 19 & 17 & 0 & 2023 Dec 21 & A & 65 & 0 & 0 & 80 \\ 
    2022 Nov 19 & A & 0 & 0 & 177 & 0 & 2023 Dec 22 & A & 0 & 0 & 320 & 0 \\ 
    2022 Nov 20 & A & 0 & 0 & 178 & 0 & 2024 Jun 12 & B & 0 & 0 & 302 & 0 \\ 
    2022 Nov 26 & A & 0 & 0 & 43 & 0 & 2024 Jul 21 & A & 33 & 0 & 0 & 34 \\ 
    2023 May 24 & A & 0 & 0 & 321 & 0 & 2024 Jul 22 & A & 44 & 0 & 0 & 42 \\ 
    2023 Jun 19 & C & 117 & 0 & 0 & 122 & 2024 Jul 29 & A & 23 & 0 & 0 & 20 \\ 
    2023 Jun 20 & C & 104 & 0 & 0 & 121 & 2024 Jul 31 & A & 89 & 0 & 0 & 89 \\ 
    2023 Jul 18 & A & 0 & 0 & 703 & 0 & 2024 Aug 05 & A & 84 & 0 & 0 & 99 \\ 
    2023 Jul 18 & D1& 180 & 0 & 0 & 0 & 2024 Aug 07 & A & 100 & 91 & 98 & 80 \\ 
    2023 Jul 21 & A & 0 & 0 & 296 & 0 & 2024 Aug 17 & A & 85 & 0 & 0 & 73 \\ 
    2023 Jul 21 & C & 195 & 0 & 0 & 202 & 2024 Aug 27 & A & 120 & 0 & 0 & 120 \\ 
    2023 Jul 24 & A & 0 & 0 & 794 & 0 & 2024 Sep 08 & A & 41 & 0 & 0 & 39 \\ 
    2023 Jul 25 & A & 0 & 0 & 363 & 0 & 2024 Oct 09 & A & 113 & 0 & 0 & 97 \\ 
    2023 Jul 26 & A & 0 & 0 & 780 & 0 & 2024 Oct 11 & A & 99 & 0 & 0 & 92 \\ 
    2023 Aug 09 & A & 21 & 0 & 0 & 23 & 2024 Nov 30 & A & 40 & 0 & 0 & 45 \\ 
    2023 Aug 13 & C & 165 & 0 & 0 & 172 & 2024 Dec 07 & A & 0 & 0 & 0 & 104 \\ 
    2023 Aug 16 & D2& 135 & 0 & 0 & 585 &   &   &   &   &   &   \\ 
 \hline
 \end{tabular}
\end{table*}

%%%%%%%%%%%%%%%%%%%%%%%%%%%%%%%%%%%%%%%%%%%%%%%%%%%%%%%%%%%%%%%%%%%%%%%%%%%%%%%%%%%%%%%%
%Statistical Analysis
%%%%%%%%%%%%%%%%%%%%%%%%%%%%%%%%%%%%%%%%%%%%%%%%%%%%%%%%%%%%%%%%%%%%%%%%%%%%%%%%%%%%%%%%
\section{Analysis of the Light Curves}
\label{sec:stat_analysis} 

Following recent studies on IDV in blazars, we employed the power-enhanced $F$-test and Nested ANOVA, which utilize multiple comparison stars to detect micro-variations within a single night. The outcomes of these tests were used to calculate the variation amplitude for each variable LC and to estimate the source’s DC. The structure function was applied to determine the minimum variability time-scale. Spectral variations were analysed alongside their corresponding SEDs. Time lags between different bands were estimated using the $z$-transformed discrete correlation function ($z$DCF), and LCs exhibiting flaring activity were modelled using a double-exponential function.

%%%%%%%%%%%%%%%%%%%%%%%%%%%%%%%%%%%%%%%%%%%%%%%%%%%%%%%%%%%%%%%%%%%%%%%%%%%%%%%%%%%%%%%%

\subsection{Power-enhanced $F$-test}
\label{sec:F-test} 

We used the power-enhanced F-test following the approach of \cite{Diego2014} and \cite{Diego2015} to explore the IDV of BL Lacertae. This test has been used as a statistical methodology to find micro-variation in blazars in previous studies of optical variability of blazars by \cite{Dhalla2010, Abdo2010, Nestoras2015, Liu2019, Pandey2020, Pandian2022} and in our recently published studies by \cite{Tripathi2024,2024PASA...41...52O,2024ApJ...971...74E}. In the power-enhanced $F$-test, the brightest comparison star No. 20 is taken as a reference/control to illustrate the differential LCs of the blazar and the rest of the comparison stars \citep[C and H;][]{Polednikova2016, Pandey2019,Pandey2020}. Based on \cite{Pandey2019}, the power-enhanced $F$-test is defined as 

\begin{equation}
    \label{eq:f-test}
    F_{\rm enh} = \frac{S_{\rm blz}^2}{S_{\rm c}^2},
\end{equation}
where $S_{\rm blz}^2$ is the estimated differential variance of the blazar and $S_{\rm c}^2$ is the combined variance of the comparison stars and the reference star given as

\begin{equation}
    \label{eq:Sc}
    S_{\rm c}^2 = \frac{1}{(\Sigma_{j=1}^k N_j) -k} \sum_{j=1}^{k} \sum_{i=1}^{N_i} S_{j,i}^2,
\end{equation}
where k is the total number of comparison stars,  $N_j$ is the number of data points of the $j^{\rm th}$ comparison star while $S_{j,i}^2$ represents its scaled square deviation defined as,

\begin{equation}
    \label{eq:Sji}
    S_{j,i}^2 = \omega_j(m_{j,i} - \bar{m}_j)^2
\end{equation}
where $\omega_j$ is a scaling factor used to scale the variance of $j^{\rm th}$ comparison star to the level of the blazar \citep{Joshi2011}, $m_{j,i}$ and $\bar{m}_j$ are the differential magnitude, and the mean magnitude of the $j^{\rm th}$ comparison star, respectively.

Because the blazar and the comparison/control stars share the same observational frame, they have an identical number of data points ($N$). Consequently, the degrees of freedom are $N - 1$ in the numerator and $k(N - 1)$ in the denominator. The resulting values from these degrees of freedom are then compared with the critical F-value ($F_{\rm critical}$). A LC is classified as variable (V) when $F_{\rm enh} \geq F_{\rm critical}$; otherwise, it is considered non-variable (NV).

%%%%%%%%%%%%%%%%%%%%%%%%%%%%%%%%%%%%%%%%%%%%%%%%%%%%%%%%%%%%%%%%%%%%%%%%%%%%%%%%%%%%%%%%

\subsection{Nested ANOVA test}
\label{sec:anova}
The nested analysis of variance (ANOVA) test is an updated ANOVA test that can generate the different LCs of blazars based on two or three reference stars. Unlike the $F$-test, in the nested ANOVA test we used the stars C, H, and 20 as comparison stars to find the differential light curve (DLCs) of the blazar target \citep{Diego2015}. To better identify the micro-variation in the blazar, the DLCs were grouped into different temporal groups with 5 points in each group. The $F$-statistic is, hence, computed by dividing the mean square due to groups ($MS_{\rm G}$) by the mean square due to nested observations in groups ($MS_{\rm O(G)}$), that is $F =(MS_{\rm G})/(MS_{\rm O(G)})$ with degrees of freedom of $(a-1)$ for the numerator and $a(b-1)$ for the denominator, where $a$ is the number of groups in the night’s observations and $b$ is the number of data points in each group. Following this technique, a LC is marked as variable (V) if $F$-value exceeded the critical
value $F^{(\alpha)}$  at a significance level of 99\,per cent ($\alpha$ = 0.01), otherwise the LC will be marked as NV or no variation.

%%%%%%%%%%%%%%%%%%%%%%%%%%%%%%%%%%%%%%%%%%%%%%%%%%%%%%%%%%%%%%%%%%%%%%%%%%%%%%%%%%%%%%%%

\subsection{Variability amplitude}
\label{sec:amplitude}

We measured the variability amplitude $A$ for a LC that shows variation using the following equation \citep{Heidt1996},
\begin{equation}
    \label{eq:amplitude}
    A = 100 \times \sqrt{(A_{\rm max}- A_{\rm min})^2 - 2\bar{e}^{2}} \quad [\rm \%]
\end{equation}
where $A_{\rm max}$ and $A_{\rm min}$ are the maximum and the minimum calibrated magnitudes of the target blazar, respectively, and $\bar{e}$ is the mean photometric error. 

%%%%%%%%%%%%%%%%%%%%%%%%%%%%%%%%%%%%%%%%%%%%%%%%%%%%%%%%%%%%%%%%%%%%%%%%%%%%%%%%%%%%%%%%

\subsection{Duty cycle}
\label{sec:DC}

The DC is a critical parameter in understanding the variability behaviour of blazars, representing the fraction of time a blazar spends in an active or flaring state compared to its quiescent state. This metric provides insights into the emission mechanisms and energy release processes governing blazar activity.
The DC is defined as \citep{Romero1999}
\begin{equation}
    \label{eq:duty_cycle}
    DC = \frac{\sum_{i=1}^{n} N_i(1/\Delta t'_i)}{\sum_{i=1}^{n} (1/\Delta t'_i)}.
\end{equation}
The variable $N_i$ is assigned a value of 1 if IDV is detected and 0 if it is not. The DC calculation is weighted by the redshift-corrected monitoring duration $\Delta t'_i = \Delta t_i / (1 + z) $ for the $i$-th observation, which typically varies between observations \citep{Tripathi2024}; here and below the primed quantities are related to the rest-frame of the blazar. 
A higher DC indicates a blazar that is active for a larger fraction of the observed period, suggesting more frequent or prolonged flux changing episodes.

%%%%%%%%%%%%%%%%%%%%%%%%%%%%%%%%%%%%%%%%%%%%%%%%%%%%%%%%%%%%%%%%%%%%%%%%%%%%%%%%%%%%%%%%

\subsection{Structure function}
\label{sec:Tscale}

The structure function (SF), first introduced by \cite{1985ApJ...296...46S}, is a widely used method for estimating the variability time-scale of a time series \citep{2003A&A...400..487C, 2005A&A...443..451F, 2015MNRAS.451.3882A, 2024ApJ...971...74E}. An important advantage of the SF is that it can be applied to unevenly sampled datasets, as it is not significantly affected by irregular sampling. The first-order SF is defined as:
\begin{equation}
    SF(\tau)=\frac{1}{N(\tau)}\sum_{i}^{N}[m(t_i)-m(t_i+\tau)]^2
\end{equation}
where $m$ is the magnitude at a given time $t$, $N(\tau)$ is the number of pairs, and $\tau$ is the time lag. The SF typically shows a power-law rise on intermediate time-scales, flattening at short lags where variations are dominated by noise, and at long lags approaching twice the variance of the LC. The characteristic time-scale of variability is identified near the turnover point, where the SF changes slope from a rising power-law regime to a plateau, and can be associated with physical processes in the emission region \citep{2001ApJ...555..775C}.

%%%%%%%%%%%%%%%%%%%%%%%%%%%%%%%%%%%%%%%%%%%%%%%%%%%%%%%%%%%%%%%%%%%%%%%%%%%%%%%%%%%%%%%%

\subsection{$z$-transformed Discrete correlation function}

The discrete correlation function (DCF) is a powerful statistical method used to examine correlations within astronomical time series data that are irregularly sampled (time lags). As introduced by \citep{Edelson1988}, the DCF is particularly advantageous in blazar studies, where observations across multiple wavelengths often suffer from irregular sampling and gaps, and it has been extensively utilized in multi-band analyses of blazars \citep[e.g.,][]{2007A&A...469..899H, 2015MNRAS.450..541A, 2022MNRAS.513.2239P}. By reducing the effects of these observational inconsistencies, the DCF enables us to investigate potential time lags and correlations between different energy bands, providing insights into the emission mechanisms and regions within blazars.

The computation begins with the unbinned discrete correlation function (UDCF) for each pair of data points $(x_i,y_i)$ from two data series, defined as:

\begin{equation}
    \label{eq:UDCF}
    UDCF_{ij} = \frac{(x_i-\bar x))(y_i-\bar y)}{\sqrt{(\sigma_x^2-e_{x_i}^2)(\sigma_y^2-e_{y_i}^2)}},
\end{equation}

Here, $\bar x$ and $\bar y$ represent the mean values of the data series $x$ and $y$, respectively; $\sigma_x$ and $\sigma_y$ denote their standard deviations; and $e_{x_i}$ and $e_{y_i}$ are the measurement errors associated with each data point. Subsequently, the UDCF values are averaged over bins of time delay $\tau$ to yield the DCF for each $\tau$:

\begin{equation}
    \label{eq:DCF}
    DCF(\tau) = \frac{1}{N}\sum UDCF_{ij},
\end{equation}

In this equation, $N$ is the number of pairs $(i,j)$ whose time delay $\Delta t_{ij} = t_j - t_i$ falls within the bin centered at 
$\tau$ with width $\Delta \tau$. The DCF quantifies the degree of correlation between the two data series: positive values indicate a positive correlation, negative values suggest an inverse correlation, and values around zero imply no significant correlation. The uncertainty associated with each DCF value is determined by the standard deviation of the UDCF values within the bin, expressed as:

\begin{equation}
    \label{eq:DCF_SDV}
    \sigma_{DCF(\tau)} = \frac{1}{N-1}\sqrt{\sum [UDCF_{ij}-DCF(\tau)]^2},
\end{equation}

This methodology allows for a significant analysis of correlations in unevenly sampled data, making the DCF a valuable tool in time-domain astronomy.

An improvement of DCF was proposed by \citet{1997ASSL..218..163A,2013arXiv1302.1508A}, the so called $z$-transformed DCF ($z$DCF). It corrects some of the DCF biases and provides more realistic error estimates and thereby is more efficient in uncovering the real correlations. In contrast to DCF, $z$DCF applies equal-population binning with a minimum number of data points per bin of 11 and Fisher’s $z$-transform.

%%%%%%%%%%%%%%%%%%%%%%%%%%%%%%%%%%%%%%%%%%%%%%%%%%%%%%%%%%%%%%%%%%%%%%%%%%%%%%%%%%%%%%%%

\subsection{Flares analysis}
\label{sec:flare:1}
For a detailed morphological analysis of individual flares, we first identify the nights exhibiting active variable state. From this subset, we exclusively selected nights suitable for detailed flare analysis, where the continuous observational session covers at least one flare event with rise, peak, and decay phases.

In general, the observed intraday flux changes of a blazar could be considered as a combination of rapid variability on intraday time-scales (flares) and an underlying smooth component, variable on night-to-night time-scales \citep[e.g.][]{2024arXiv241201592M}. Therefore, to properly derive the flare's intrinsic properties, we need to correct for the low-frequency component. In this study, we shall assume that it is a result of a Doppler factor change on interday time-scales. The correction itself is done following the approach of \citet{Agarwal2023}; see also \citet{2024arXiv241201592M}. In brief, we assume that the underlying component locally vary in a linear fashion. We then fitted a linear function to the LC parts assumed to represent the low-frequency component. After the correction (or deboosting), the LC under consideration is characterized by a single, but still unknown value of the Doppler factor.

The individual flares were modelled with a double exponential function \citep[DE;][]{Abdo2010}:
\begin{equation}
    F(t) = F_{\rm base} + \frac{F_0}{\exp\left(\frac{t_0-t}{\tau_{\rm r}}\right)+\exp\left(\frac{t-t_0}{\tau_{\rm d}}\right)},
\end{equation}
where $F_{\rm base}$ is the constant base level, $F_0$ is twice the flare amplitude (with respect to the base level), $t_0$ is the approximate position of the flare peak, and $\{\tau_{\rm r},\tau_{\rm d}\}$ are the $e$-folding rise and decay time-scales, respectively.
We use a linear combination of DE functions to decompose the LCs into individual flares.
If the LC has been corrected for the smooth component (or deboosted for short), then the base level is set to the minimal value of the linear function. If no correction has been done, then the base level is left free.

We use symmetric DE function in decomposition of the LCs, that is, we set $\tau_{\rm r} = \tau_{\rm d}$, if either the LC is sparsely sampled, or the flare is partially sampled (e.g. missing rise or decay phases), or individual flares overlap substantially, or the rise and decay time-scales we got are equal to each other to within the uncertainties, or the uncertainties we got are larger than the parameter values themselves.
The best-fitting parameters for DE function(s) are obtained through a least-squares technique after an initial parameter values estimation. Given the DE function, some additional flare characteristics can be inferred, namely the exact time of the flux maximum, $t_{\rm max}$, the approximate flare duration, $\Delta \tau$, and the asymmetry parameter, $\xi$. The latter parameter quantifies the difference of the rise and decay time-scales and equals zero for symmetric flares \citep[see][]{Abdo2010}. For such flares one also gets $t_{\rm max}=t_0$.

%%%%%%%%%%%%%%%%%%%%%%%%%%%%%%%%%%%%%%%%%%%%%%%%%%%%%%%%%%%%%%%%%%%%%%%%%%%%%%%%%%%%%%%%
%Results and Discussion
%%%%%%%%%%%%%%%%%%%%%%%%%%%%%%%%%%%%%%%%%%%%%%%%%%%%%%%%%%%%%%%%%%%%%%%%%%%%%%%%%%%%%%%%

\section{Results and Discussion}

Over the period from 2020 September 30 to 2024 December 07, we obtained single- and multi-band optical photometric observations of BL Lacertae across 62 nights using the four telescopes listed in Table~\ref{tbl:telescopes}. The dataset comprises 6 nights of four-band ($BVRI$) observations, 4 nights with three bands, 29 nights with two bands, and 23 nights with a single band. These observations together form a comprehensive dataset, enabling an in-depth analysis of the IDV behaviour of BL Lacertae.

\begin{figure}
\centering
\includegraphics[width=0.5\textwidth,clip=true]{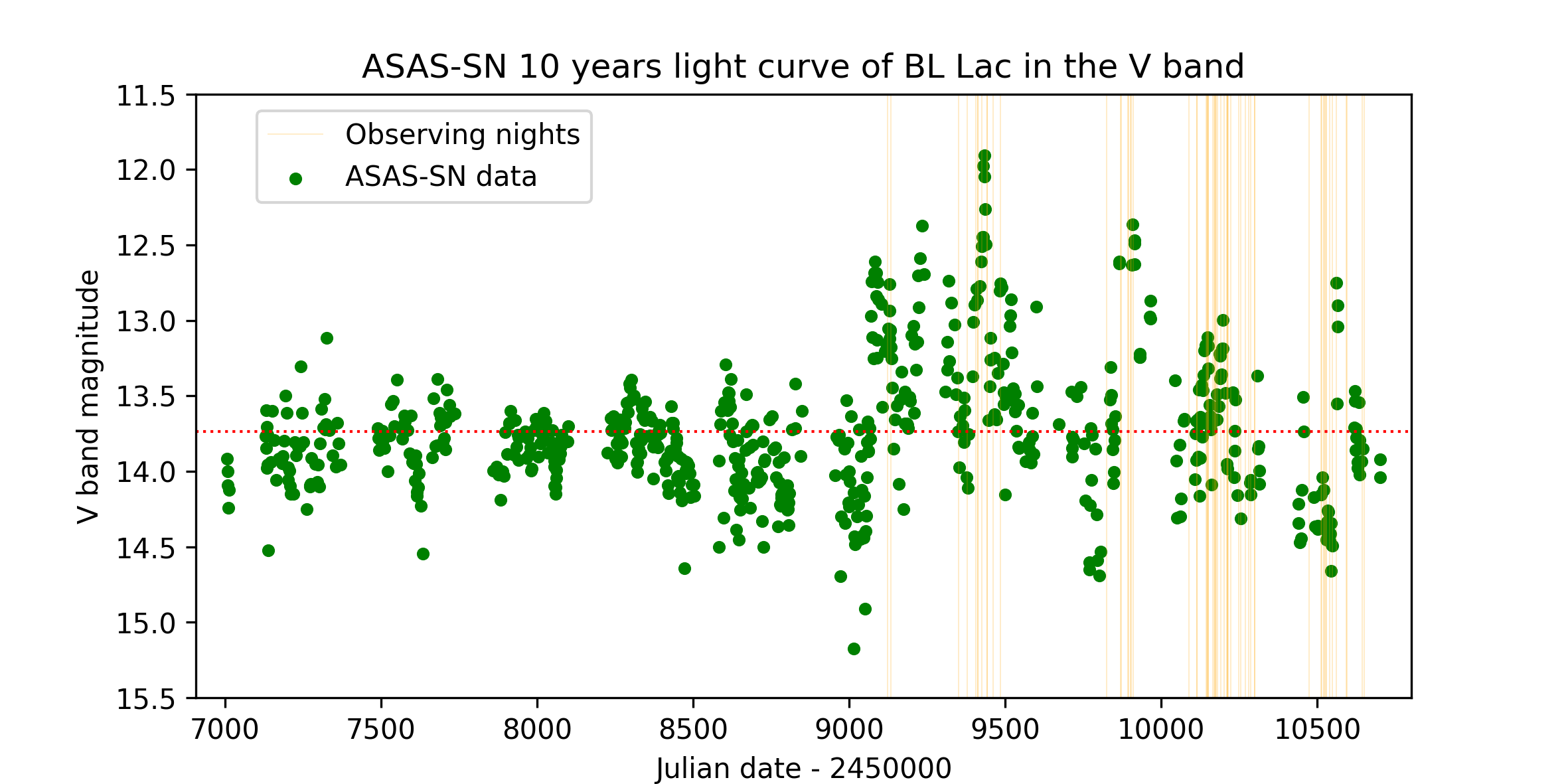}
\caption{Long-term (10-year) LC of BL Lacertae (green dots) based on ASAS-SN $V$-band measurements, with the survey’s $g$-band data converted to the $V$ band using an appropriate offset. Solid vertical lines mark the Julian dates of the 62 observing nights used in this study. The dotted horizontal line indicates the average $V$-band magnitude (13.737\,mag) over the entire LC spanning 2014 December to 2025 January.}
\label{fig:LTV_ASAS_SN}
\end{figure}

It is worth noting that our observing nights span all brightness states of BL Lacertae~-- low, intermediate, and high~-- over the four year span, and include coverage of the historical peak ($R = 11.14$) reported in August 2021 by \citet{2023MNRAS.522..102R}. Fig.~\ref{fig:LTV_ASAS_SN} displays the distribution of our 62 observing nights over a 10-year (December 2014–January 2025) long-term LC extracted from the All-Sky Automated Survey for Supernovae (ASAS-SN; \citealt{Hart2023}) in the $V$ band. The survey’s $g$-band measurements of BL Lacertae were converted to the $V$ band using an offset derived from the median difference of a comparison star in the same field with overlapping $V$- and $g$-band observations. Over the full 10-year span, the mean $V$-band magnitude of BL Lacertae is 13.737\,mag. The results of our intraday flux and spectral variability analysis, based on these observations, are presented and discussed in the following subsections.

%%%%%%%%%%%%%%%%%%%%%%%%%%%%%%%%%%%%%%%%%%%%%%%%%%%%%%%%%%%%%%%%%%%%%%%%%%%%%%%%%%%%%%%%

\subsection{Intraday flux variability}
\label{sec:idv}
%\textcolor{red}{THE following paragraph is partially rewritten:}
To assess the presence of IDV, we employ the power-enhanced $F$-test and the nested ANOVA test, as described in Sections~\ref{sec:F-test} and \ref{sec:anova}, respectively. These two methods were applied to 117 LCs obtained over 62 nights and across multiple optical bands. The results of the variability analysis are summarized in Table~\ref{tab:var_results} (Appendix~\ref{app:a}). For a given band, a LC is classified as variable (V) only when both statistical tests indicate variability; otherwise, it is considered non-variable (NV; see `Band status' column of Table~\ref{tab:var_results}). In the `Night status' column of Table~\ref{tab:var_results}, V, PV, and NV denote LCs of variable, probably variable, and non-variable status, respectively. For nights with observations in multiple photometric bands, the overall variability status was determined based on the results from all available bands and the `Night status' is written  in each corresponding row in the table. A night was classified as V if all or the majority of the observed bands exhibited variability. If exactly half of the bands were variable while the other half were not, the night was designated as PV. Conversely, if all or most bands showed no variability, the night was considered NV. Finally, we note that variability status of LCs and, hence, DC estimate based on it depend on the variability detection test(s) used (e.g. more vs less conservative); discussion on this topic, however, is beyond the scope of this paper \citep[see in this context][and references therein] {2017MNRAS.467..340Z, 2020MNRAS.498.3013Z}. 

Based on this classification, out of the 62 monitoring nights, 44 nights show clear variability (V status), 8 nights show probable variability (PV status), and 10 nights show no variability (NV status). This indicates that 71\,per cent of the observing nights exhibit variability in their LCs, increasing to 84\,per cent when probable variable nights are included.  
Across all observing bands, 88 of the 117 LCs (75\,per cent) are classified as V, while the remaining 29 are categorized as NV. \citet{Agarwal2025} found variability in 39 of 53 $BVRI$-band LCs (74\,per cent) for the period mid-July to mid-September in 2020.

\begin{figure}
\centering
\includegraphics[width=0.5\textwidth,clip=true]{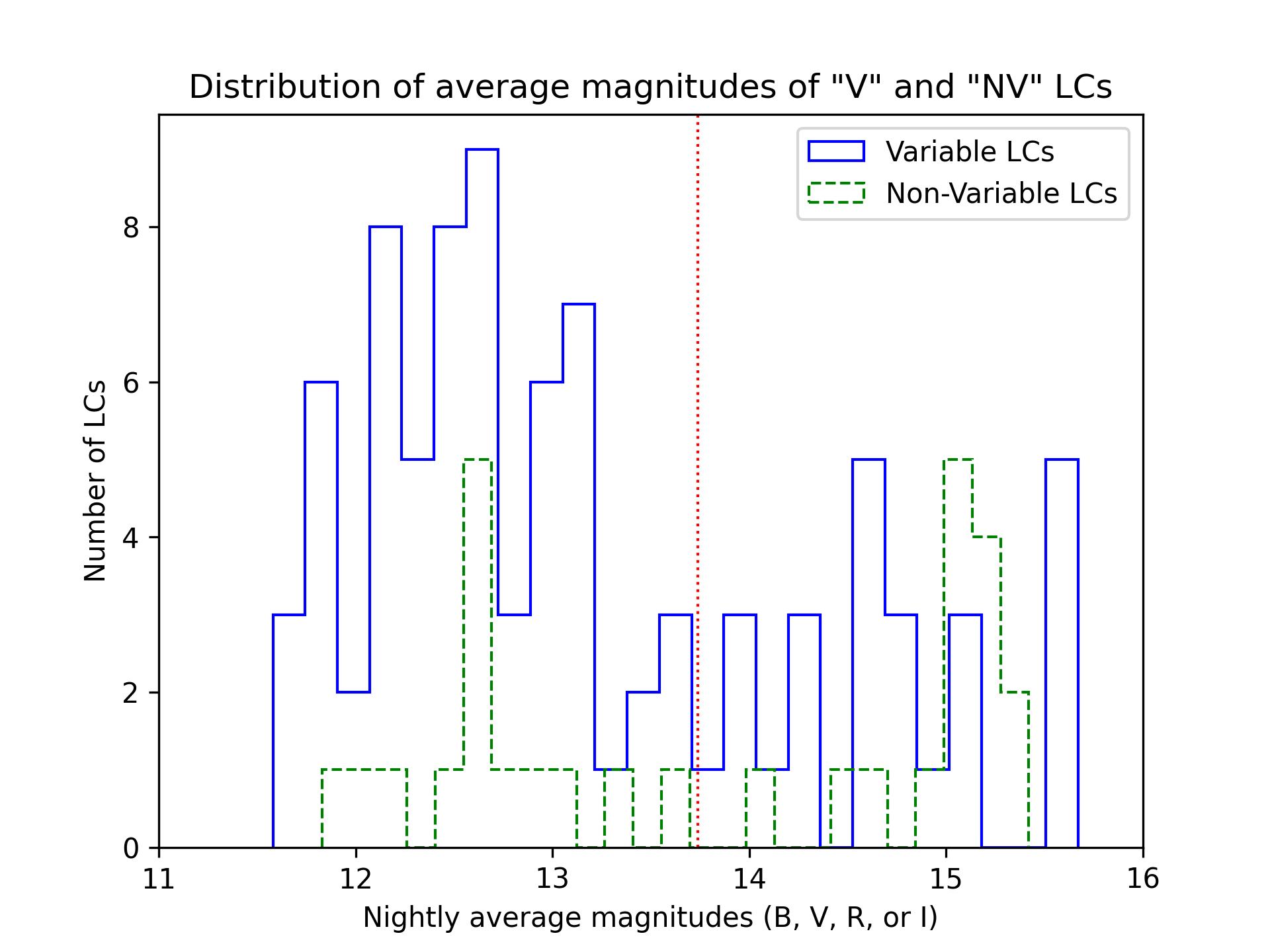}
\caption{Distribution of nightly average magnitudes for variable LCs (blue solid) and non-variable LCs (green dashed). The vertical dotted line denotes the 10-year average $V$-band magnitude from the ASAS-SN data.}
\label{fig:Hist_BL_BVRI_mag_Var_status}
\end{figure}

As the brightness of BL Lacertae has exhibited pronounced long-term variations in recent years~-- rising and decaying phases clearly visible in the ASAS-SN LC in Fig.~\ref{fig:LTV_ASAS_SN}~-- it is useful to identify whether each LC was obtained during a relatively low or high brightness state. As an approximate classification, we compare the nightly average magnitude of each LC (in the $B$, $V$, $R$, or $I$ bands) with the 10-year mean $V$-band magnitude derived from the ASAS-SN data. This comparison indicates that roughly two-thirds of our LCs were obtained during a high state, that is, brighter than the long-term average magnitude. Fig.~\ref{fig:Hist_BL_BVRI_mag_Var_status} shows histograms of the nightly average magnitudes for variable and non-variable LCs. The larger number of the high state LCs compared to the low state ones does not allow us to investigate the dependence of the LC variability status on the BL Lacertae brightness state.

The calibrated $B$-, $V$-, $R$-, and $I$-band LCs of BL Lacertae are shown in Fig.~\ref{fig:all_IDV_LC} (Appendix~\ref{app:a}), plotted in blue, green, red, and gray, respectively. Each panel indicates the observation date, the available photometric bands, and the variability status for that night. Visual inspection of these LCs reveals strong consistency between the statistical variability classifications and the qualitative appearance of the curves. In addition to smoothly varying LCs exhibiting monotonic rises or decays, several nights display more complex behaviour, including single or multiple flaring events and quasi-periodic features. These intriguing cases merit further analysis, which is pursued in a subsequent subsection.

For each LC in each band that is classified as variable, we estimated the corresponding variability amplitude using the method described in Section~\ref{sec:amplitude}; the resulting values are listed in Table~\ref{tab:var_results}. The variability amplitudes of the individual LCs (in the $B$, $V$, $R$, and $I$ bands) range from 3.0 to 44.5\,per cent. 
Notably, the maximum amplitude of 44.5\,per cent was detected in the $B$-band LC on 2023 July 21, during a flaring episode whose peak was below the historical maximum recorded in 2021 August. Comparable maximum variation amplitudes of $\sim$43\,per cent were reported by \citet{Agarwal2025} and \citet{Yuan2023} for the $R$-band LC on 2020 September 21 and the $r$-band LC on 2020 September 8, respectively. A higher amplitude of $\sim$64\,per cent in the $B$ band was observed on 2020 August 26 by \citet{Agarwal2023}. This date is earlier than our observational campaign that began on 2020 September 30.

\begin{figure}
\centering
\includegraphics[width=0.5\textwidth,clip=true]{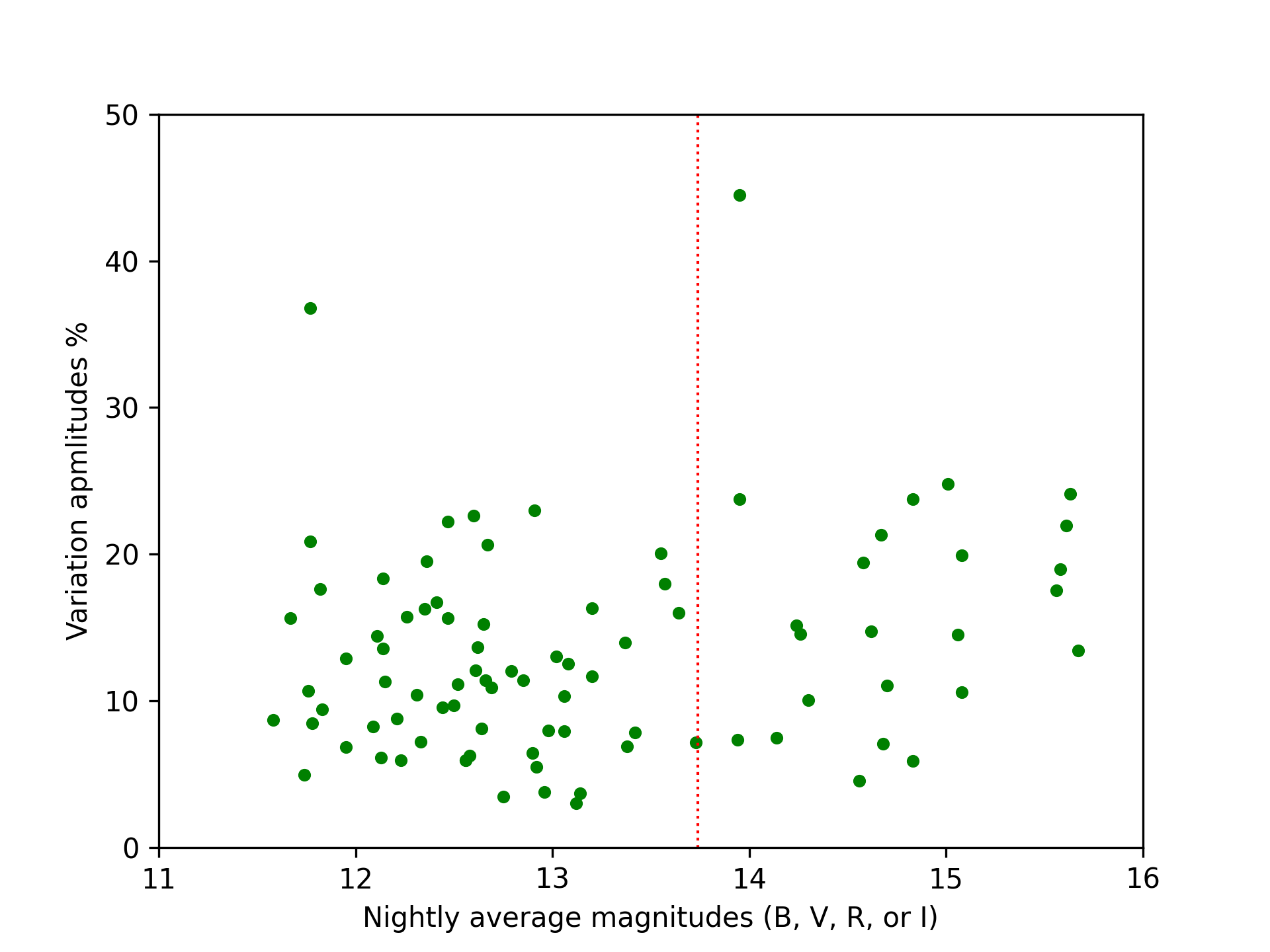}
\caption{Variability amplitudes of the variable LCs plotted against the nightly average magnitudes in the $B$, $V$, $R$, and $I$ bands. The vertical dotted line indicates the 10-year average $V$-band magnitude from the ASAS-SN data.}
\label{fig:Amplitudes_ave_mags}
\end{figure}

Fig.~\ref{fig:Amplitudes_ave_mags} presents the distribution of variability amplitudes as a function of nightly average magnitudes for all variable LCs. Our data suggest no dependence of the variability amplitude $A$ on the brightness state (Fig.~\ref{fig:Amplitudes_ave_mags}). Furthermore, for 15 nights out of 23 we found an increase of the amplitude with increasing frequency. The remaining nights exhibit more complex behaviour, with amplitudes sometimes decreasing in one band before rising again. Such patterns have been reported in previous IDV campaigns \citep[e.g.,][]{Kalita2023}.

%%%%%%%%%%%%%%%%%%%%%%%%%%%%%%%%%%%%%%%%%%%%%%%%%%%%%%%%%%%%%%%%%%%%%%%%%%%%%%%%%%%%%%%%

\subsection{Duty cycle}

Based on the variability tests for each night and band listed in Table~\ref{tab:var_results}, we detected variability in 24/39, 8/9, 30/33, and 26/36 LCs in the $B$, $V$, $R$, and $I$ bands, respectively. We note that a single LC is marked as V only if both statistical tests detect variability, as described above. 
The corresponding DCs of each band were calculated using the standard definition described in Section~\ref{sec:DC}, and the results are presented in Table~\ref{tab:DC_results}. The DC values differ across bands, with $B$, $V$, $R$, and $I$ bands showing 46, 89, 94, and 58\,per cent, respectively. These variations likely reflect differences in observing nights and the brightness states of the source. BL Lacertae has exhibited pronounced variability in recent years, as evident from the long-term LC in Fig.~\ref{fig:LTV_ASAS_SN}. We also note that our observing campaigns did not use the same filters each night, as indicated by the total number of LCs in each band and detailed in the observation log (Table~\ref{tab:obs_log}).

Another factor that may affect the DC is the duration of each LC. To investigate this, we divided the LCs in each band into two subsamples: those with monitoring durations of 3\,h or more, and those with less than 3\,h. The DCs for these subsamples were calculated and are listed in Table~\ref{tab:DC_results}. Comparing the longer-duration subsample with the full sample, we find that increasing the monitoring duration generally increases the DC in the $B$, $V$, and $I$ bands, while it decreases slightly in the $R$ band. In all bands except $R$, the DCs of the longer-duration subsample are higher than those of the shorter-duration subsample. The deviation observed in the $R$ band may result from the presence of many individual $R$-band LCs distributed over the entire observing period, which sample different brightness states. 

Finally, the DC may vary depending on the variability detection test(s) applied (as we mentioned in brief in Section~\ref{sec:idv}) and on the way used to calculate the DC itself \citep[see e.g.][]{2021A&A...645A.137A,2025MNRAS.541..732M}.

\begin{table}
\centering
\tabcolsep=4pt
\caption{Duty cycle results of BL Lacertae for each band.}
\label{tab:DC_results}
\begin{tabular}{ccccccc}
\hline
Band & \multicolumn{2}{c}{All $t_{\rm obs}$} & \multicolumn{2}{c}{$t_{\rm obs} \ge 3$~h} & \multicolumn{2}{c}{$t_{\rm obs} < 3$~h} \\
\cmidrule(lr){2-3} \cmidrule(lr){4-5} \cmidrule(lr){6-7}
    & $N$ & DC [per cent] & $N$ & DC [per cent] & $N$ & DC [per cent] \\
(1) & (2) & (3) & (4) & (5) & (6) & (7) \\
\hline
$B$ & 39 & 46 & 23 &  80 & 16 & 25 \\
$V$ &  9 & 89 &  3 & 100 &  6 & 87 \\
$R$ & 33 & 94 & 15 &  85 & 18 & 96 \\
$I$ & 36 & 58 & 22 &  80 & 14 & 44 \\
\hline
\end{tabular}
    \flushleft{{\em Note.} Columns~2 and 3 list the total number of LCs and the corresponding DCs for all observations. Columns~4 and 5 provide the number of LCs and DCs for observations with durations $t_{\rm obs} \ge 3$~h, while columns~6 and 7 correspond to observations with $t_{\rm obs} < 3$~h.}
\end{table}

%%%%%%%%%%%%%%%%%%%%%%%%%%%%%%%%%%%%%%%%%%%%%%%%%%%%%%%%%%%%%%%%%%%%%%%%%%%%%%%%%%%%%%%%

\subsection{Variability time-scales}

\begin{figure*}
\centering
\includegraphics[width=\textwidth,clip=true]{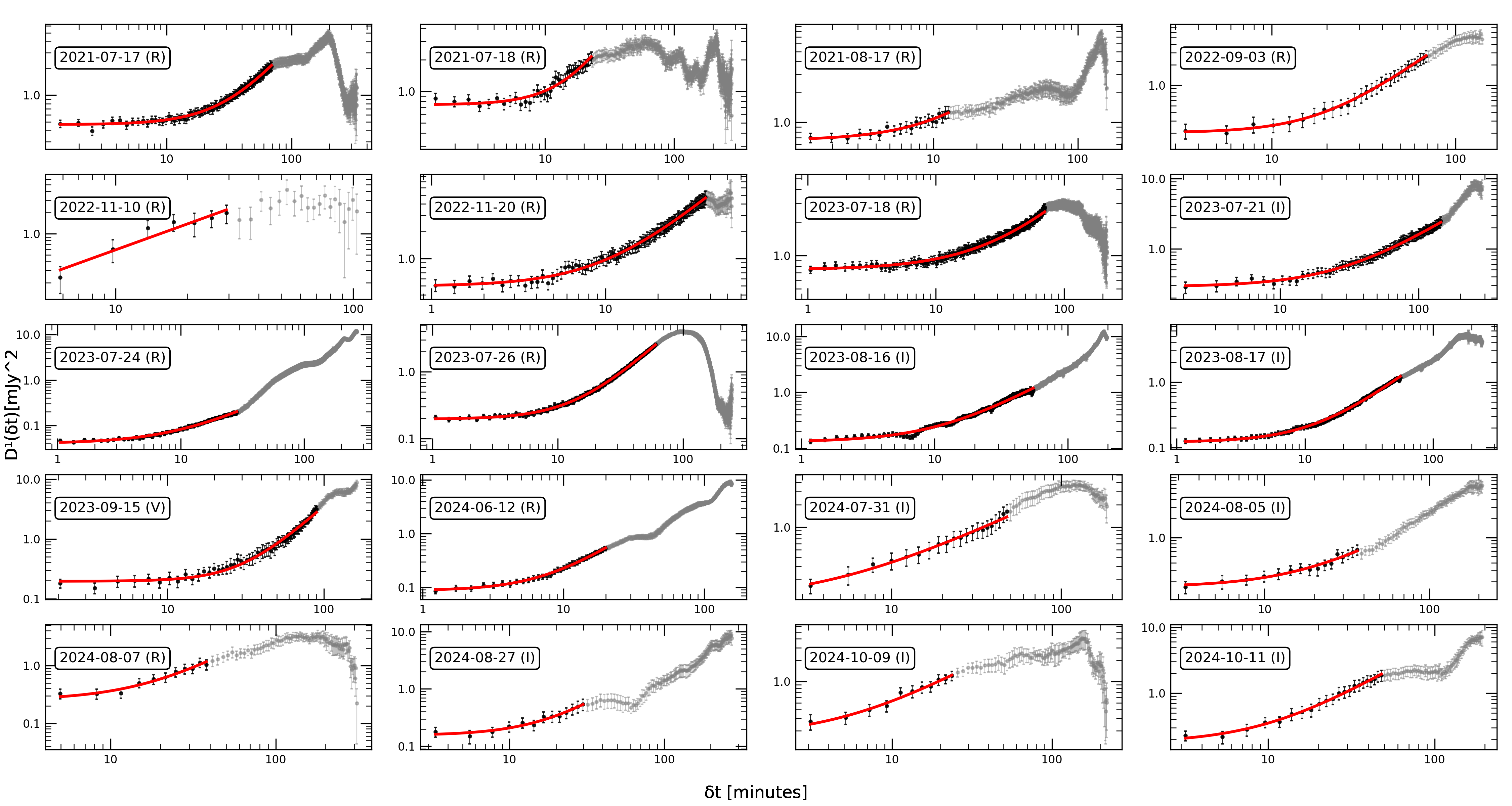}
\caption{Structure function plots of the selected variable intraday LCs. The observation dates and bands are indicated at the top of each each panel.}
\label{fig:sf}
\end{figure*}

\begin{table}
%\scriptsize
\caption{The results of SF analysis.}
\label{tab:sf_results}
\centering
\begin{tabular}{lccc}
\toprule
Date  &  Band & $\beta$ & $\delta t_{to}$ \\
\midrule
2021 Jul 17 & $R$ & 1.71 $\pm$ 0.07 & 69.2 \\[0.1cm]
2021 Jul 18 & $R$ & 2.00 $\pm$ 0.30 & 22.8 \\[0.1cm]
2021 Aug 17 & $R$ & 1.37 $\pm$ 0.61 & 12.7 \\[0.1cm]
2022 Sep 03 & $B$ & 2.10 $\pm$ 0.25 & 48.8 \\
           & $V$ & 1.89 $\pm$ 0.19 & 66.9 \\
           & $R$ & 1.97 $\pm$ 0.20 & 69.2 \\[0.1cm]
2022 Nov 10 & $B$ & 0.97 $\pm$ 0.91 & 33.2 \\
           & $V$ & 1.48 $\pm$ 0.58 & 37.1 \\
           & $R$ & 1.22 $\pm$ 0.86 & 29.3 \\[0.1cm]
2022 Nov 20 & $R$ & 1.64 $\pm$ 0.07 & 37.5 \\[0.1cm]
2023 Jul 18 & $B$ & 1.47 $\pm$ 0.07 & 74.3 \\
           & $R$ & 1.14 $\pm$ 0.04 & 70.5 \\[0.1cm]
2023 Jul 21 & $B$ & 1.87 $\pm$ 0.08 & 150.1 \\
           & $I$ & 1.27 $\pm$ 0.06 & 144.6 \\[0.1cm]
2023 Jul 24 & $R$ & 1.28 $\pm$ 0.06 & 28.4 \\[0.1cm]
2023 Jul 26 & $R$ & 1.72 $\pm$ 0.02 & 60.3 \\[0.1cm]
2023 Aug 16 & $I$ & 1.32 $\pm$ 0.03 & 54.9 \\[0.1cm]
2023 Aug 17 & $B$ & 1.47 $\pm$ 0.14 & 49.3 \\
           & $V$ & 2.38 $\pm$ 0.14 & 89.8 \\
           & $I$ & 1.50 $\pm$ 0.02 & 55.6 \\[0.1cm]
2023 Sep 15 & $B$ & 1.80 $\pm$ 0.23 & 37.7 \\[0.1cm]
% 2023-09-15 & $I$ & 1.421 $\pm$ 1.46 & 32.2 \\
2024 Jun 12 & $R$ & 1.72 $\pm$ 0.12 & 19.9 \\[0.1cm]
2024 Jul 31 & $B$ & 1.30 $\pm$ 0.19 & 59.0 \\
           & $I$ & 1.07 $\pm$ 0.26 & 47.9 \\[0.1cm]
2024 Aug 05 & $I$ & 1.45 $\pm$ 0.49 & 36.5 \\[0.1cm]
2024 Aug 07 & $B$ & 2.03 $\pm$ 0.68 & 47.9 \\
           & $R$ & 1.57 $\pm$ 0.56 & 38.0 \\
           & $I$ & 1.12 $\pm$ 0.68 & 47.9 \\[0.1cm]
% 2024-08-07 & $V$ & 3.418 $\pm$ 2.43 & 38.0 \\
2024 Aug 27 & $I$ & 1.78 $\pm$ 0.71 & 29.9 \\[0.1cm]
2024 Oct 09 & $B$ & 1.03 $\pm$ 0.30 & 23.6 \\
           & $I$ & 1.24 $\pm$ 0.44 & 23.6 \\[0.1cm]
2024 Oct 11 & $B$ & 1.42 $\pm$ 0.43 & 32.8 \\
           & $I$ & 1.48 $\pm$ 0.20 & 47.9 \\
\bottomrule
\end{tabular}
\end{table}

\begin{figure}
\centering
\includegraphics[width=0.45\textwidth,clip=true]{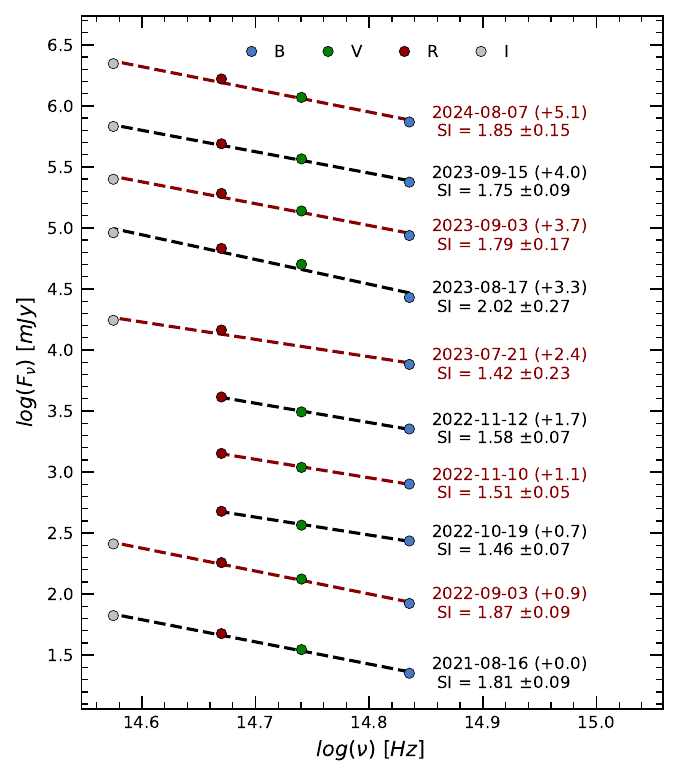}
\caption{Optical SEDs for 10 variable nights with three or four bands of data. The blue, green, red, and gray circles correspond to the $B$, $V$, $R$, and $I$ bands, respectively. Dates and SIs (slopes) are given beside each SED. The best fit lines are overplotted  in dashed lines} on the SEDs.
\label{fig:SED_average}
\end{figure}

We applied structure function analysis described in Section~\ref{sec:Tscale} to intraday LCs that exhibited statistically variability, using the code described in \cite{2018MNRAS.478.2557G}. The SF plots of the selected LCs exhibiting well-defined features are presented in Fig.~\ref{fig:sf}. The first-order SF was computed for each night in order to examine variability behaviour in the time-lag domain. For each SF profile, the rising portion above the noise-dominated regime was identified and fitted with a power-law model of the form $SF(\tau) \propto \tau^{\,\beta}$. The fitting range was selected where the SF exhibited stable power-law behaviour in log–log space, in accordance with the methods of \citet{1992ApJ...396..469H}. The lower bound of the fit, $\tau_{\rm min}$, corresponds to the beginning of the rising SF profile, while the upper bound, $\tau_{\rm char}$, marks the point at which the profile turns over. For the analysed nights $\tau_{\rm char}$ lies in the range 22 to 300\,min.

The fitted power-law slopes $\beta$ range from  0.97 to 2.38, indicating varying degrees of temporal correlation in the intraday flux variations Table~\ref{tab:sf_results}. These values are in good correspondence with the results of \citet{Agarwal2023}. SF slopes in the range of 0.5–2 have commonly been reported in optical, intra-night studies of blazars \citep{1992ApJ...396..469H, 2010MNRAS.404..931E}.
These slopes are generally interpreted as signatures of stochastic, correlated variability, rather than of deterministic processes. For nights with multi-band observations, the SF slopes derived from different bands on the same night are generally consistent with each other. This suggests that the variability behaviour is similar across filters on intraday time-scales. The SF results provide a complementary time-domain characterization that broadly aligns with the trends inferred from the DCF analysis.

%%%%%%%%%%%%%%%%%%%%%%%%%%%%%%%%%%%%%%%%%%%%%%%%%%%%%%%%%%%%%%%%%%%%%%%%%%%%%%%%%%%%%%%%

\subsection{Spectral energy distribution and spectral variation}

% \onecolumn
\begin{table*}
%\scriptsize
\caption{SED and spectral variability results for the variable nights of BL Lacertae with multi-band observations. The first three columns present the observation date, the available photometric bands, and the average spectral index derived from their SEDs. The results of the linear fitting of individual spectral indices (SI) against the intermediate flux are provided in the subsequent columns. Nights exhibiting a BWB trend are indicated in the colour trend column.}
\label{tab:Spec_Var}
\centering
% \begin{tabular}{ccccccccccccc}
\begin{tabular}{ccccccccc}
\toprule
 Date  &  Bands & $\langle \rm SI \rangle_{wm}$ &  \multicolumn{6}{c}{SI vs Flux}\\
\cmidrule(lr){4-9} 
 &   &   &  Band  & $m$ & $c$ & $r$ & $p$ & Color trend \\

\midrule
2022 Sep 03 & $BVRI$ & 1.872 $\pm$ 0.011 & $(V+R)/2$ & $-0.077 \pm 0.007$ & $3.393$ & $-0.804$ & $2.1 \times 10^{-15}$ & BWB\\ 
2022 Oct 19 & $BVR$ & 1.480 $\pm$ 0.004 & $V$ & $-0.003 \pm 0.002$ & $1.666$ & $-0.252$ & $2.8 \times 10^{-1}$ & ---\\ 
2022 Nov 10 & $BVR$ & 1.521 $\pm$ 0.007 & $V$ & $-0.007 \pm 0.005$ & $2.123$ & $-0.344$ & $1.5 \times 10^{-1}$ & ---\\ 
2022 Nov 12 & $BVR$ & 1.622 $\pm$ 0.010 & $V$ & $-0.014 \pm 0.006$ & $2.448$ & $-0.533$ & $3.4 \times 10^{-2}$ & BWB\\ 
2023 Jun 20 & $BI$ & 1.674 $\pm$ 0.010 & $(B+I)/2$ & $+0.008 \pm 0.012$ & $1.516$ & $+0.064$ & $5.2 \times 10^{-1}$ & ---\\ 
2023 Jul 18 & $BR$ & 1.996 $\pm$ 0.006 & $(B+R)/2$ & $-0.019 \pm 0.009$ & $2.511$ & $-0.163$ & $3.3 \times 10^{-2}$ & ---\\ 
2023 Jul 21 & $BRI$ & 1.439 $\pm$ 0.027 & $R$ & $-0.006 \pm 0.004$ & $1.773$ & $-0.188$ & $1.4 \times 10^{-1}$ & ---\\ 
2023 Aug 13 & $BI$ & 1.552 $\pm$ 0.009 & $(B+I)/2$ & $+0.031 \pm 0.007$ & $0.703$ & $+0.321$ & $3.1 \times 10^{-5}$ & ---\\ 
2023 Aug 16 & $BI$ & 1.942 $\pm$ 0.004 & $(B+I)/2$ & $-0.013 \pm 0.002$ & $2.379$ & $-0.455$ & $2.9 \times 10^{-8}$ & ---\\ 
2023 Aug 17 & $BVRI$ & 2.022 $\pm$ 0.027 & $(V+R)/2$ & $+0.000 \pm 0.002$ & $2.018$ & $+0.008$ & $9.4 \times 10^{-1}$ & ---\\ 
2023 Aug 20 & $BI$ & 1.714 $\pm$ 0.005 & $(B+I)/2$ & $-0.021 \pm 0.008$ & $2.366$ & $-0.378$ & $8.8 \times 10^{-3}$ & ---\\ 
2023 Sep 03 & $BVRI$ & 1.791 $\pm$ 0.046 & $(V+R)/2$ & $-0.012 \pm 0.020$ & $2.175$ & $-0.184$ & $5.7 \times 10^{-1}$ & ---\\ 
2023 Sep 15 & $BVRI$ & 1.758 $\pm$ 0.009 & $(V+R)/2$ & $-0.006 \pm 0.003$ & $2.017$ & $-0.214$ & $4.9 \times 10^{-2}$ & ---\\ 
2023 Sep 23 & $BI$ & 1.924 $\pm$ 0.003 & $(B+I)/2$ & $-0.027 \pm 0.006$ & $2.516$ & $-0.457$ & $1.8 \times 10^{-5}$ & ---\\ 
2023 Oct 06 & $BI$ & 1.814 $\pm$ 0.003 & $(B+I)/2$ & $-0.015 \pm 0.002$ & $2.223$ & $-0.499$ & $1.6 \times 10^{-9}$ & BWB\\ 
2023 Nov 06 & $BI$ & 2.079 $\pm$ 0.004 & $(B+I)/2$ & $-0.091 \pm 0.034$ & $3.294$ & $-0.260$ & $8.4 \times 10^{-3}$ & ---\\ 
2024 Jul 31 & $BI$ & 1.707 $\pm$ 0.003 & $(B+I)/2$ & $-0.041 \pm 0.004$ & $2.548$ & $-0.754$ & $6.4 \times 10^{-16}$ & BWB\\ 
2024 Aug 05 & $BI$ & 1.847 $\pm$ 0.003 & $(B+I)/2$ & $-0.108 \pm 0.012$ & $3.227$ & $-0.707$ & $4.7 \times 10^{-13}$ & BWB\\ 
2024 Aug 07 & $BVRI$ & 1.853 $\pm$ 0.018 & $(V+R)/2$ & $-0.094 \pm 0.016$ & $2.906$ & $-0.570$ & $2.1 \times 10^{-7}$ & BWB\\ 
2024 Aug 17 & $BI$ & 1.797 $\pm$ 0.005 & $(B+I)/2$ & $-0.112 \pm 0.027$ & $3.233$ & $-0.451$ & $8.8 \times 10^{-5}$ & ---\\ 
2024 Aug 27 & $BI$ & 1.860 $\pm$ 0.003 & $(B+I)/2$ & $-0.025 \pm 0.012$ & $2.174$ & $-0.197$ & $3.6 \times 10^{-2}$ & ---\\ 
2024 Oct 09 & $BI$ & 1.535 $\pm$ 0.003 & $(B+I)/2$ & $-0.011 \pm 0.002$ & $2.136$ & $-0.557$ & $8.2 \times 10^{-9}$ & BWB\\ 
2024 Oct 11 & $BI$ & 1.767 $\pm$ 0.003 & $(B+I)/2$ & $-0.010 \pm 0.003$ & $2.219$ & $-0.380$ & $2.9 \times 10^{-4}$ & ---\\ 
\bottomrule
\end{tabular}
\end{table*}
% \twocolumn

For nights with multi-band IDV data of BL Lacertae, we computed their average spectral index (SI) from their SEDs. To obtain reliable SEDs, the data points for available filters on each night were cross-matched within a 2\,min interval. These simultaneous or quasi-simultaneous measurements were averaged in each band. 
For nights with three or four band data, we construct their SEDs by plotting the logarithmic flux density $\log(F_\nu)$ versus logarithmic frequency $\log(\nu)$. To compute the flux density, the calibrated magnitudes in the $B$, $V$, $R$, and $I$ bands were first corrected for Galactic extinction using the reddening values from \cite{Schlafly2011}, adopting an extinction law from \cite{Cardelli1989} with $R_V = 3.1$. Then, de-reddened magnitudes were converted to flux densities using standard zero-point values provided by \cite{1998A&A...333..231B}. Since the host galaxy contributes 1.297, 2.888, 4.229, and 5.903 mJy in the $B$, $V$, $R$, and $I$ bands, respectively, we subtracted 60\,per cent of these values from BL Lacertae fluxes, corresponding to the expected contamination for an aperture radius $8$ arcsec \citep{Raiteri2009,2023MNRAS.522..102R}.  

To derive the SI, we first constructed averaged SEDs for each night with data in at least three bands, as illustrated in Fig.~\ref{fig:SED_average}. The mean flux density measurements in the  $B$, $V$, $R$, and $I$ bands are shown as blue, green, red, and gray circles, respectively. These SEDs were then fitted with dashed lines using a linear regression of the form $y = mx + c$, where $y$ denotes $\log(F_\nu)$ and $x$ is $\log(\nu)$. The fitting was performed with the \texttt{linregress} function from the \textit{SciPy} Python module \citep{Virtanen2020}, yielding the slope ($m \equiv \rm SI$), intercept $c$, correlation coefficient $r$, and $p$-value $p$.
The observation date of each SED and its associated SI are reported in Table~\ref{tab:Spec_Var} and, where available, are also indicated in Fig.~\ref{fig:SED_average}. For nights with data in only two bands, the SI was calculated directly from the ratio of the flux densities to the corresponding frequency ratio. Overall, the spectral indices obtained from nights with two, three, or four bands range from 1.439 to 2.079, with a weighted mean of 1.783 and a weighted standard deviation of 0.156. 

Spectral variability in the BL Lacertae's variable IDV LCs can be examined by correlating the spectral indices with the fluxes after host galaxy correction. To this end, we adopt an intermediate flux derived from the available band data. For nights with two-band observations, the SI of each data pair is correlated with the mean flux, $(F_1 + F_2)/2$. For nights with three or four bands, the SI is correlated with (i) $F_2$ when three fluxes ($F_1, F_2, F_3$) are available, or (ii) the average $(F_2 + F_3)/2$ when four fluxes ($F_1, F_2, F_3, F_4$) are present.
Fig.~\ref{fig:SI_vs_flux} (Appendix~\ref{app:b}) presents the correlation between the spectral indices and the intermediate flux, together with linear fits derived using the \texttt{linregress} function. The resulting fit parameters are summarized in Table~\ref{tab:Spec_Var}. We regard a correlation as significant when the correlation coefficient is greater than 0.5 and the $p$-value is less than 0.01. When a significant correlation is present, a negative slope means a BWB trend, while a positive slope indicates a redder-when-brighter trend. According to the fitting results, ten out of the 23 variable nights display tight correlations and all of them show a BWB behaviour.

We additionally analyzed the spectral-index (SI) light curves using the mean observation durations of the intermediate optical bands, as illustrated in Fig.~\ref{fig:SI_vs_time} (Appendix~\ref{app:b}). The results indicate that the SI LCs do not exhibit a uniform trend of either increase or decrease over time. Instead, the observed behaviour includes phases of increasing trends, decreasing trends, a combination of both, or periods with no clear trend.

%%%%%%%%%%%%%%%%%%%%%%%%%%%%%%%%%%%%%%%%%%%%%%%%%%%%%%%%%%%%%%%%%%%%%%%%%%%%%%%%%%%%%%%%

%%%%%%%%%%%%%%%%%%%%%%%%%%%%%%%%%%%%%%%%%%%%%%%%%%%%%%%%%%%%%%%%%%%%%%%%%%%%%%%%%%%%%%%%

\subsection{Correlation analysis}
\label{sec:ccorr}

\begin{figure*}
\centering
\includegraphics[width=\textwidth,clip=true]{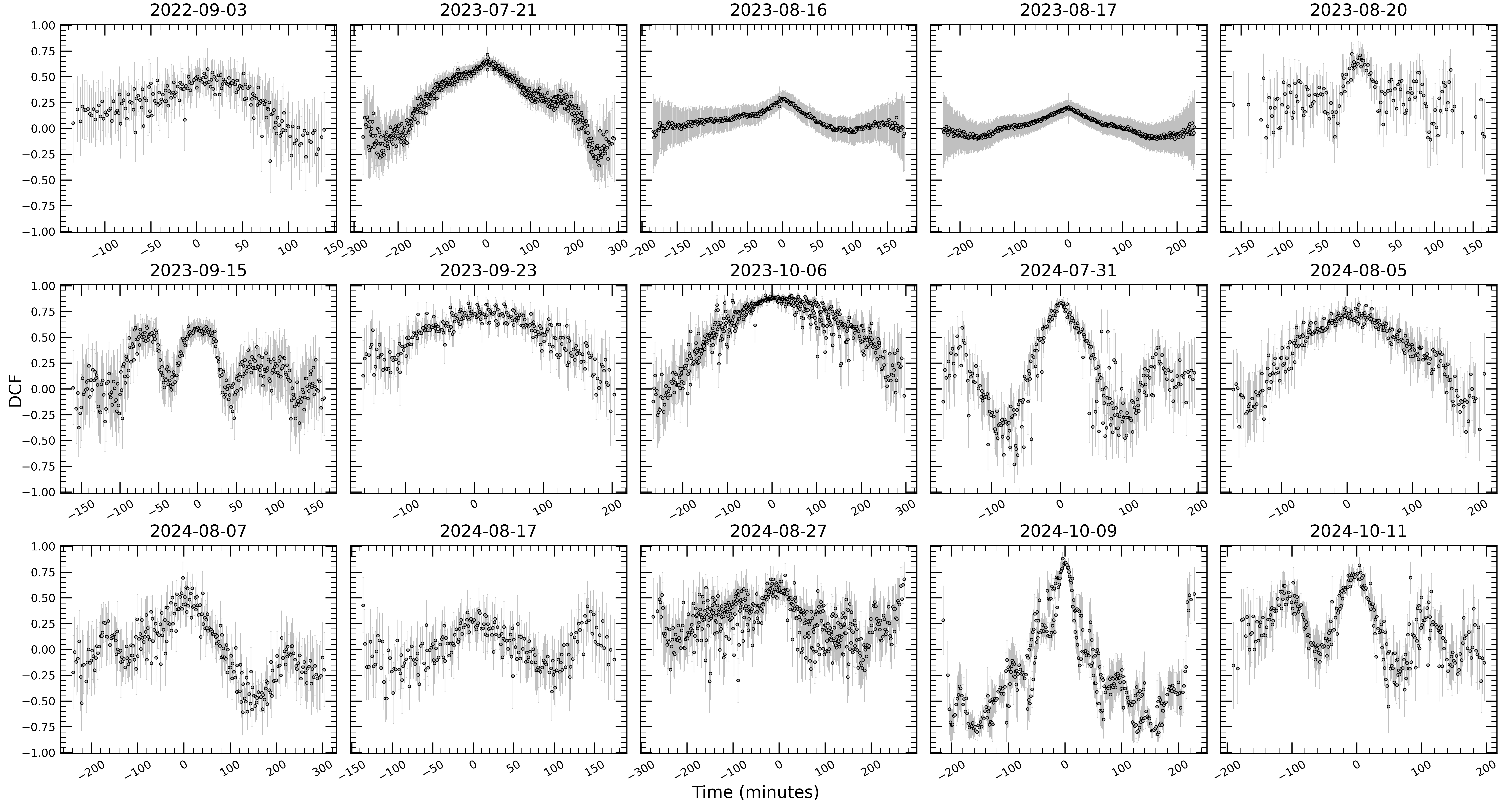}
\caption{Discrete correlation function plots for the LCs in the $B$ vs $I$ bands showing variability. The observation dates are indicated at the top of each each panel.}
\label{fig:dcf}
\end{figure*}

We applied the $z$DCF analysis to the intraday light curves in the $B$- and $I$-band for nights showing significant variability, using the pyZDCF implementation \citep{2022zndo...7253034J}. The DCF plots exhibiting clear profile are presented in Fig.~\ref{fig:dcf}.  The resulting DCFs exhibited clear correlation profiles, with dominant peaks indicating a strong temporal connection between flux variations in the $B$ and $I$ optical bands. In most cases, the DCF maxima are found at time lags consistent with zero within the associated uncertainties. This suggests that the $B$- and $I$-band variations are closely coupled and likely originate from the same or closely related emission regions. Where small non-zero lags are present, they are not statistically significant and may reflect subtle energy-dependent effects such as particle cooling, light-travel time differences or stratification within the emission region. However, given the uncertainties inherent in intraday datasets, these interpretations remain tentative, and the observed correlations primarily support a scenario of nearly simultaneous optical variability across bands.

Discrete correlation function profiles for some nights (e.g. 2023 September 15 and 2024 October 11) show more than one correlation peak. The presence of multiple peaks likely reflects the complex temporal structure of the intraday light curves rather than multiple distinct physical delays. Such behaviour can arise from repetitive flux variation, the superposition of variable components with comparable amplitudes, or the finite duration of the light curves. In this context, the strongest peak represents the dominant correlation time-scale, while the additional peaks are artifacts of correlated sub-structures within the same emission episode. Due to the uncertainties and limited time coverage of intraday observations, these secondary peaks are not considered to be physically distinct or independent lags. The multiple peaks may also indicate that optical variability is governed by a combination of processes operating on similar time-scales, such as localized disturbances propagating through a turbulent emission region. However, the current dataset does not allow these possibilities to be distinguished.

Previous observational studies have reported intraday QPOs in several blazars \citep{2021PASP..133g4101Y, 2021RAA....21..138Y, 2024MNRAS.533..120C}, including BL Lacertae \citep{2023ApJS..269...60Y}. We emphasize that the two nights displaying variability (with amplitudes of $\sim$24 and $\sim$16\,per cent, respectively) exhibit periodic behaviour of 46 and 101\,min on 9 and 11 October 2024. These periods coincide with a phase of intense very-high-energy (VHE) gamma-ray flaring activity reported by \citet{Majumdar2025}. In that study, the authors reported the detection of the second most energetic photon, with an energy of 175.7\,GeV, recorded by the Fermi Large Area Telescope (LAT) on 9 October 2024 from BL Lacertae, following the historically highest-energy photon of 238\,GeV reported by \citet{Prince2021}. Subsequently, the MAGIC telescopes detected a VHE gamma-ray flux comparable to that of the Crab Nebula above 250\,GeV on 10 October 2024 \citep{Paneque2024}. Their analysis revealed pronounced IDV, with flux changes by a factor of two on hour-long time-scales. Notably, our two observing nights fell one day before and one day after the flaring episode reported by \citet{Paneque2024}. While investigating periodic flaring across multiple wavelengths would be valuable, this is beyond the scope of the present work.

%%%%%%%%%%%%%%%%%%%%%%%%%%%%%%%%%%%%%%%%%%%%%%%%%%%%%%%%%%%%%%%%%%%%%%%%%%%%%%%%%%%%%%%%

%%%%%%%%%%%%%%%%%%%%%%%%%%%%%%%%%%%%%%%%%%%%%%%%%%%%%%%%%%%%%%%%%%%%%%%%%%%%%%%%%%%%%%%%

\subsection{Flares analysis}
\label{sec:flare:2}

\begin{figure*}
\centering
\includegraphics[width=0.326\textwidth,clip=true]{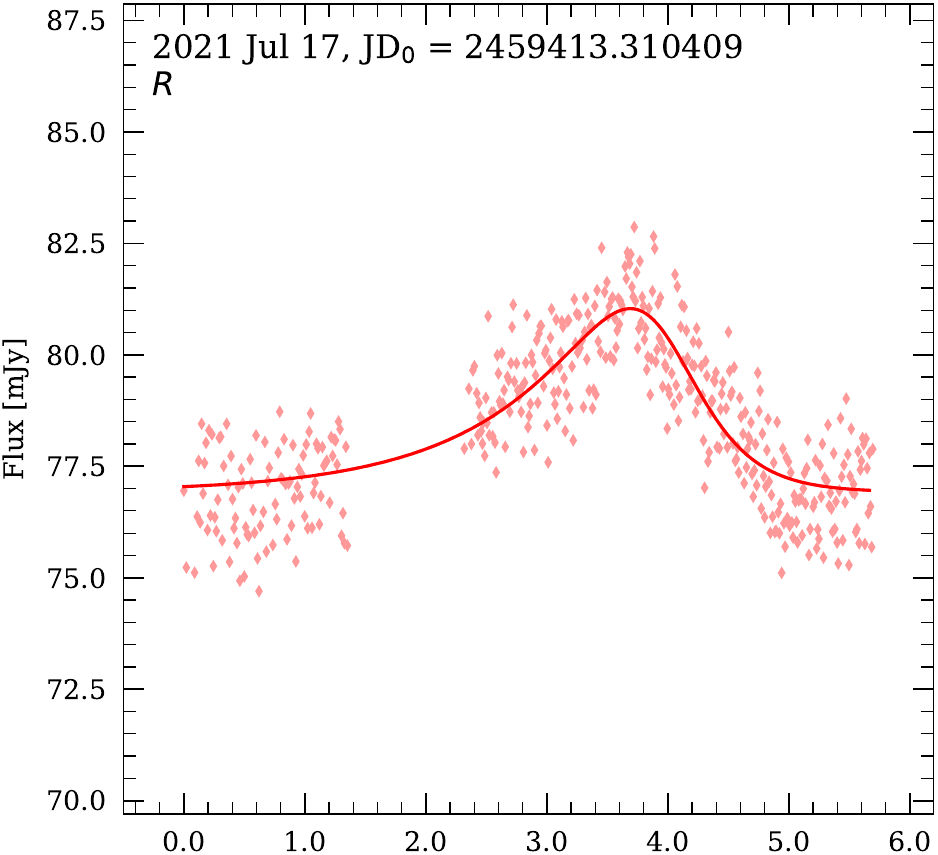}
\includegraphics[width=0.31\textwidth,clip=true]{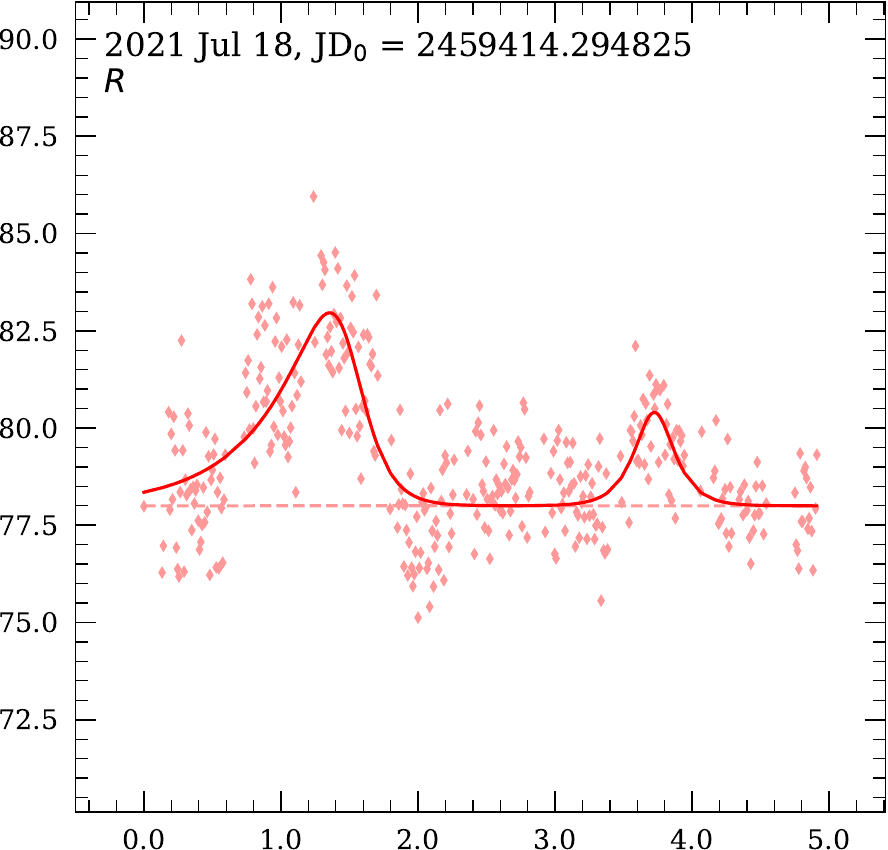}
\includegraphics[width=0.31\textwidth,clip=true]{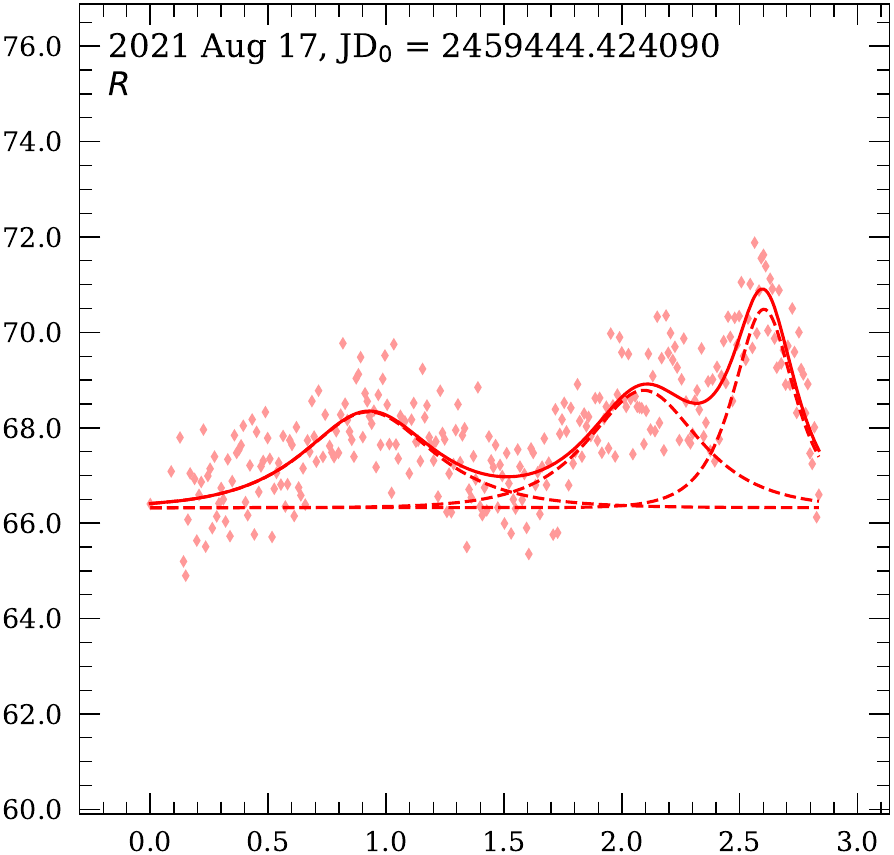}\vfill
\includegraphics[width=0.326\textwidth,clip=true]{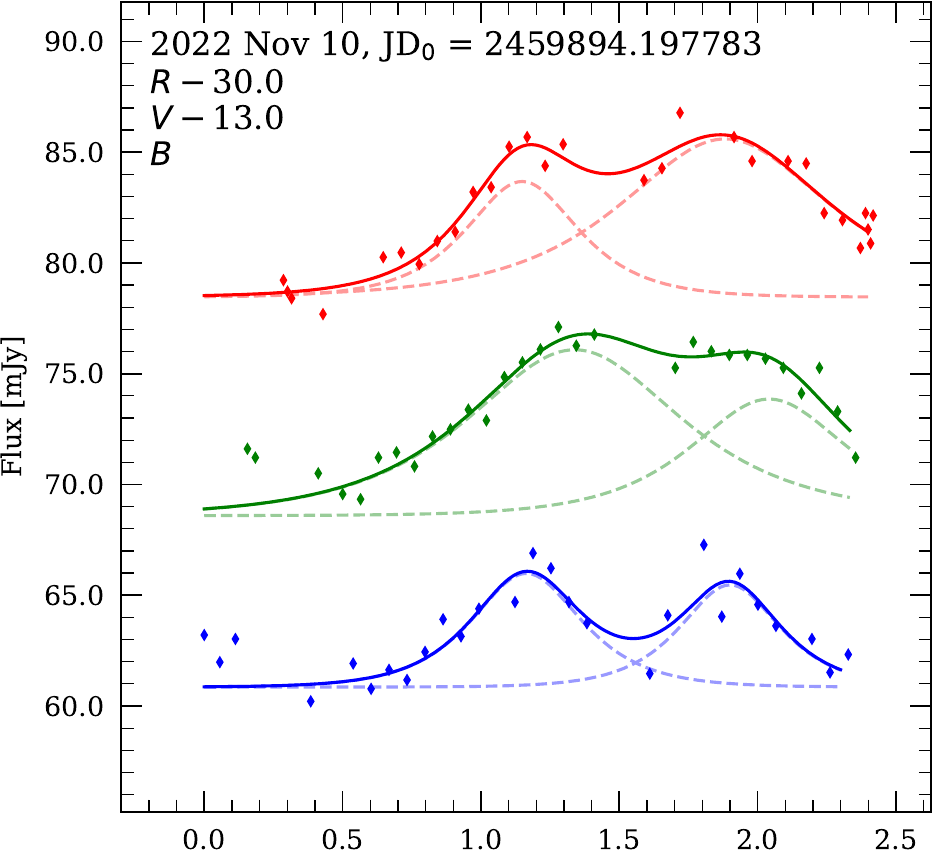}
\includegraphics[width=0.31\textwidth,clip=true]{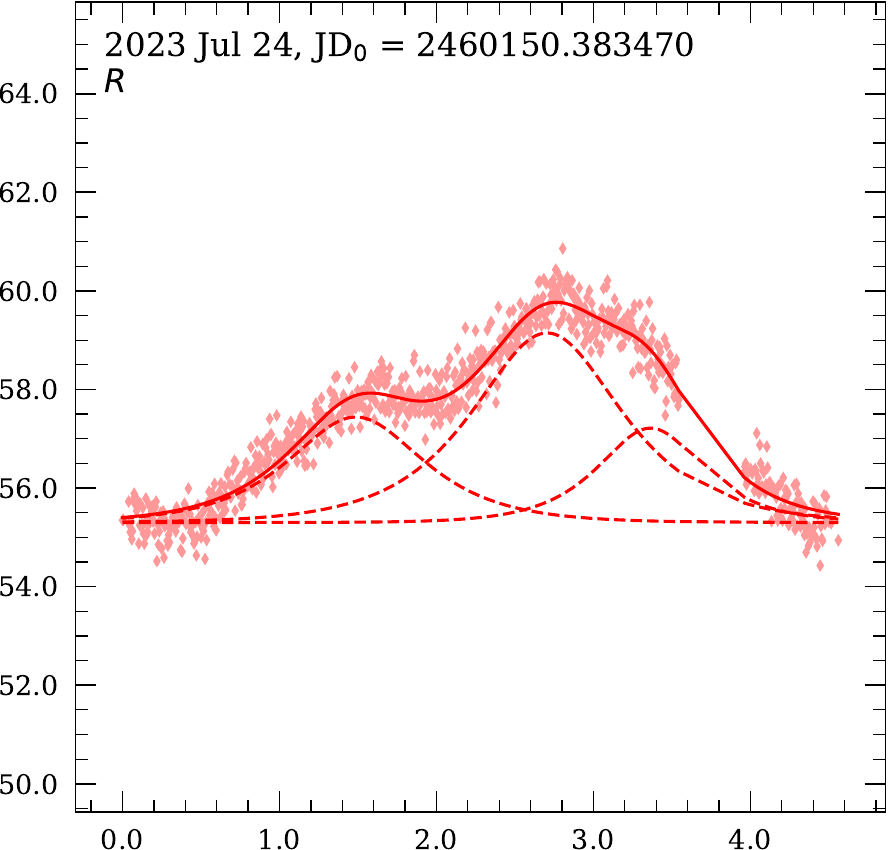}
\includegraphics[width=0.31\textwidth,clip=true]{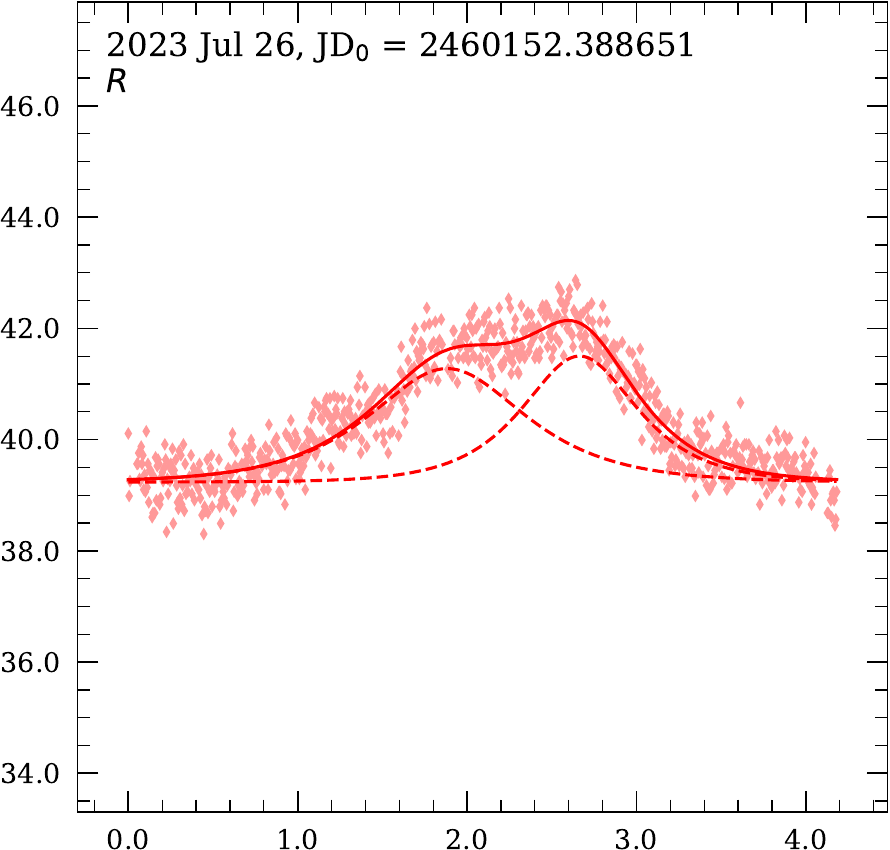}\vfill
\includegraphics[width=0.326\textwidth,clip=true]{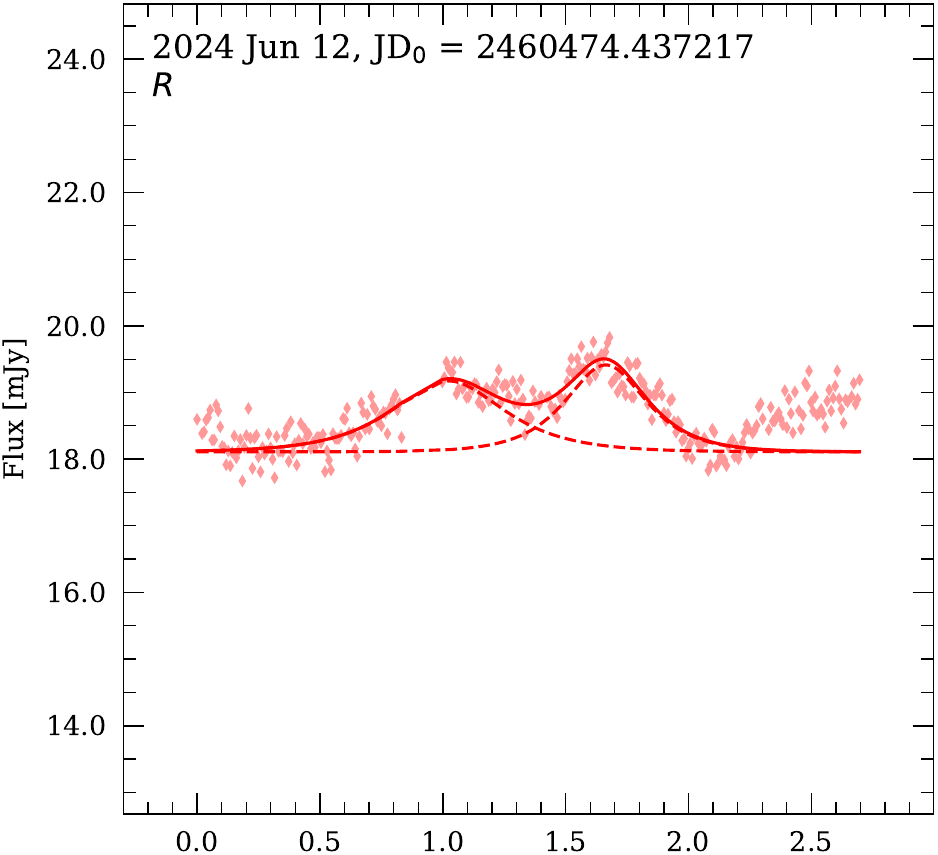}
\includegraphics[width=0.31\textwidth,clip=true]{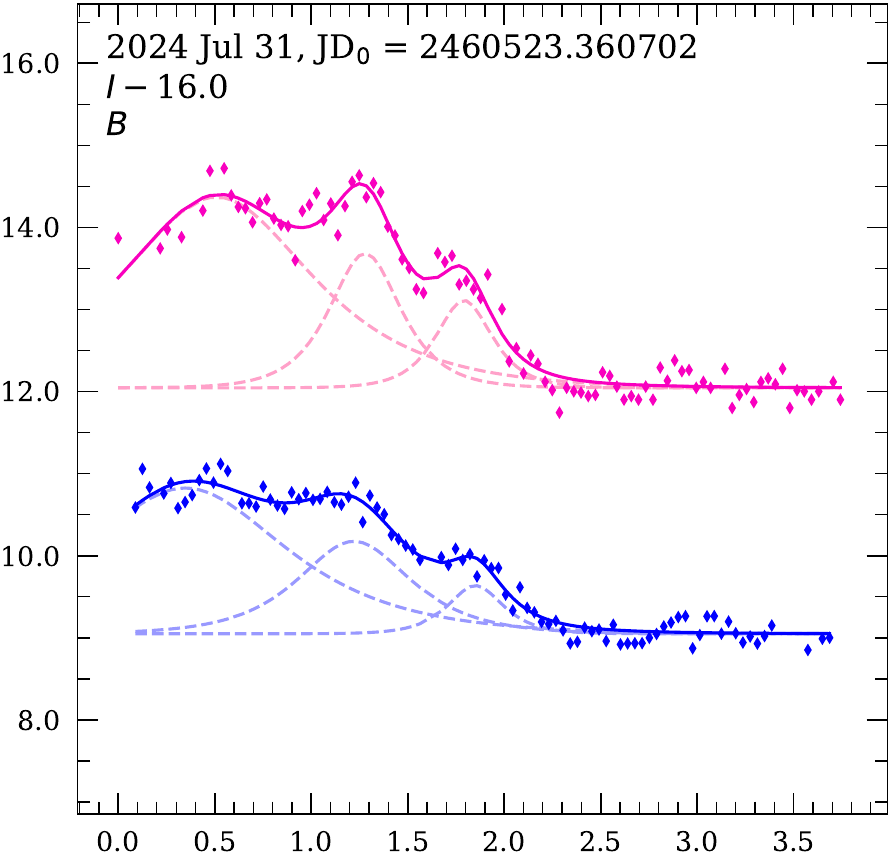}
\includegraphics[width=0.31\textwidth,clip=true]{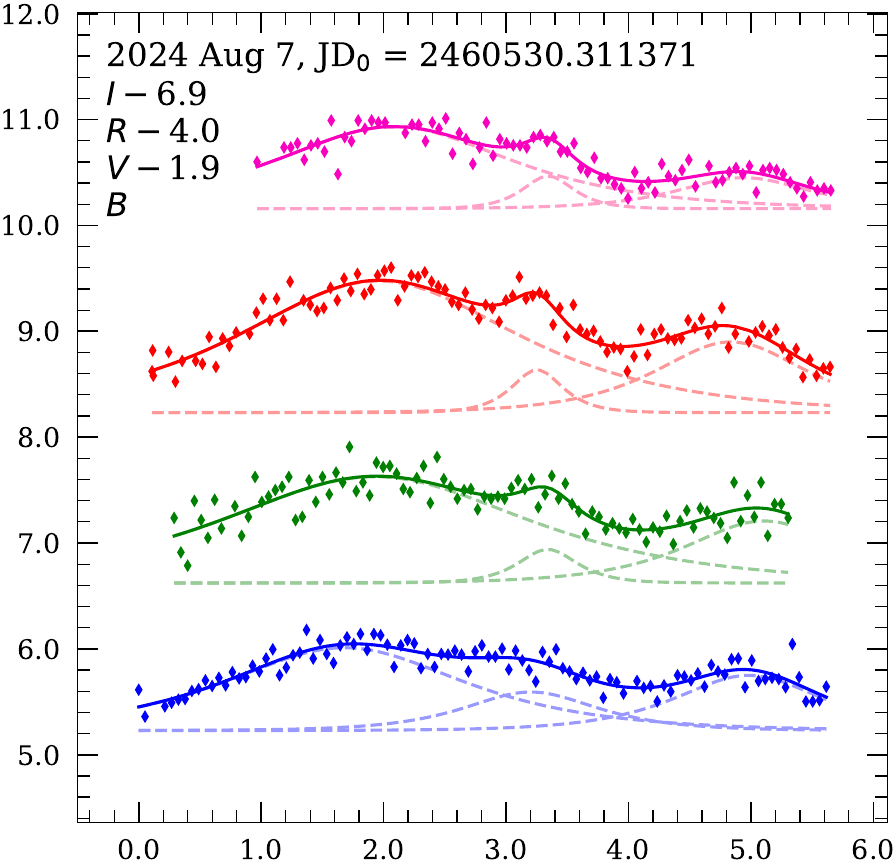}\vfill
\includegraphics[width=0.3415\textwidth,clip=true]{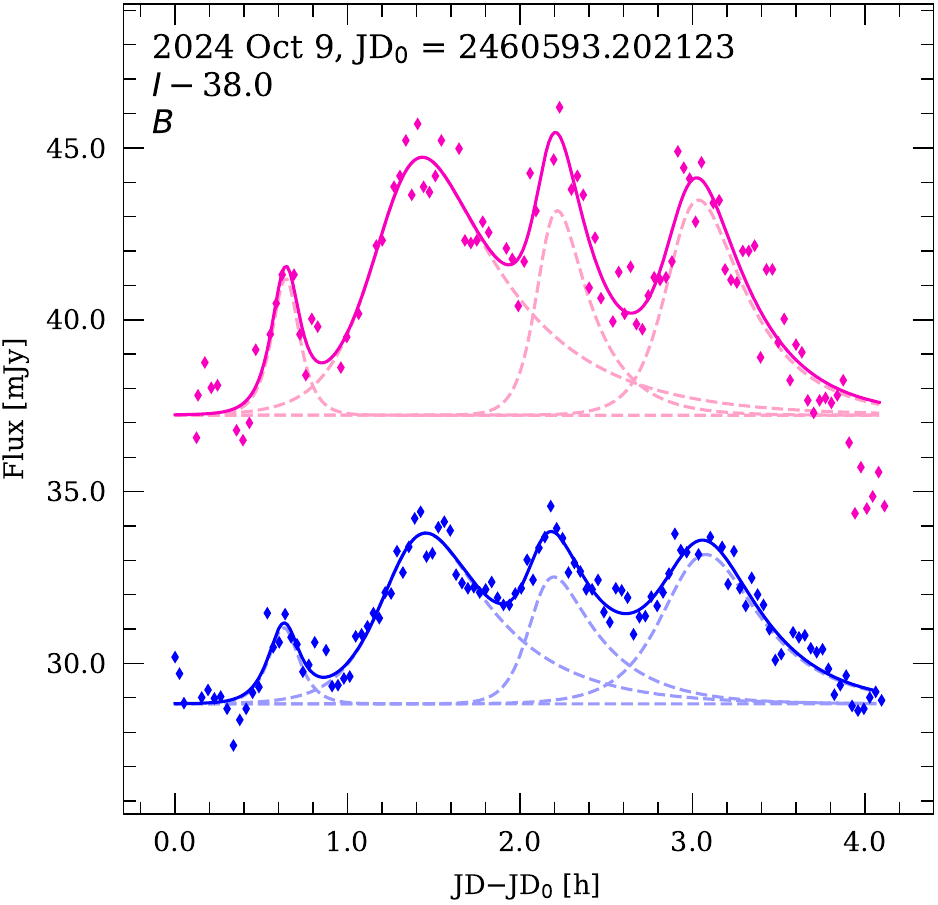}
\includegraphics[width=0.326\textwidth,clip=true]{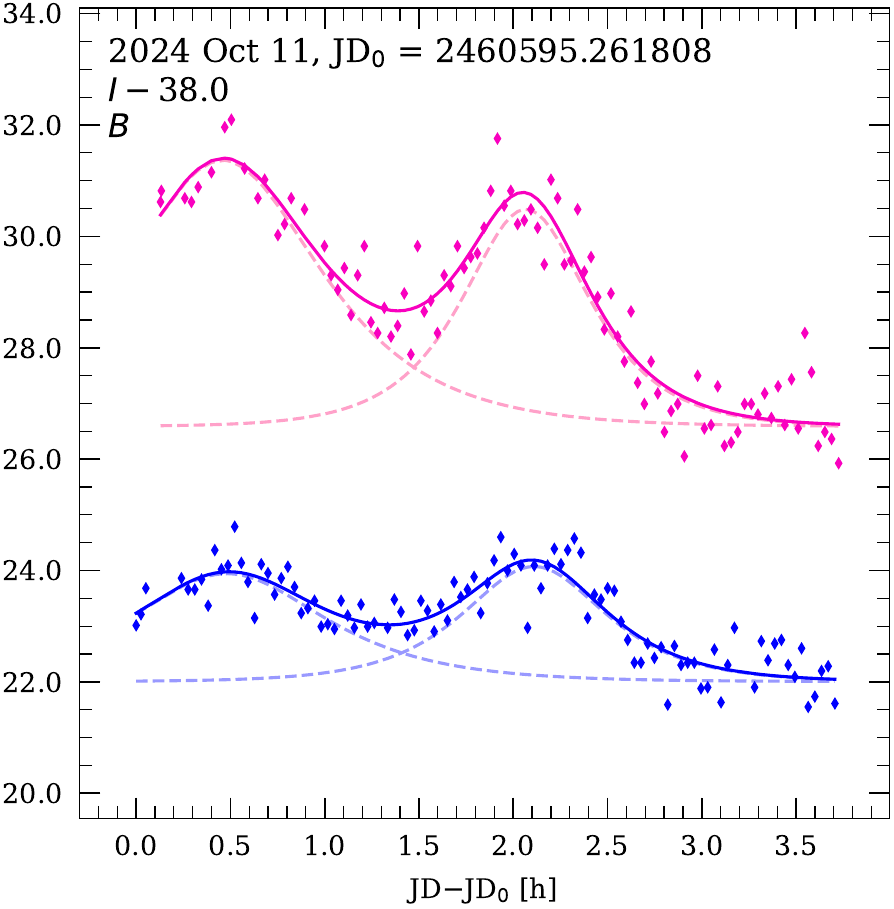}
\caption{Intra-day LCs decomposition. In each panel the civil date, the JD moment of the session start, $\rm JD_0$, the bands plotted, and the LC offsets applied are shown. The solid lines denote the composite LC fit, while the dashed lines represent the individual flares. Error bars are not shown for the sake of clarity.}
\label{fig:decompo}
\end{figure*}

\begin{table*}
	\centering
	\caption{Results from the intra-day variable LC decompositions.}
	\label{tab:decompo:pars}
	\begin{tabular}{lcc r@{~$\pm$~}r r@{~$\pm$~}r r@{~$\pm$~}r r@{~$\pm$~}r r@{~$\pm$~}r  c c} % four columns, alignment for each
		\hline
        Date & Band & Flare ID & \multicolumn{2}{c}{$F_{0}$} & \multicolumn{2}{c}{$t_{0}$} & \multicolumn{2}{c}{$\tau_{\rm r}$} &\multicolumn{2}{c}{$\tau_{\rm d}$} & \multicolumn{2}{c}{$\Delta \tau$} & $\xi$  & $\chi^2_{\rm df}$ \\
             &      &          & \multicolumn{2}{c}{[mJy]} & \multicolumn{2}{c}{[min]} & \multicolumn{2}{c}{[min]} & \multicolumn{2}{c}{[min]} & \multicolumn{2}{c}{[min]} & & \\
		\hline
		2021 Jul 17$^{a}$ & $R$ & 1a & 7.3 &   0.5 &  237.3 &   5.1 &   58.2 &   8.0 &   19.9 &   2.9 &  156.2 &  17.0 & $-0.491 \pm 0.003$ & 0.51 \\[0.1cm]
        2021 Jul 18$^{a}$ & $R$ & 2a & 8.5 &   0.9 &   89.1 &   3.1 &   27.9 &   4.8 &    8.3 &   1.9 &   72.4 &  10.3 & $-0.55 \pm 0.02$ & 0.81 \\
                    &     & 2b & 4.8 &   0.7 &  223.4 &   1.5 &    7.7 &   1.8 &    7.7 &   1.8 &   30.8 &   5.1 & & \\[0.1cm]
        2021 Aug 17 & $R$ & 3a & 4.0 &   0.7 &   55.6 &   2.1 &   14.5 &   4.1 &   14.5 &   4.1 &   58.0 &  11.6 & & 0.49 \\
                    &     & 3b & 4.9 &   0.7 &  125.6 &   2.4 &   12.4 &   2.6 &   12.4 &   2.6 &   49.6 &   7.4 & & \\
                    &     & 3c & 8.3 &   0.8 &  156.2 &   0.9 &    6.9 &   1.1 &    6.9 &   1.1 &   27.6 &   3.1 & & \\[0.1cm]
        2022 Nov 10 & $B^{b}$ & 4a &10.3 &   1.5 &   69.9 &   2.1 &   10.6 &   2.7 &   10.6 &   2.7 &   42.4 &   7.6 & & 1.02 \\
                    &     & 4b & 9.2 &   1.6 &  114.0 &   2.3 &    9.7 &   2.5 &    9.7 &   2.5 &   38.8 &   7.1 & & \\[0.1cm]
                    & $V^{b}$ & 4a &15.0 &   2.9 &   80.4 &   4.8 &   20.6 &   5.1 &   20.6 &   5.1 &   82.4 &  14.4 & & 0.97 \\
                    &     & 4b &10.5 &   2.0 &  122.5 &   4.3 &   15.2 &   3.9 &   15.2 &   3.9 &   60.8 &  11.0 & & \\[0.1cm]
                    & $R$ & 4a &10.4 &   2.0 &   68.9 &   2.9 &   10.8 &   3.7 &   10.8 &   3.7 &   43.2 &  10.5 & & 0.43 \\
                    &     & 4b &14.3 &   1.6 &  113.0 &   4.0 &   20.2 &   4.3 &   20.2 &   4.3 &   80.8 &  12.2 & & \\[0.1cm]
        2023 Jul 24$^{a}$ & $R^{b}$ & 5a & 4.3 &   0.3 &   89.2 &   2.0 &   23.0 &   1.7 &   23.0 &   1.7 &   92.0 &   4.8 & & 0.39 \\
                    &     & 5b & 7.7 &   0.9 &  162.3 &   3.7 &   25.4 &   3.5 &   25.4 &   3.5 &  101.6 &   9.9 & & \\
                    &     & 5c & 3.8 &   1.3 &  201.8 &   3.7 &   18.0 &   3.9 &   18.0 &   3.9 &   72.0 &  11.0 & & \\[0.1cm]
        2023 Jul 26 & $R^{b}$ & 6a & 4.1 &   0.2 &  113.0 &   2.3 &   24.6 &   1.8 &   24.6 &   1.8 &   98.4 &   5.1 & & 0.64 \\
                    &     & 6b & 4.5 &   0.3 &  159.9 &   1.3 &   18.1 &   1.2 &   18.1 &   1.2 &   72.4 &   3.4 & & \\[0.1cm]
        2024 Jun 12$^{a}$ & $R^{b}$ & 7a & 2.1 &   0.1 &   61.3 &   1.3 &   12.2 &   1.2 &   12.2 &   1.2 &   48.8 &   3.4 & & 0.57 \\
                    &     & 7b & 2.6 &   0.1 &   99.8 &   0.7 &    8.8 &   0.7 &    8.8 &   0.7 &   35.2 &   2.0 & & \\[0.1cm]
        2024 Jul 31$^{a}$ & $B$ & 8a & 3.6 &   0.1 &   20.8 &   2.7 &   28.0 &   4.7 &   28.0 &   4.7 &  112.0 &  13.3 & & 1.28 \\
                    &     & 8b & 2.3 &   0.4 &   73.3 &   1.7 &   15.6 &   3.1 &   15.6 &   3.1 &   62.4 &   8.8 & & \\
                    &     & 8c & 1.2 &   0.2 &  111.1 &   1.3 &    7.7 &   1.6 &    7.7 &   1.6 &   30.8 &   4.5 & & \\[0.1cm]
                    & $I$ & 8a & 4.6 &   0.2 &   30.5 &   2.3 &   26.6 &   3.5 &   26.6 &   3.5 &  106.4 &   9.9 & & 0.84 \\
                    &     & 8b & 3.3 &   0.4 &   76.7 &   1.2 &    9.9 &   1.8 &    9.9 &   1.8 &   39.6 &   5.1 & & \\
                    &     & 8c & 2.1 &   0.2 &  107.5 &   1.5 &    7.8 &   1.5 &    7.8 &   1.5 &   31.2 &   4.2 & & \\[0.1cm]
        2024 Aug  7 & $B$ & 9a & 1.6 &   0.4 &  100.9 &   6.7 &   52.3 &  13.9 &   52.3 &  13.9 &  209.2 &  39.3 & & 1.26 \\
                    &     & 9b & 0.7 &   0.2 &  192.1 &   6.3 &   29.1 &   9.0 &   29.1 &   9.0 &  116.4 &  25.5 & & \\
                    &     & 9c & 1.0 &   0.3 &  299.0 &   3.5 &   32.3 &   8.3 &   32.3 &   8.3 &  129.2 &  23.5 & & \\
                    & $V$ & 9a & 2.0 &   0.6 &  115.8 &   3.9 &   67.6 &  19.1 &   67.6 &  19.1 &  270.4 &  54.0 & & 2.85 \\
                    &     & 9b & 0.6 &   0.1 &  200.4 &   2.4 &   12.9 &   3.6 &   12.9 &   3.6 &   51.6 &  10.2 & & \\
                    &     & 9c & 1.2 &   0.4 &  306.4 &   7.0 &   34.3 &   9.0 &   34.3 &   9.0 &  137.2 &  25.5 & & \\
                    & $R$ & 9a & 2.5 &   0.5 &  117.5 &   4.9 &   61.4 &  13.6 &   61.4 &  13.6 &  245.6 &  38.5 & & 0.57 \\
                    &     & 9b & 0.8 &   0.2 &  195.2 &   2.9 &   11.2 &   4.6 &   11.2 &   4.6 &   44.8 &  13.0 & & \\
                    &     & 9c & 1.3 &   0.3 &  290.4 &   5.5 &   33.2 &   7.1 &   33.2 &   7.1 &  132.8 &  20.1 & & \\
                    & $I$ & 9a & 1.5 &   1.0 &  124.9 &   6.8 &   51.8 &  25.9 &   51.8 &  25.9 &  207.2 &  73.3 & & 0.46 \\
                    &     & 9b & 0.6 &   0.2 &  200.6 &   4.0 &   11.2 &   6.0 &   11.2 &   6.0 &   44.8 &  17.0 & & \\
                    &     & 9c & 0.6 &   0.8 &  296.5 &  12.7 &   28.8 &  33.9 &   28.8 &  33.9 &  115.2 &  95.9 & & \\[0.1cm]
        2024 Oct  9 & $B$ & 10a & 4.4 &   0.5 &   37.8 &   0.7 &    4.7 &   0.8 &    4.7 &   0.8 &   18.8 &   2.3 & & 1.86\\
                    &     & 10b & 9.1 &   0.9 &   80.7 &   3.5 &   10.3 &   1.7 &   25.2 &   6.7 &   71.0 &  13.8 & $0.41 \pm 0.06$ & \\
                    &     & 10c & 6.3 &   1.2 &  126.7 &   2.4 &    5.1 &   1.7 &   18.2 &   6.5 &   46.6 &  13.4 & $0.56 \pm 0.09$ & \\
                    &     & 10d & 8.3 &   0.8 &  180.4 &   2.8 &   11.2 &   3.0 &   20.1 &   2.8 &   62.6 &   8.2 & $0.30 \pm 0.07$ & \\[0.1cm]
                    & $I^{b}$ & 10a & 8.0 &   1.2 &   38.5 &   0.7 &    4.3 &   1.0 &    4.3 &   1.0 &   17.2 &   2.8 & & 1.67 \\
                    &     & 10b &13.3 &   1.8 &   77.6 &   5.2 &   10.8 &   2.8 &   31.1 &  10.4 &   83.8 &  21.5 & $0.47 \pm 0.07$ & \\
                    &     & 10c &10.6 &   2.0 &  129.5 &   2.8 &    4.6 &   1.8 &   12.9 &   4.3 &   35.0 &   9.3 & $0.5 \pm 0.1$ & \\
                    &     & 10d &11.3 &   1.2 &  177.1 &   2.3 &    7.3 &   1.9 &   18.9 &   3.2 &   52.4 &   7.4 & $0.45 \pm 0.04$ & \\[0.1cm]
        2024 Oct 11 & $B$ & 11a & 3.9 &   0.3 &   28.9 &   2.5 &   27.8 &   3.4 &   27.8 &   3.4 &  111.2 &   9.6 & & 1.43 \\
                    &     & 11b & 4.1 &   0.2 &  126.7 &   1.5 &   20.4 &   2.1 &   20.4 &   2.1 &   81.6 &   5.9 & & \\[0.1cm]
                    & $I$ & 11a & 9.6 &   0.5 &   27.5 &   2.4 &   27.8 &   3.1 &   27.8 &   3.1 &  111.2 &   8.8 & & 0.94 \\
                    &     & 11b & 7.8 &   0.4 &  124.0 &   1.3 &   18.0 &   1.7 &   18.0 &   1.7 &   72.0 &   4.8 & & \\
		\hline
	\end{tabular}
    \flushleft{{\em Note.} Flares are labelled in the order of their appearance on the LC for the given night. Description of the flare parameters is given in Section~\ref{sec:flare:1}. The figure-of-merit of each decomposition, namely $\chi^2$ per degree of freedom, is listed in the last column.\\ $^{a}$Light curves for these dates are deboosted.\\ $^{b}$There are data points for these bands excluded from the decomposition.}
\end{table*}

Following the steps, outlined in Section~\ref{sec:flare:1}, we performed a detailed flare analysis by means of LC decomposition for 11 nights, comprising 19 intraday LCs (5, 2, 8, and 4 in $B$, $V$, $R$, and $I$ bands, respectively), which satisfied our selection criteria. Deboosting was needed for the LCs taken during 6 of the nights.
The decomposition of the LCs was performed by means of a nonlinear least-squares technique implemented into the {\sc mpfit} fitter \citep{2009ASPC..411..251M}. The decomposed LCs are shown in Fig.~\ref{fig:decompo} and the best-fitting flare parameters are listed in Table~\ref{tab:decompo:pars}; the deboosted LCs are marked accordingly. On multi-band nights, the shape of the flares occasionally differed between filters, likely reflecting variations in data quality among the bands. 

The temporal profiles of the detected flares are well described by DE model. Out of 27 flares, 5 were adequately described by asymmetric profiles, while the remaining events were modelled using symmetric DEs. The decision to use a symmetric DE function is made on a case-by-case basis following the prescriptions in Section~\ref{sec:flare:1}. Regarding the asymmetry of the flares, two of them are of negative asymmetry (longer rise time-scale), and three are of positive asymmetry (longer decay time-scale). Following the classification of \citet{Abdo2010} the flares with non-zero $\xi$ are of moderate asymmetry. We calculated the weighted mean asymmetry parameter as follows: $\langle \xi \rangle_{\rm wm} = -0.492 \pm 0.003$ (s.d. 0.042) and $\langle \xi \rangle_{\rm wm} = 0.44 \pm 0.03$ (s.d. 0.08) for negative and positive flare asymmetry, respectively. Regarding the positive asymmetry, similar weighted mean asymmetry parameter was obtained by \citet{Agarwal2023}.

Considering the 2024 August 7 LCs, we note a systematic lag of the first flare centre as the observed frequency decreases; as far as the flare is modelled symmetric, we have $t_{\rm max}  \equiv t_0$. We shall consider this issue in details in Section~\ref{sec:soft}.

%%%%%%%%%%%%%%%%%%%%%%%%%%%%%%%%%%%%%%%%%%%%%%%%%%%%%%%%%%%%%%%%%%%%%%%%%%%%%%%%%%%%%%%%

\subsubsection{Emitting region parameters}
\label{sec:pars}
We shall interpret the flaring activity of BL Lacertae on intraday time-scales using the turbulent jet model \citep{2013A&A...558A..92B,2021Galax...9..114W,2023Galax..11..108W}. In the framework of this model, either a plane shock hits a turbulent cell or the cell passage through a standing shock. The cell is characterized with a radius ${\cal R}$ and a tangled magnetic field of strength ${\cal B}$. The electrons within the cell are accelerated by the shock and then are cooled via synchrotron and inverse Compton radiation.

We assumed for simplicity a rapid acceleration of electrons within a time-scale $t'_{\rm acc} \lesssim {\mathcal R}/c$ (here $c$ stands for the speed of light); here and below the primed quantities, are in the rest frame. These electrons cool and lose half of their energy within the cooling time-scale, $t_{\rm cool}(\nu)$:

\begin{equation}\label{eq:tcool}
t_{\rm cool}(\nu) \simeq \frac{4.73\!\times\!10^4}{1+q}\,{\mathcal B}^{\,-3/2}\, \nu_{15}^{-1/2}\, \left(\frac{\delta}{1+z}\right)^{-1/2} \quad [\rm s],
\end{equation}
where $\nu_{15}$ is the observed photon frequency (in units of $10^{15}$ Hz, $\nu = 10^{15}\nu_{15}$ Hz), ${\mathcal B}$ is the co-moving magnetic field strength (in units of Gauss), $\delta$ is the Doppler factor, and $q$ is the Compton dominance parameter. The latter parameter is the ratio of the co-moving energy densities of the radiation and magnetic fields, $q={\mathcal U}'_{\rm rad}/{\mathcal U}'_{\mathcal B}$.

In the case of symmetric flares, we shall assume that the dominant time-scale is the light-crossing one, $t'_{\rm cros}$. This means that the rise time-scale is its upper limit, that is $t'_{\rm cros} \lesssim {\tau_{\rm r}}\, \delta / (1+z)$. Thus, we can relate $\tau_{\rm r}$ to the upper limit (or the maximum value) of the emitting region radius:
\begin{equation} 
    {\mathcal R} \lesssim {\mathcal R}_{\rm max} = c\,{\tau_{\rm r}} \left(\frac{\delta}{1+z}\right) \quad \rm [cm];
\end{equation}
here and below the time-scales are in seconds.
If the flares possess negative asymmetry, then we shall assume that the rise time-scale is dominated by the acceleration one, $t'_{\rm acc}$. This means that in the corresponding emitting regions conditions are created for a gradual particle acceleration. For example, if turbulence is developed in the post-shock region, then the electrons could be energized by second-order Fermi process instead of cooled off.

Regarding the decay time-scale, we shall assume that it is an upper limit of the rest-frame synchrotron cooling time of the emitting electrons, $t'_{\rm cool} \lesssim {\tau_{\rm d}}\, \delta / (1+z)$. Rewriting equation (\ref{eq:tcool}) accordingly, we can obtain an estimate of the lower limit (or the minimum value) of the magnetic field strength, ${\mathcal B_{\rm min}}$, related to the emitting region \citep[e.g.][]{2021RAA....21..302F,2025MNRAS.541..732M}:
\begin{equation}
\begin{split}
    {\mathcal B} &\gtrsim {\mathcal B}_{\rm min}, \\    
    {\mathcal B}_{\rm min} &\simeq 1307\, {\tau_{\rm d}}^{\,-2/3}\, (1+q)^{\,-2/3}\, \nu_{15}^{-1/3}\, \left(\frac{1+z}{\delta}\right)^{1/3} \quad [\rm G].
    \end{split}
\end{equation}

\begin{figure}
\centering
\includegraphics[width=1\linewidth,clip=true]{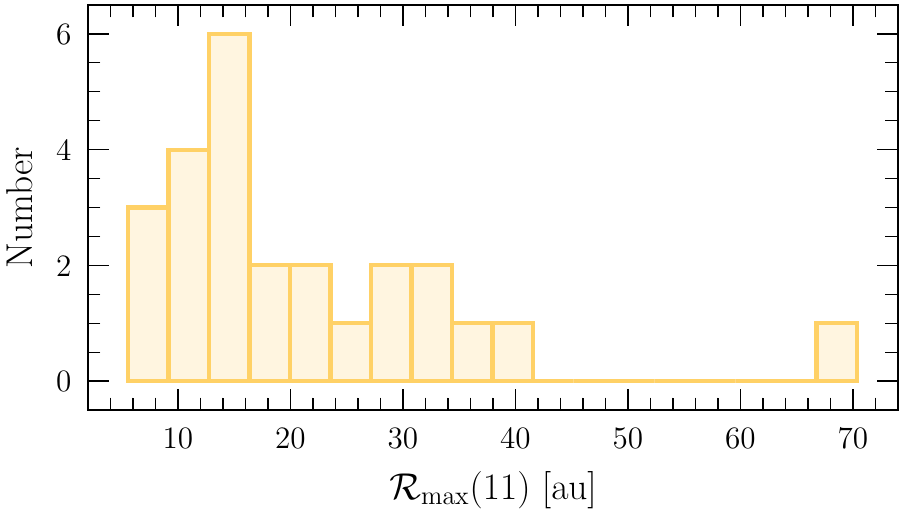}\vspace{0.15cm}
\includegraphics[width=1\linewidth,clip=true]{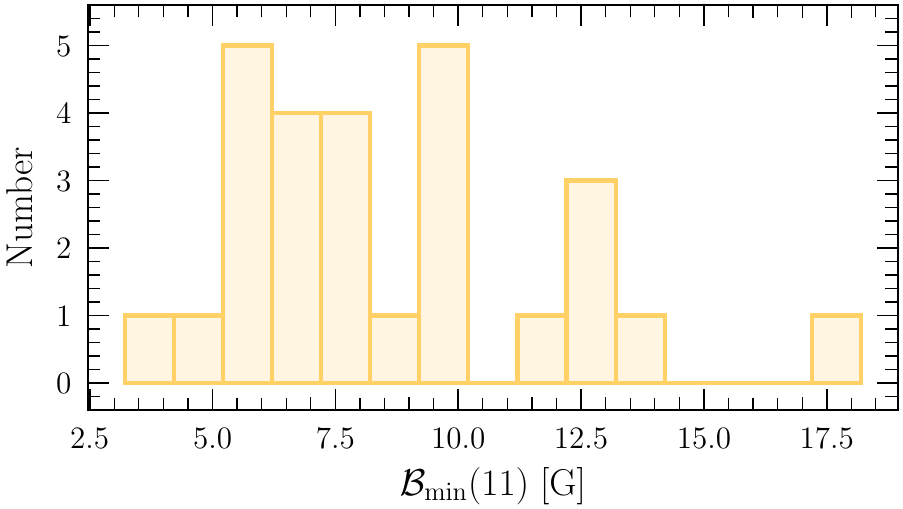}\vspace{0.15cm}
\includegraphics[width=1\linewidth,clip=true]{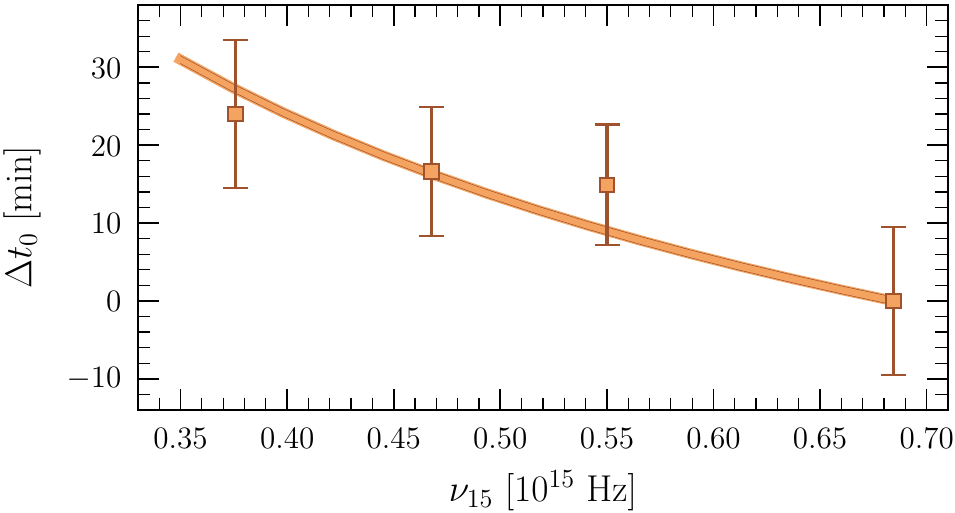}
\caption{Distribution of the radius maximum values (top panel) and magnetic field strength minimum values (middle panel). Limits are calculated using $\delta=11$. Bottom panel: time lag of the $VRI$-band variations (with respect to the $B$-band ones) as a function the frequency of the corresponding bands (squares). The solid curve is an approximation of this dependence for $\{\delta,{\mathcal B}\}=\{23.4,1.9\,\rm G\}$; see text for details.}
\label{fig:pars}
\end{figure}

Using the above equations, we obtained the limits for $\mathcal R$ and $\mathcal B$ for each flare and band; we assume $\delta=11$ and $q=0$. Next, for multi-band flares the limits were weight-averaged over bands. Distributions of $\mathcal R$ and $\mathcal B$ are shown in Fig.~\ref{fig:pars}. On the basis of these distributions, we calculated median values for the limits:
\begin{align}
    \nonumber
    \langle {\mathcal R_{\rm max}} \rangle_{\rm med} =\,~~~~~~~~~\, \delta \, 1.4^{+\,1.6}_{-0.6} &\stackrel{\text{($\delta$=11)}}{=} 15.8^{+18.0}_{-~6.2} \quad \rm au, \\
    \nonumber
    \langle {\mathcal B_{\rm min}} \rangle_{\rm med} = \delta^{-1/3} \, 17.3^{+10.1}_{-~5.2} &\stackrel{\text{($\delta$=11)}}{=} 7.8^{+\,4.5}_{-2.3} \quad \rm G;
\end{align}
uncertainties are based on the 16th and 84th percentiles of the corresponding distributions. These median values are in good correspondence with the ones obtained for BL Lacertae for the period 2020 mid-July to mid-September \citep{Agarwal2023}. During that period the blazar showed an intense flaring activity on intraday time-scales, which is possibly related to the current-driven kink instability developed in the jet of BL Lacertae in 2020 August \citep{2022Natur.609..265J}.

In the framework of the adopted turbulent jet model, the smallest spatial scales are related to the Kolmogorov scale \citep[e.g.][]{2019ApJ...884...92X,Agarwal2023}. This scale is independent of the large-scale structure of the turbulent flow and at this scale the kinetic energy of turbulence dissipates into heat. Understanding Kolmogorov scale is critical in turbulence studies of jets and in this context the accumulation of more information about the turbulent cell sizes and their distribution is of great importance. For example, strong turbulence has smaller Kolmogorov scale owing to the higher energy dissipation rate.

Alternatively, IDV could be explained within the geometric scenario \citep[e.g.][]{2021APh...12902577B}. According to \citet{2021APh...12902577B}, the IDV flares are produced by sub-components of the main emitting component. The sub-components move in random directions and this movement can lead, for a certain time interval, to a higher Doppler factor of a sub-component compared to that of the main component, thus producing an IDV flare. The main component is responsible for the short- and long-term variability, which could be geometrically explained, for example, by a helical jet \citep[the so called `lighthouse effect';][]{1993A&A...278..391S}, a twisted inhomogeneous jet \citep{2017Natur.552..374R}, a wiggling filamentary jet \citep{2024A&A...692A..48R}.

\begin{figure}
\centering
\includegraphics[width=1\linewidth,clip=true]{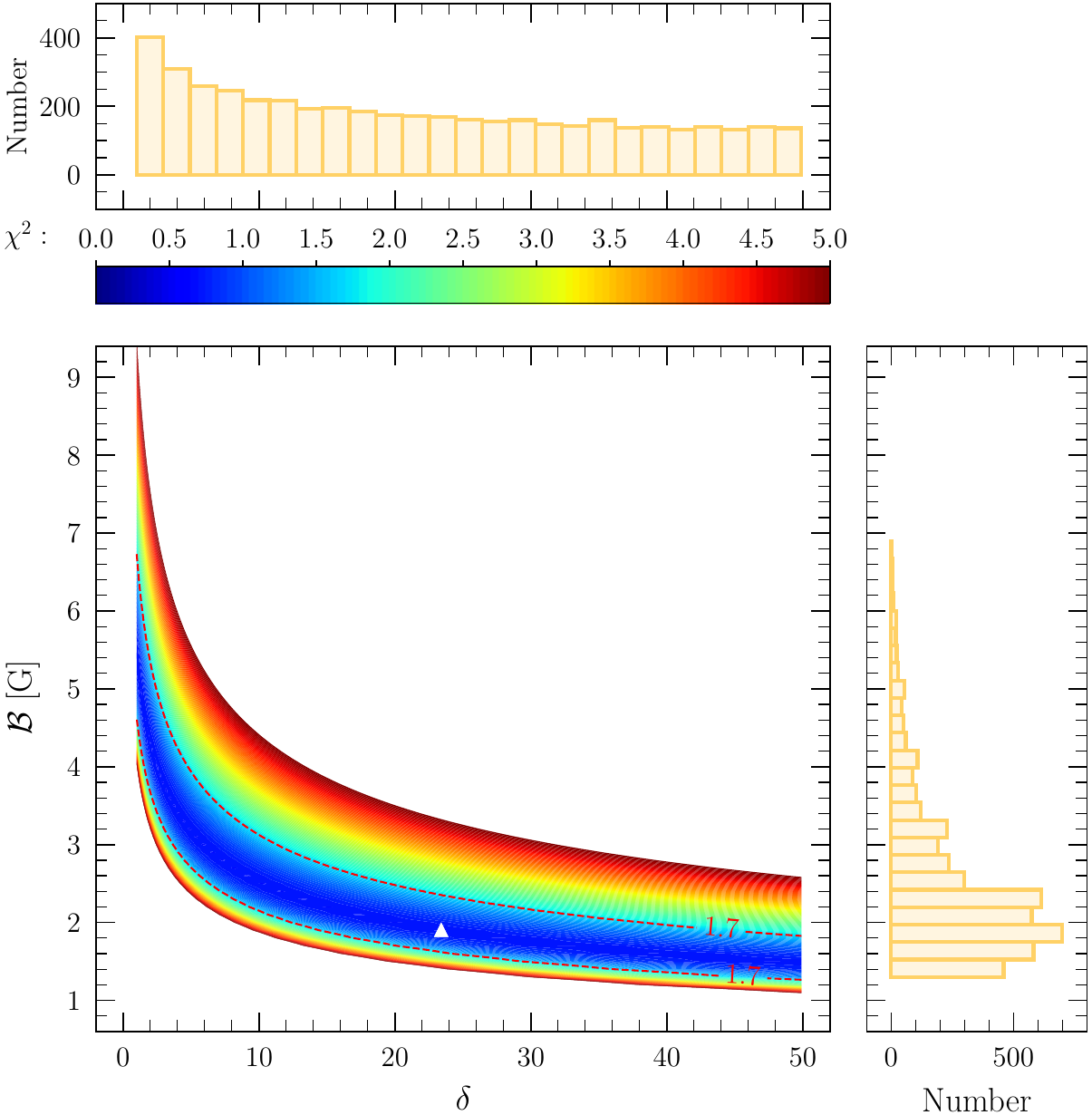}
\caption{Map of the $\chi^2$ values computed onto a grid of $\{\delta,\mathcal{B}\}$ pairs; we restricted the map to values $\chi^2 \leq 5$. The contour corresponding to $1\sigma$ confidence region is drawn (see text for details). The minimum $\chi^2$ value at $\{\delta,{\mathcal B}\}=\{23.4,1.9\,\rm G\}$ is marked with a white triangle. The histograms depict the distributions of $\delta$ (at the top) and $\mathcal B$ (to the right).}
\label{fig:pars2}
\end{figure}

%%%%%%%%%%%%%%%%%%%%%%%%%%%%%%%%%%%%%%%%%%%%%%%%%%%%%%%%%%%%%%%%%%%%%%%%%%%%%%%%%%%%%%%%

\subsubsection{Soft time lag}
\label{sec:soft}

If the flaring in a blazar LC is related to the processes of acceleration and cooling of electrons, then the so called soft and hard time lags could be observed. For the soft lag, the longer wavelength radiation (less energetic) lags the shorter one (more energetic), while for the hard lag is vice versa. Considering the soft lag, it results from the faster cooling of higher energy electrons ($t'_{\rm cool} \propto \gamma^{-1}_{\rm e}$, where $\gamma_{\rm e}$ is the electron energy in units of $m_{\rm e}c^2$) and assumes quasi-instantaneous injection of high energy, monoenergetic electron population across the emitting region. The hard lag results from the faster acceleration of lower energy electrons \citep[$t'_{\rm acc} \propto \gamma_{\rm e}$ for Fermi processes; e.g. ][]{2008ApJ...686..181F} and assumes gradual building of the high energy electron population from the low energy one via acceleration processes. In the optical range, examples of the soft lag could be found in \citet{Agarwal2023}, while the detection of hard lags is reported in \citet{2007ApJ...670..968B} and \citet{2011A&A...528L..10B}; generally, the hard lag is less commonly observed than the soft one in the optical.

In this context, the results from the LC decompositions revealed the presence of soft lags among the bands for the first flare of the 2024 August 7 LCs~-- the longer wavelength radiation lags the shorter wavelength one. The lags were calculated as $\Delta t_0(k,B) = t_0(k)-t_0(B)$, $k=B,V,R,I$ (see Table~\ref{tab:decompo:pars}): we got lags 0.0$\,\pm\,$9.5, 14.9$\,\pm\,$7.8, 16.6$\,\pm\,$8.3, and 24.0$\,\pm\,$9.5\,min for $BVRI$ against $B$ bands, respectively.
In Fig.~\ref{fig:pars}, we plot $\Delta t_0$ as a function of the observed frequency.

The cross-correlation analysis of the 2024 August 7 LCs, however, resulted in no lags (see Section~\ref{sec:ccorr}). If we consider the entire $BI$-band LCs, having largest wavelength separation, then we got a time lag of $2.2^{+\,15.5}_{-10.0}$\,min. This time lag is consistent with zero to within the quoted uncertainties. The lag uncertainties are estimated by means of the maximum likelihood peak location algorithm \citep[{\sc plike \oldstylenums{4.0}};][]{2013arXiv1302.1508A}. 
As previously discussed by, for example \citet{2019ApJ...884...92X}, \citet{Agarwal2023}, and \citet{2025MNRAS.541..732M}, the cross-correlation of the entire LCs results in a time lag, which is a kind of weight-average over the individual flares. Therefore, for the given LC of flaring behaviour the global time lag may differ from time lags obtained for individual flares. Considering our case, if we restrict ourselves to the first flare, then we got a $BI$-band time lag of $28.6^{+~~7.6}_{-15.5}$\,min, which agrees with the decomposition result to within the uncertainties, but having somewhat worse precision.

Generally, the presence of lags among the multi-band LCs allows us to estimate $\delta$ and $\mathcal B$ in an alternative way.
The soft time lag between two frequencies $\nu_1$ and $\nu_2$ ($\nu_1>\nu_2$) is $\tau(\nu_1,\nu_2) = t_{\rm cool}(\nu_2)-t_{\rm cool}(\nu_1)$, where $t_{\rm cool}(\nu)$ is given by equation~(\ref{eq:tcool}) with $q=0$. In our case we have:
\begin{equation}\label{eq:lag}
    \tau(\nu_B,\nu_k) = t_{\rm cool}(\nu_k)-t_{\rm cool}(\nu_B) \propto \frac{1}{\mathcal{B}^{3/2}\delta^{1/2}},
\end{equation}
where $k=V,R,I$.
In this notation the lags $\tau(\nu_B,\nu_k)$ are positive and match the signs of the decomposition results.

The estimate of $\{\delta,\mathcal{B}\}$ with the help of lags was done by means of the grid search. We varied $\delta$ and ${\mathcal B}$ with fixed steps in predefined ranges. %; we used a step of 0.1 for both variables.
In each step, we calculated $\tau(\nu_B,\nu_k)$ via equation~(\ref{eq:lag}) and then the corresponding value of the $\chi^2$ figure-of-merit. The so obtained $\chi^2$ surface is shown in Fig.~\ref{fig:pars2}. The grid search resulted in the following pair of parameters that minimizes $\chi^2$: $\{\delta,{\mathcal B}\}=\{23.4,1.9\,\rm G\}$. Inserting these parameters into equation~(\ref{eq:lag}), we obtained an approximation of the observed frequency dependence of time lags as shown in Fig.~\ref{fig:pars}.

Considering the $\chi^2$ surface along with the equation~(\ref{eq:lag}) helps us to understand better the parameters behaviour. Firstly, the parameters are strongly correlated and so, instead of a well-defined global $\chi^2$ minimum, we got a narrow valley of values $\chi^2 \simeq \chi^2_{\rm min}$. The correlation follows from equation~(\ref{eq:lag}) and, hence, we can constrain tightly only the product, but not the individual parameters themselves.
Consequently, the $1\sigma$ confidence region, limited by a $\chi^2$ value of $\chi^2_{\rm min} + 1$ for one degree of freedom, is an open curve\footnote{In our case $\chi^2_{\rm min} + 1 \simeq 1.7$.}. This prevents us of reliable estimation of the parameter uncertainties.
Secondly, the time lags depend stronger on $\mathcal B$ rather than on $\delta$, that is, large deviations of $\delta$ result in small deviations of $\mathcal B$ for fixed time lags. Hence, the values of the magnetic field strength are well constrained~-- most of them lie in the range 1--4\,G for almost all values of $\delta$. Using the distribution of magnetic field strengths, we estimated their median and modal values and 16th and 84th percentiles and obtained: $\langle \mathcal{B} \rangle_{\rm med}=2.1^{+\,1.3}_{-0.4}$\,G and $\langle \mathcal{B} \rangle_{\rm mod}=1.9^{+\,1.5}_{-0.2}$\,G, respectively. We can fix Doppler factor at certain values to break the degeneracy and to estimate the magnetic field strength and its uncertainty. If we set $\delta=23.4$, as estimated from the grid search, then the least-squares fit of the time lags frequency dependence resulted in $\mathcal{B}|_{\delta=23.4}=1.9 \pm 0.2$\,G. Setting $\delta=11.0$, as used in Section~\ref{sec:pars}, we got $\mathcal{B}|_{\delta=11.0}=2.4 \pm 0.2$\,G. The later value agrees with the weighted-mean value for that flare calculated using $\tau_{\rm d}$, namely 3.2\,G (s.d. 0.4\,G).

%%%%%%%%%%%%%%%%%%%%%%%%%%%%%%%%%%%%%%%%%%%%%%%%%%%%%%%%%%%%%%%%%%%%%%%%%%%%%%%%%%%%%%%%

%%%%%%%%%%%%%%%%%%%%%%%%%%%%%%%%%%%%%%%%%%%%%%%%%%%%%%%%%%%%%%%%%%%%%%%%%%%%%%%%%%%%%%%%

\section{Summary}

We have presented the results of an extensive four-year (2020–2024) optical IDV monitoring campaign of the archetypal blazar BL Lacertae. This period encompassed the source's historical peak brightness and various flux states, enabling us to investigate the characteristics and mechanisms of its rapid variability in detail. Our analysis of 117 LCs obtained from telescopes in Egypt, T\"{u}rkiye, and Bulgaria across 62 nights leads to the following principal conclusions:

$-$ Applying the power-enhanced F-test and nested ANOVA, we found significant IDV in 88 out of 117 LCs, corresponding to $\sim$75\,per cent of our monitored datasets. Considering entire observing nights, $\sim$70\,per cent (44 out of 62 nights) showed clear variability, increasing to $\sim$84\,per cent when probable variable nights were included. There is an indication that the variability detection probability increased markedly during high-states. 

$-$ The measured variability amplitudes ranged from 3.0 to 44.5\,per cent, with a maximum amplitude of 44.5\,per cent observed in the B band during a flaring episode on 2023 July 21. No clear correlation was found between variability amplitude and the brightness state of the source (high or low), or with the nightly average magnitude. Structure function analysis of variable LCs revealed power-law slopes ($\beta$) between 0.97 and 2.38, with characteristic variability time-scales ($\tau_{\rm char}$) ranging from 12 to 150\,min.

$-$ The results of DC analysis showed dependence on both the photometric band and the duration of the observations. For the full sample, the DC values were 46\,per cent ($B$ band), 89\,per cent ($V$ band), 94\,per cent ($R$ band), and 58\,per cent ($I$ band). Subsamples with longer monitoring durations ($\geq$3\,h) generally exhibited higher DC values, highlighting the importance of sufficient temporal coverage for robust variability detection. The lower DC in the $B$ and $I$ bands may be related to fewer observing nights and to the sampling of different activity states.

$-$ Analysis of SEDs constructed from multi-band observations at night yielded optical spectral indices ($\alpha$) ranging from 1.44 to 2.08 (mean $\sim$1.76). Crucially, our investigation of intraday spectral variability revealed a consistent BWB trend during flares. Of the 23 variable nights with multi-band data, 10 showed tight correlations and exhibited BWB behaviour, which aligns with the dominant trend reported in the literature for BL Lacertae during active states \citep{2023ApJS..269...60Y, 2024MNRAS.528.6823L, Agarwal2025}, and for other BL Lac objects as well \citep[e.g.][]{2023MNRAS.524.4333D, 2022ApJ...933...42A}.

$-$ We applied the $z$DCF analysis to the $B$- and $I$-band light curves on nights with significant variability and found strong correlations, with the DCF peaks consistent with zero lag within uncertainties. This indicates that flux variations in the two bands are nearly simultaneous, likely originating from the same or closely co-spatial emission regions. Small non-zero lags, where present, are not statistically significant and thus do not provide robust evidence for energy-dependent delays.

$-$ A detailed morphological analysis reveals a wide diversity of flare characteristics. Rise and decay time-scales span from minutes to hours, and amplitudes, $F_0/2$, varying from 0.3 to 7.5\,mJy. Our flare modelling was restricted in most of the cases to usage of a symmetric DE function. For the rest, we found moderate positive (3 flares) and negative (2 flares) asymmetries. We also detected soft time lags among the bands for the first flare of 2024 August 7 by means of both cross-correlation and decomposition techniques ($BI$-band lags in the range 24--29\,min).

$-$ On the basis of the derived time-scales, we estimated limits on the radii and magnetic field strengths of the emitting regions. The median value for the upper limit $\mathcal R_{\rm max}$ is 15.8\,au and the median value for the lower limit $\mathcal B_{\rm min}$ is 7.8\,G. The soft time lags modelling resulted in the following pair of parameters $\{\delta,{\mathcal B}\}=\{23.4,1.9\,\rm G\}$, with $\mathcal B$ being better constrained, for the emitting region that produces the first flare of 2024 August 7.

Our long-term, high-cadence campaign confirms BL Lacertae to be an exceptionally active and dynamic source, with IDV emerging as an almost universal feature during the high-state epoch of 2020–2024. Consistent BWB spectral behaviour supports models in which IDV is driven by particle acceleration and cooling processes within a turbulent jet, likely involving shocks or magnetic reconnection events. The absence of a clear correlation between amplitude and brightness state, alongside the high DC observed during extended monitoring, implies that rapid variability is an inherent characteristic of the jet's emission region, which remains active but is more easily detectable during periods of increased flux.

The minimum observed time-scale of tens of minutes places stringent upper limits on the size of the emitting region, implying extreme Doppler factors and reinforcing the role of relativistic beaming. Future multi-band campaigns conducted alongside intensive optical monitoring, such as that performed here, are essential in order to build a complete picture of the evolution of the spectral energy distribution during flares, and to definitively identify the dominant particle acceleration mechanisms at work in the jet of this quintessential blazar.

%%%%%%%%%%%%%%%%%%%%%%%%%%%%%%%%%%%%%%%%%%%%%%%%%%%%%%%%%%%%%%%%%%%%%%%%%%%%%%%%%%%%%%%%%%

\section*{Acknowledgements}

This work was supported by the Bulgarian Academy of Sciences under grant number IC-EG/09/2022-2024 and the Egyptian Academy of Scientific Research and Technology (ASRT) under grant number 10125 within the project titled 'Study of blazar jets through optical microvariability based on coordinated astronomical observations in Bulgaria and Egypt'.
We thank T\"{U}B\.{I}TAK National Observatory for partial support in using the TUG100 telescope at the TUG (T\"{U}B\.{I}TAK National Observatory, Antalya) site under the T\"{u}rkiye National Observatories with project number 19AT100-1486.
We also thank The Scientiﬁc Research Project Coordination Unit of Istanbul University for partial support with Project ID FAB-2026-42672.
This study was also supported by Scientific and Technological Research Council of T\"{u}rkiye (T\"{U}B\.{I}TAK) under the grant No. 121F427.
The research that led to these results was partially carried out with the help of infrastructure purchased/renovated under the National Roadmap for Research Infrastructure (2020--2027), financially coordinated by the Ministry of Education and Science of Republic of Bulgaria.
A. O. acknowledges financial support from the BAGEP Award of the Science Academy.

%%%%%%%%%%%%%%%%%%%%%%%%%%%%%%%%%%%%%%%%%%%%%%%%%%%%%%%%%%%%%%%%%%%%%%%%%%%%%%%%%%%%%%%%%%

\section*{Data Availability}
The data used in this article will be shared on reasonable request to the first or the corresponding author.

%%%%%%%%%%%%%%%%%%%%%%%%%%%%%%%%%%%%%%%%%%%%%%%%%%%%%%%%%%%%%%%%%%%%%%%%%%%%%%%%%%%%%%%%
% REFERENCES 
%%%%%%%%%%%%%%%%%%%%%%%%%%%%%%%%%%%%%%%%%%%%%%%%%%%%%%%%%%%%%%%%%%%%%%%%%%%%%%%%%%%%%%%%

\bibliographystyle{mnras}
\bibliography{refbib} % if your bibtex file is called example.bib

@ARTICLE{1993A&A...278..391S,
       author = {{Schramm}, K.-J. and {Borgeest}, U. and {Camenzind}, M. and {Wagner}, S.~J. and {Bade}, N. and {Dreissigacker}, O. and {Heidt}, J. and {Hoff}, W. and {Kayser}, R. and {Kuhl}, D. and {von Linde}, J. and {Linnert}, M.~D. and {Pelt}, J. and {Schramm}, T. and {Sillanpaa}, A. and {Takalo}, L.~O. and {Valtaoja}, E. and {Vigotti}, M.},
        title = "{Recent activity in the optical and radio lightcurves of the blazar 3C 345 : indications for a 'lighthouse effect' due to jet rotation.}",
      journal = {\aap},
         year = 1993,
        month = nov,
       volume = {278},
        pages = {391-405},
       adsurl = {https://ui.adsabs.harvard.edu/abs/1993A&A...278..391S}
}

@ARTICLE{2017Natur.552..374R,
       author = {{Raiteri}, C.~M. and {Villata}, M. and {Acosta-Pulido}, J.~A. and {Agudo}, I. and {Arkharov}, A.~A. and {Bachev}, R. and {Baida}, G.~V. and {Ben{\'\i}tez}, E. and {Borman}, G.~A. and {Boschin}, W. and {Bozhilov}, V. and {Butuzova}, M.~S. and {Calcidese}, P. and {Carnerero}, M.~I. and {Carosati}, D. and {Casadio}, C. and {Castro-Segura}, N. and {Chen}, W.-P. and {Damljanovic}, G. and {D'Ammando}, F. and {di Paola}, A. and {Echevarr{\'\i}a}, J. and {Efimova}, N.~V. and {Ehgamberdiev}, Sh. A. and {Espinosa}, C. and {Fuentes}, A. and {Giunta}, A. and {G{\'o}mez}, J.~L. and {Grishina}, T.~S. and {Gurwell}, M.~A. and {Hiriart}, D. and {Jermak}, H. and {Jordan}, B. and {Jorstad}, S.~G. and {Joshi}, M. and {Kopatskaya}, E.~N. and {Kuratov}, K. and {Kurtanidze}, O.~M. and {Kurtanidze}, S.~O. and {L{\"a}hteenm{\"a}ki}, A. and {Larionov}, V.~M. and {Larionova}, E.~G. and {Larionova}, L.~V. and {L{\'a}zaro}, C. and {Lin}, C.~S. and {Malmrose}, M.~P. and {Marscher}, A.~P. and {Matsumoto}, K. and {McBreen}, B. and {Michel}, R. and {Mihov}, B. and {Minev}, M. and {Mirzaqulov}, D.~O. and {Mokrushina}, A.~A. and {Molina}, S.~N. and {Moody}, J.~W. and {Morozova}, D.~A. and {Nazarov}, S.~V. and {Nikolashvili}, M.~G. and {Ohlert}, J.~M. and {Okhmat}, D.~N. and {Ovcharov}, E. and {Pinna}, F. and {Polakis}, T.~A. and {Protasio}, C. and {Pursimo}, T. and {Redondo-Lorenzo}, F.~J. and {Rizzi}, N. and {Rodriguez-Coira}, G. and {Sadakane}, K. and {Sadun}, A.~C. and {Samal}, M.~R. and {Savchenko}, S.~S. and {Semkov}, E. and {Skiff}, B.~A. and {Slavcheva-Mihova}, L. and {Smith}, P.~S. and {Steele}, I.~A. and {Strigachev}, A. and {Tammi}, J. and {Thum}, C. and {Tornikoski}, M. and {Troitskaya}, Yu. V. and {Troitsky}, I.~S. and {Vasilyev}, A.~A. and {Vince}, O.},
        title = "{Blazar spectral variability as explained by a twisted inhomogeneous jet}",
      journal = {\nat},
         year = 2017,
        month = dec,
       volume = {552},
       number = {7685},
        pages = {374-377},
          doi = {10.1038/nature24623},
archivePrefix = {arXiv},
       eprint = {1712.02098},
 primaryClass = {astro-ph.HE},
       adsurl = {https://ui.adsabs.harvard.edu/abs/2017Natur.552..374R}
}

@ARTICLE{2024A&A...692A..48R,
       author = {{Raiteri}, C.~M. and {Villata}, M. and {Carnerero}, M.~I. and {Kurtanidze}, S.~O. and {Mirzaqulov}, D.~O. and {Ben{\'\i}tez}, E. and {Bonnoli}, G. and {Carosati}, D. and {Acosta-Pulido}, J.~A. and {Agudo}, I. and {Andreeva}, T.~S. and {Apolonio}, G. and {Bachev}, R. and {Borman}, G.~A. and {Bozhilov}, V. and {Brown}, L.~F. and {Carbonell}, W. and {Casadio}, C. and {Chen}, W.~P. and {Damljanovic}, G. and {Ehgamberdiev}, S.~A. and {Elsaesser}, D. and {Escudero}, J. and {Feige}, M. and {Fuentes}, A. and {Gabellini}, D. and {Gazeas}, K. and {Giroletti}, M. and {Grishina}, T.~S. and {Gupta}, A.~C. and {Gurwell}, M.~A. and {Hagen-Thorn}, V.~A. and {Hamed}, G.~M. and {Hiriart}, D. and {Hodges}, M. and {Ivanidze}, R.~Z. and {Ivanov}, D.~V. and {Joner}, M.~D. and {Jorstad}, S.~G. and {Jovanovic}, M.~D. and {Kiehlmann}, S. and {Kimeridze}, G.~N. and {Kopatskaya}, E.~N. and {Kovalev}, Yu. A. and {Kovalev}, Y.~Y. and {Kurtanidze}, O.~M. and {Kurtenkov}, A. and {Larionova}, E.~G. and {Lessing}, A. and {Lin}, H.~C. and {L{\'o}pez}, J.~M. and {Lorey}, C. and {Ludwig}, J. and {Marchili}, N. and {Marchini}, A. and {Marscher}, A.~P. and {Matsumoto}, K. and {Max-Moerbeck}, W. and {Mihov}, B. and {Minev}, M. and {Mingaliev}, M.~G. and {Modaressi}, A. and {Morozova}, D.~A. and {Mortari}, F. and {Mufakharov}, T.~V. and {Myserlis}, I. and {Nikolashvili}, M.~G. and {Pearson}, T.~J. and {Popkov}, A.~V. and {Rahimov}, I.~A. and {Readhead}, A.~C.~S. and {Reinhart}, D. and {Reeves}, R. and {Righini}, S. and {Romanov}, F.~D. and {Savchenko}, S.~S. and {Semkov}, E. and {Shishkina}, E.~V. and {Sigua}, L.~A. and {Slavcheva-Mihova}, L. and {Sotnikova}, Yu. V. and {Steineke}, R. and {Stojanovic}, M. and {Strigachev}, A. and {Takey}, A. and {Traianou}, E. and {Troitskaya}, Yu. V. and {Troitskiy}, I.~S. and {Tsai}, A.~L. and {Valcheva}, A. and {Vasilyev}, A.~A. and {Verna}, G. and {Vince}, O. and {Vrontaki}, K. and {Weaver}, Z.~R. and {Webb}, J. and {Yuldoshev}, Q.~X. and {Zaharieva}, E. and {Zhovtan}, A.~V.},
        title = "{A wiggling filamentary jet at the origin of the blazar multi-wavelength behaviour}",
      journal = {\aap},
         year = 2024,
        month = dec,
       volume = {692},
          eid = {A48},
        pages = {A48},
          doi = {10.1051/0004-6361/202452311},
archivePrefix = {arXiv},
       eprint = {2410.22319},
 primaryClass = {astro-ph.HE},
       adsurl = {https://ui.adsabs.harvard.edu/abs/2024A&A...692A..48R}
}

@ARTICLE{2021APh...12902577B,
       author = {{Butuzova}, M.~S.},
        title = "{A geometrical interpretation for the properties of multiband optical variability of the blazar S5 0716+714}",
      journal = {Astroparticle Physics},
         year = 2021,
        month = may,
       volume = {129},
          eid = {102577},
        pages = {102577},
          doi = {10.1016/j.astropartphys.2021.102577},
archivePrefix = {arXiv},
       eprint = {2005.08161},
 primaryClass = {astro-ph.GA},
       adsurl = {https://ui.adsabs.harvard.edu/abs/2021APh...12902577B}
}

@ARTICLE{2026A&A...708A..30N,
       author = {{Nikolov}, Yanko and {Borisov}, Galin and {Bagnulo}, Stefano and {Nikolov}, Plamen and {Bogdanovski}, Rumen and {Bonev}, Tanyu},
        title = "{Polarimetric and spectropolarimetric observations with FoReRo2: Instrument overview and standard star monitoring}",
      journal = {\aap},
         year = 2026,
        month = mar,
       volume = {708},
          eid = {A30},
        pages = {A30},
          doi = {10.1051/0004-6361/202558243},
archivePrefix = {arXiv},
       eprint = {2512.05670},
 primaryClass = {astro-ph.SR},
       adsurl = {https://ui.adsabs.harvard.edu/abs/2026A&A...708A..30N}
}

@ARTICLE{2024arXiv241201592M,
       author = {{Mihov}, B.~M. and {Slavcheva-Mihova}, L.~S.},
        title = "{Multi-band intra-night variability of the blazar CTA 102 during its 2016 December outburst}",
      journal = {arXiv e-prints},
         year = 2024,
        month = dec,
          eid = {arXiv:2412.01592},
        pages = {arXiv:2412.01592},
          doi = {10.48550/arXiv.2412.01592},
archivePrefix = {arXiv},
       eprint = {2412.01592},
 primaryClass = {astro-ph.HE},
       adsurl = {https://ui.adsabs.harvard.edu/abs/2024arXiv241201592M}
}

@ARTICLE{2021A&A...645A.137A,
       author = {{Agarwal}, A. and {Mihov}, B. and {Andruchow}, I. and {Cellone}, S.~A. and {Anupama}, G.~C. and {Agrawal}, V. and {Zola}, S. and {Slavcheva-Mihova}, L. and {{\"O}zd{\"o}nmez}, A. and {Ege}, Erg{\"u}n and {Raj}, A. and {Mammana}, L. and {Zibecchi}, L. and {Fern{\'a}ndez-Laj{\'u}s}, E.},
        title = "{Multi-band behaviour of the TeV blazar PG 1553+113 in optical range on diverse timescales. Flux and spectral variations}",
      journal = {\aap},
         year = 2021,
        month = jan,
       volume = {645},
          eid = {A137},
        pages = {A137},
          doi = {10.1051/0004-6361/202039301},
archivePrefix = {arXiv},
       eprint = {2011.04074},
 primaryClass = {astro-ph.HE},
       adsurl = {https://ui.adsabs.harvard.edu/abs/2021A&A...645A.137A}
}

@ARTICLE{2020MNRAS.498.3013Z,
       author = {{Zibecchi}, L. and {Andruchow}, I. and {Cellone}, S.~A. and {Carpintero}, D.~D.},
        title = "{Microvariability in AGNs: study of different statistical methods - II. Light curves from simulated images}",
      journal = {\mnras},
         year = 2020,
        month = oct,
       volume = {498},
       number = {2},
        pages = {3013-3022},
          doi = {10.1093/mnras/staa2544},
archivePrefix = {arXiv},
       eprint = {2008.08137},
 primaryClass = {astro-ph.GA},
       adsurl = {https://ui.adsabs.harvard.edu/abs/2020MNRAS.498.3013Z}
}

@ARTICLE{2017MNRAS.467..340Z,
       author = {{Zibecchi}, L. and {Andruchow}, I. and {Cellone}, S.~A. and {Carpintero}, D.~D. and {Romero}, G.~E. and {Combi}, J.~A.},
        title = "{Microvariability in AGNs: study of different statistical methods - I. Observational analysis}",
      journal = {\mnras},
         year = 2017,
        month = may,
       volume = {467},
       number = {1},
        pages = {340-352},
          doi = {10.1093/mnras/stx054},
archivePrefix = {arXiv},
       eprint = {1702.04643},
 primaryClass = {astro-ph.GA},
       adsurl = {https://ui.adsabs.harvard.edu/abs/2017MNRAS.467..340Z}
}

@ARTICLE{2023ApJS..268...54X,
       author = {{Xu}, Jingran and {Hu}, Shaoming and {Chen}, Xu and {Jiang}, Yunguo and {Alexeeva}, Sofya},
        title = "{A Small-scale Structure Model of a Jet Based on Observations of Microvariability}",
      journal = {\apjs},
         year = 2023,
        month = oct,
       volume = {268},
       number = {2},
          eid = {54},
        pages = {54},
          doi = {10.3847/1538-4365/aceda8},
archivePrefix = {arXiv},
       eprint = {2309.09193},
 primaryClass = {astro-ph.HE},
       adsurl = {https://ui.adsabs.harvard.edu/abs/2023ApJS..268...54X}
}

@ARTICLE{2021Galax...9..114W,
       author = {{Webb}, James R. and {Arroyave}, Viviana and {Laurence}, Douglas and {Revesz}, Stephen and {Bhatta}, Gopal and {Hollingsworth}, Hal and {Dhalla}, Sarah and {Howard}, Emily and {Cioffi}, Michael},
        title = "{The Nature of Micro-Variability in Blazars}",
      journal = {Galaxies},
         year = 2021,
        month = dec,
       volume = {9},
       number = {4},
          eid = {114},
        pages = {114},
          doi = {10.3390/galaxies9040114},
       adsurl = {https://ui.adsabs.harvard.edu/abs/2021Galax...9..114W}
}

@ARTICLE{2023Galax..11..108W,
       author = {{Webb}, James R. and {Sanz}, Ivan Parra},
        title = "{The Structure of Micro-Variability in the WEBT BL Lacertae Observation}",
      journal = {Galaxies},
         year = 2023,
        month = nov,
       volume = {11},
       number = {6},
          eid = {108},
        pages = {108},
          doi = {10.3390/galaxies11060108},
       adsurl = {https://ui.adsabs.harvard.edu/abs/2023Galax..11..108W}
}

@ARTICLE{2013arXiv1302.1508A,
       author = {{Alexander}, Tal},
        title = "{Improved AGN light curve analysis with the z-transformed discrete correlation function}",
      journal = {arXiv e-prints},
         year = 2013,
        month = feb,
          eid = {arXiv:1302.1508},
        pages = {arXiv:1302.1508},
          doi = {10.48550/arXiv.1302.1508},
archivePrefix = {arXiv},
       eprint = {1302.1508},
 primaryClass = {astro-ph.IM},
       adsurl = {https://ui.adsabs.harvard.edu/abs/2013arXiv1302.1508A}
}

@ARTICLE{2008ApJ...686..181F,
       author = {{Finke}, Justin D. and {Dermer}, Charles D. and {B{\"o}ttcher}, Markus},
        title = "{Synchrotron Self-Compton Analysis of TeV X-Ray-Selected BL Lacertae Objects}",
      journal = {\apj},
         year = 2008,
        month = oct,
       volume = {686},
       number = {1},
        pages = {181-194},
          doi = {10.1086/590900},
archivePrefix = {arXiv},
       eprint = {0802.1529},
 primaryClass = {astro-ph},
       adsurl = {https://ui.adsabs.harvard.edu/abs/2008ApJ...686..181F}
}

@ARTICLE{2007ApJ...670..968B,
       author = {{B{\"o}ttcher}, M. and {Basu}, S. and {Joshi}, M. and {Villata}, M. and {Arai}, A. and {Aryan}, N. and {Asfandiyarov}, I.~M. and {Bach}, U. and {Bachev}, R. and {Berduygin}, A. and {Blaek}, M. and {Buemi}, C. and {Castro-Tirado}, A.~J. and {de Ugarte Postigo}, A. and {Frasca}, A. and {Fuhrmann}, L. and {Hagen-Thorn}, V.~A. and {Henson}, G. and {Hovatta}, T. and {Hudec}, R. and {Ibrahimov}, M. and {Ishii}, Y. and {Ivanidze}, R. and {Jel{\'\i}nek}, M. and {Kamada}, M. and {Kapanadze}, B. and {Katsuura}, M. and {Kotaka}, D. and {Kovalev}, Y.~Y. and {Kovalev}, Yu. A. and {Kub{\'a}nek}, P. and {Kurosaki}, M. and {Kurtanidze}, O. and {L{\"a}hteenm{\"a}ki}, A. and {Lanteri}, L. and {Larionov}, V.~M. and {Larionova}, L. and {Lee}, C.-U. and {Leto}, P. and {Lindfors}, E. and {Marilli}, E. and {Marshall}, K. and {Miller}, H.~R. and {Mingaliev}, M.~G. and {Mirabal}, N. and {Mizoguchi}, S. and {Nakamura}, K. and {Nieppola}, E. and {Nikolashvili}, M. and {Nilsson}, K. and {Nishiyama}, S. and {Ohlert}, J. and {Osterman}, M.~A. and {Pak}, S. and {Pasanen}, M. and {Peters}, C.~S. and {Pursimo}, T. and {Raiteri}, C.~M. and {Robertson}, J. and {Robertson}, T. and {Ryle}, W.~T. and {Sadakane}, K. and {Sadun}, A. and {Sigua}, L. and {Sohn}, B.-W. and {Strigachev}, A. and {Sumitomo}, N. and {Takalo}, L.~O. and {Tamesue}, Y. and {Tanaka}, K. and {Thorstensen}, J.~R. and {Tosti}, G. and {Trigilio}, C. and {Umana}, G. and {Vennes}, S. and {Vitek}, S. and {Volvach}, A. and {Webb}, J. and {Yamanaka}, M. and {Yim}, H.-S.},
        title = "{The WEBT Campaign on the Blazar 3C 279 in 2006}",
      journal = {\apj},
         year = 2007,
        month = dec,
       volume = {670},
       number = {2},
        pages = {968-977},
          doi = {10.1086/522583},
archivePrefix = {arXiv},
       eprint = {0708.2291},
 primaryClass = {astro-ph},
       adsurl = {https://ui.adsabs.harvard.edu/abs/2007ApJ...670..968B}
}

@ARTICLE{2011A&A...528L..10B,
       author = {{Bachev}, R. and {Semkov}, E. and {Strigachev}, A. and {Mihov}, B. and {Gupta}, A.~C. and {Peneva}, S. and {Ovcharov}, E. and {Valcheva}, A. and {Lalova}, A.},
        title = "{Intranight variability of 3C 454.3 during its 2010 November outburst}",
      journal = {\aap},
         year = 2011,
        month = apr,
       volume = {528},
          eid = {L10},
        pages = {L10},
          doi = {10.1051/0004-6361/201116637},
archivePrefix = {arXiv},
       eprint = {1102.3307},
 primaryClass = {astro-ph.HE},
       adsurl = {https://ui.adsabs.harvard.edu/abs/2011A&A...528L..10B}
}

@ARTICLE{2022Natur.609..265J,
       author = {{Jorstad}, S.~G. and {Marscher}, A.~P. and {Raiteri}, C.~M. and {Villata}, M. and {Weaver}, Z.~R. and {Zhang}, H. and {Dong}, L. and {G{\'o}mez}, J.~L. and {Perel}, M.~V. and {Savchenko}, S.~S. and {Larionov}, V.~M. and {Carosati}, D. and {Chen}, W.~P. and {Kurtanidze}, O.~M. and {Marchini}, A. and {Matsumoto}, K. and {Mortari}, F. and {Aceti}, P. and {Acosta-Pulido}, J.~A. and {Andreeva}, T. and {Apolonio}, G. and {Arena}, C. and {Arkharov}, A. and {Bachev}, R. and {Banfi}, M. and {Bonnoli}, G. and {Borman}, G.~A. and {Bozhilov}, V. and {Carnerero}, M.~I. and {Damljanovic}, G. and {Ehgamberdiev}, S.~A. and {Els{\"a}sser}, D. and {Frasca}, A. and {Gabellini}, D. and {Grishina}, T.~S. and {Gupta}, A.~C. and {Hagen-Thorn}, V.~A. and {Hallum}, M.~K. and {Hart}, M. and {Hasuda}, K. and {Hemrich}, F. and {Hsiao}, H.~Y. and {Ibryamov}, S. and {Irsmambetova}, T.~R. and {Ivanov}, D.~V. and {Joner}, M.~D. and {Kimeridze}, G.~N. and {Klimanov}, S.~A. and {Kn{\"o}tt}, J. and {Kopatskaya}, E.~N. and {Kurtanidze}, S.~O. and {Kurtenkov}, A. and {Kuutma}, T. and {Larionova}, E.~G. and {Leonini}, S. and {Lin}, H.~C. and {Lorey}, C. and {Mannheim}, K. and {Marino}, G. and {Minev}, M. and {Mirzaqulov}, D.~O. and {Morozova}, D.~A. and {Nikiforova}, A.~A. and {Nikolashvili}, M.~G. and {Ovcharov}, E. and {Papini}, R. and {Pursimo}, T. and {Rahimov}, I. and {Reinhart}, D. and {Sakamoto}, T. and {Salvaggio}, F. and {Semkov}, E. and {Shakhovskoy}, D.~N. and {Sigua}, L.~A. and {Steineke}, R. and {Stojanovic}, M. and {Strigachev}, A. and {Troitskaya}, Y.~V. and {Troitskiy}, I.~S. and {Tsai}, A. and {Valcheva}, A. and {Vasilyev}, A.~A. and {Vince}, O. and {Waller}, L. and {Zaharieva}, E. and {Chatterjee}, R.},
        title = "{Rapid quasi-periodic oscillations in the relativistic jet of BL Lacertae}",
      journal = {\nat},
         year = 2022,
        month = sep,
       volume = {609},
       number = {7926},
        pages = {265-268},
          doi = {10.1038/s41586-022-05038-9},
       adsurl = {https://ui.adsabs.harvard.edu/abs/2022Natur.609..265J}
}

@ARTICLE{2025MNRAS.541..732M,
       author = {{Mihov}, B.~M. and {Elhosseiny}, E.~G. and {Slavcheva-Mihova}, L.~S. and {Takey}, Ali and {Ismail}, M.~N. and {Mawad}, Ramy},
        title = "{Multiwavelength variability analysis of the blazar S5 0716+714 during a long-lasting period of low activity}",
      journal = {\mnras},
         year = 2025,
        month = aug,
       volume = {541},
       number = {2},
        pages = {732-749},
          doi = {10.1093/mnras/staf1019},
archivePrefix = {arXiv},
       eprint = {2507.01184},
 primaryClass = {astro-ph.HE},
       adsurl = {https://ui.adsabs.harvard.edu/abs/2025MNRAS.541..732M}
}

@ARTICLE{2021RAA....21..302F,
       author = {{Fan}, Xu-Liang and {Yan}, Da-Hai and {Wu}, Qing-Wen and {Chen}, Xu},
        title = "{Constraining evolution of magnetic field strength in the dissipation region of two BL Lac objects}",
      journal = {Research in Astronomy and Astrophysics},
         year = 2021,
        month = dec,
       volume = {21},
       number = {12},
          eid = {302},
        pages = {302},
          doi = {10.1088/1674-4527/ac299e},
archivePrefix = {arXiv},
       eprint = {2109.11229},
 primaryClass = {astro-ph.HE},
       adsurl = {https://ui.adsabs.harvard.edu/abs/2021RAA....21..302F}
}

@ARTICLE{2019ApJ...884...92X,
       author = {{Xu}, Jingran and {Hu}, Shaoming and {Webb}, James R. and {Bhatta}, Gopal and {Jiang}, Yunguo and {Chen}, Xu and {Alexeeva}, Sofya and {Li}, Yutong},
        title = "{Statistical Analysis of Microvariability Properties of the Blazar S5 0716+714}",
      journal = {\apj},
         year = 2019,
        month = oct,
       volume = {884},
       number = {1},
          eid = {92},
        pages = {92},
          doi = {10.3847/1538-4357/ab3e50},
archivePrefix = {arXiv},
       eprint = {1909.03129},
 primaryClass = {astro-ph.HE},
       adsurl = {https://ui.adsabs.harvard.edu/abs/2019ApJ...884...92X}
}

@ARTICLE{2013A&A...558A..92B,
       author = {{Bhatta}, G. and {Webb}, J.~R. and {Hollingsworth}, H. and {Dhalla}, S. and {Khanuja}, A. and {Bachev}, R. and {Blinov}, D.~A. and {B{\"o}ttcher}, M. and {Bravo Calle}, O.~J.~A. and {Calcidese}, P. and {Capezzali}, D. and {Carosati}, D. and {Chigladze}, R. and {Collins}, A. and {Coloma}, J.~M. and {Efimov}, Y. and {Gupta}, A.~C. and {Hu}, S.-M. and {Kurtanidze}, O. and {Lamerato}, A. and {Larionov}, V.~M. and {Lee}, C.-U. and {Lindfors}, E. and {Murphy}, B. and {Nilsson}, K. and {Ohlert}, J.~M. and {Oksanen}, A. and {P{\"a}{\"a}kk{\"o}nen}, P. and {Pollock}, J.~T. and {Rani}, B. and {Reinthal}, R. and {Rodriguez}, D. and {Ros}, J.~A. and {Roustazadeh}, P. and {Sagar}, R. and {Sanchez}, A. and {Shastri}, P. and {Sillanp{\"a}{\"a}}, A. and {Strigachev}, A. and {Takalo}, L. and {Vennes}, S. and {Villata}, M. and {Villforth}, C. and {Wu}, J. and {Zhou}, X.},
        title = "{The 72-h WEBT microvariability observation of blazar S5 0716 + 714 in 2009}",
      journal = {\aap},
         year = 2013,
        month = oct,
       volume = {558},
          eid = {A92},
        pages = {A92},
          doi = {10.1051/0004-6361/201220236},
archivePrefix = {arXiv},
       eprint = {1310.4670},
 primaryClass = {astro-ph.HE},
       adsurl = {https://ui.adsabs.harvard.edu/abs/2013A&A...558A..92B}
}

@INPROCEEDINGS{2009ASPC..411..251M,
       author = {{Markwardt}, C.~B.},
        title = "{Non-linear Least-squares Fitting in IDL with MPFIT}",
    booktitle = {Astronomical Data Analysis Software and Systems XVIII},
         year = 2009,
       editor = {{Bohlender}, D.~A. and {Durand}, D. and {Dowler}, P.},
       series = {Astronomical Society of the Pacific Conference Series},
       volume = {411},
        month = sep,
        pages = {251},
          doi = {10.48550/arXiv.0902.2850},
archivePrefix = {arXiv},
       eprint = {0902.2850},
 primaryClass = {astro-ph.IM},
       adsurl = {https://ui.adsabs.harvard.edu/abs/2009ASPC..411..251M}
}

@INPROCEEDINGS{1997ASSL..218..163A,
       author = {{Alexander}, Tal},
        title = "{Is AGN Variability Correlated with Other AGN Properties? ZDCF Analysis of Small Samples of Sparse Light Curves}",
    booktitle = {Astronomical Time Series},
         year = 1997,
       editor = {{Maoz}, D. and {Sternberg}, A. and {Leibowitz}, E.~M.},
       series = {Astrophysics and Space Science Library},
       volume = {218},
        month = jan,
        pages = {163},
          doi = {10.1007/978-94-015-8941-3_14},
       adsurl = {https://ui.adsabs.harvard.edu/abs/1997ASSL..218..163A}
}

@ARTICLE{1996A&AS..116..403F,
       author = {{Fiorucci}, M. and {Tosti}, G.},
        title = "{VRI photometry of stars in the fields of 12 BL Lacertae objects.}",
      journal = {\aaps},
         year = 1996,
        month = may,
       volume = {116},
        pages = {403-407},
       adsurl = {https://ui.adsabs.harvard.edu/abs/1996A&AS..116..403F}
}

@INPROCEEDINGS{1995ASPC...77..437L,
       author = {{Landsman}, W.~B.},
        title = "{The IDL Astronomy User's Library}",
    booktitle = {Astronomical Data Analysis Software and Systems IV},
         year = 1995,
       editor = {{Shaw}, R.~A. and {Payne}, H.~E. and {Hayes}, J.~J.~E.},
       series = {Astronomical Society of the Pacific Conference Series},
       volume = {77},
        month = jan,
        pages = {437},
       adsurl = {https://ui.adsabs.harvard.edu/abs/1995ASPC...77..437L}
}

@ARTICLE{Kalita2023,
       author = {{Kalita}, Nibedita and {Yuan}, Yuhai and {Gu}, Minfeng and {Fan}, Junhui and {Mizuno}, Yosuke and {Jiang}, Peng and {Gupta}, Alok C. and {Zhou}, Hongyan and {Pan}, Xiang and {Strigachev}, Anton A. and {Bachev}, Rumen S. and {Cui}, Lang},
        title = "{Optical Flux and Spectral Variability of BL Lacertae during Its Historical High Outburst in 2020}",
      journal = {\apj},
         year = 2023,
        month = feb,
       volume = {943},
       number = {2},
          eid = {135},
        pages = {135},
          doi = {10.3847/1538-4357/aca801},
archivePrefix = {arXiv},
       eprint = {2212.04181},
 primaryClass = {astro-ph.HE},
       adsurl = {https://ui.adsabs.harvard.edu/abs/2023ApJ...943..135K}
}

@ARTICLE{Virtanen2020,
  author  = {Virtanen, Pauli and Gommers, Ralf and Oliphant, Travis E. and
            Haberland, Matt and Reddy, Tyler and Cournapeau, David and
            Burovski, Evgeni and Peterson, Pearu and Weckesser, Warren and
            Bright, Jonathan and {van der Walt}, St{\'e}fan J. and
            Brett, Matthew and Wilson, Joshua and Millman, K. Jarrod and
            Mayorov, Nikolay and Nelson, Andrew R. J. and Jones, Eric and
            Kern, Robert and Larson, Eric and Carey, C J and
            Polat, {\.I}lhan and Feng, Yu and Moore, Eric W. and
            {VanderPlas}, Jake and Laxalde, Denis and Perktold, Josef and
            Cimrman, Robert and Henriksen, Ian and Quintero, E. A. and
            Harris, Charles R. and Archibald, Anne M. and
            Ribeiro, Ant{\^o}nio H. and Pedregosa, Fabian and
            {van Mulbregt}, Paul and {SciPy 1.0 Contributors}},
  title   = {{{SciPy} 1.0: Fundamental Algorithms for Scientific
            Computing in Python}},
  journal = {Nature Methods},
  year    = {2020},
  volume  = {17},
  pages   = {261--272},
  adsurl  = {https://rdcu.be/b08Wh},
  doi     = {10.1038/s41592-019-0686-2},
}

@ARTICLE{Paneque2024,
       author = {{Paneque}, David and {Nozaki}, Seiya and {Bonnoli}, Giacomo and {Arbet-Engels}, Axel and {Garcia Soto}, Silvia and {Imazawa}, Ryo},
        title = "{Detection of flaring very-high-energy gamma-ray emission from BL Lacertae with the MAGIC telescopes}",
      journal = {The Astronomer's Telegram},
         year = 2024,
        month = oct,
       volume = {16861},
        pages = {1},
       adsurl = {https://ui.adsabs.harvard.edu/abs/2024ATel16861....1P}
}

@ARTICLE{Prince2021,
       author = {{Prince}, Raj},
        title = "{Broad-band study of BL Lac during flare of 2020: spectral evolution and emergence of HBL component}",
      journal = {\mnras},
         year = 2021,
        month = nov,
       volume = {507},
       number = {4},
        pages = {5602-5612},
          doi = {10.1093/mnras/stab2486},
archivePrefix = {arXiv},
       eprint = {2105.00221},
 primaryClass = {astro-ph.HE},
       adsurl = {https://ui.adsabs.harvard.edu/abs/2021MNRAS.507.5602P}
}

@ARTICLE{Majumdar2025,
       author = {{Majumdar}, Joysankar and {Maurya}, Sakshi and {Prince}, Raj},
        title = "{BL Lacertae under the flare of 2024: Probing temporal and spectral dynamics}",
      journal = {Journal of High Energy Astrophysics},
         year = 2025,
        month = aug,
       volume = {48},
          eid = {100402},
        pages = {100402},
          doi = {10.1016/j.jheap.2025.100402},
archivePrefix = {arXiv},
       eprint = {2505.18666},
 primaryClass = {astro-ph.HE},
       adsurl = {https://ui.adsabs.harvard.edu/abs/2025JHEAp..4800402M}
}

@ARTICLE{Agarwal2023,
       author = {{Agarwal}, A. and {Mihov}, B. and {Agrawal}, V. and {Zola}, S. and {{\"O}zd{\"o}nmez}, Aykut and {Ege}, Erg{\"u}n and {Slavcheva-Mihova}, L. and {Reichart}, D.~E. and {Caton}, D.~B. and {Das}, Avik Kumar},
        title = "{Analysis of the Intranight Variability of BL Lacertae during Its 2020 August Flare}",
      journal = {\apjs},
         year = 2023,
        month = apr,
       volume = {265},
       number = {2},
          eid = {51},
        pages = {51},
          doi = {10.3847/1538-4365/acbcbd},
archivePrefix = {arXiv},
       eprint = {2302.10177},
 primaryClass = {astro-ph.HE},
       adsurl = {https://ui.adsabs.harvard.edu/abs/2023ApJS..265...51A}
}

@ARTICLE{Jockers2000,
       author = {{Jockers}, K. and {Credner}, T. and {Bonev}, T. and {Kisele}, V.~N. and {Korsun}, P. and {Kulyk}, I. and {Rosenbush}, V. and {Andrienko}, A. and {Karpov}, N. and {Sergeev}, A. and {Tarady}, V.},
        title = "{Exploration of the solar system with the Two-Channel Focal Reducer at the 2m-RCC telescope of Pik Terskol Observatory}",
      journal = {Kinematika i Fizika Nebesnykh Tel Supplement},
         year = 2000,
        month = sep,
       volume = {3},
        pages = {13-18},
       adsurl = {https://ui.adsabs.harvard.edu/abs/2000KFNTS...3...13J}
}

@ARTICLE{Azzam2022,
       author = {{Azzam}, Yosry A. and {Elnagahy}, F.~I.~Y. and {Ali}, Gamal B. and {Essam}, A. and {Saad}, Somaya and {Ismail}, Hamed and {Zead}, I. and {Ahmed}, Nasser M. and {Yoshida}, Michitoshi and {Kawabata}, Koji S. and {Akitaya}, Hiroshi and {Shokry}, A. and {Hendy}, Y.~H.~M. and {Takey}, Ali and {Hamed}, G.~M. and {Mack}, Peter},
        title = "{Kottamia Faint Imaging Spectro-Polarimeter (KFISP): opto-mechanical design, software control and performance analysis}",
      journal = {Experimental Astronomy},
         year = 2022,
        month = feb,
       volume = {53},
       number = {1},
        pages = {45-70},
          doi = {10.1007/s10686-021-09802-z},
       adsurl = {https://ui.adsabs.harvard.edu/abs/2022ExA....53...45A}
}

@ARTICLE{Hart2023,
       author = {{Hart}, K. and {Shappee}, B.~J. and {Hey}, D. and {Kochanek}, C.~S. and {Stanek}, K.~Z. and {Lim}, L. and {Dobbs}, S. and {Tucker}, M. and {Jayasinghe}, T. and {Beacom}, J.~F. and {Boright}, T. and {Holoien}, T. and {Ong}, J.~M. Joel and {Prieto}, J.~L. and {Thompson}, T.~A. and {Will}, D.},
        title = "{ASAS-SN Sky Patrol V2.0}",
      journal = {arXiv e-prints},
         year = 2023,
        month = apr,
          eid = {arXiv:2304.03791},
        pages = {arXiv:2304.03791},
          doi = {10.48550/arXiv.2304.03791},
archivePrefix = {arXiv},
       eprint = {2304.03791},
 primaryClass = {astro-ph.IM},
       adsurl = {https://ui.adsabs.harvard.edu/abs/2023arXiv230403791H}
}

@ARTICLE{Stetson1987,
       author = {{Stetson}, Peter B.},
        title = "{DAOPHOT: A Computer Program for Crowded-Field Stellar Photometry}",
      journal = {\pasp},
         year = 1987,
        month = mar,
       volume = {99},
        pages = {191},
          doi = {10.1086/131977},
       adsurl = {https://ui.adsabs.harvard.edu/abs/1987PASP...99..191S}
}

@ARTICLE{Gonzalez2001,
       author = {{Gonz{\'a}lez-P{\'e}rez}, Jos{\'e} Nicol{\'a}s and {Kidger}, Mark R. and {Mart{\'\i}n-Luis}, Fabiola},
        title = "{Optical and Near-Infrared Calibration of AGN Field Stars: An All-Sky Network of Faint Stars Calibrated on the Landolt System}",
      journal = {\aj},
         year = 2001,
        month = oct,
       volume = {122},
       number = {4},
        pages = {2055-2098},
          doi = {10.1086/322129},
       adsurl = {https://ui.adsabs.harvard.edu/abs/2001AJ....122.2055G}
}

@software{2021ascl.soft12006K,
       author = {{Karpov}, Sergey},
        title = "{STDPipe: Simple Transient Detection Pipeline}",
 howpublished = {Astrophysics Source Code Library, record ascl:2112.006},
         year = 2021,
        month = dec,
          eid = {ascl:2112.006},
archivePrefix = {ascl},
       eprint = {2112.006},
       adsurl = {https://ui.adsabs.harvard.edu/abs/2021ascl.soft12006K}
}

@ARTICLE{Romero1999,
       author = {{Romero}, G.~E. and {Cellone}, S.~A. and {Combi}, J.~A.},
        title = "{Optical microvariability of southern AGNs}",
      journal = {\aaps},
         year = 1999,
        month = mar,
       volume = {135},
        pages = {477-486},
          doi = {10.1051/aas:1999184},
       adsurl = {https://ui.adsabs.harvard.edu/abs/1999A&AS..135..477R}
}

@INPROCEEDINGS{Miller1978,
       author = {{Miller}, J.~S. and {French}, H.~B. and {Hawley}, S.~A.},
        title = "{Optical spectra of BL Lacertae objects.}",
    booktitle = {BL Lac Objects},
         year = 1978,
       editor = {{Wolfe}, A.~M.},
        month = jan,
        pages = {176-187},
       adsurl = {https://ui.adsabs.harvard.edu/abs/1978bllo.conf..176M}
}

@ARTICLE{Yuan2023,
       author = {{Yuan}, Y.~H. and {Du}, G.~J. and {Fan}, J.~H. and {Liu}, Y. and {Yang}, J.~H. and {Ding}, G.~Z. and {Pei}, Z.~Y.},
        title = "{Optical Monitoring and Intraday Variabilities of BL Lacertae}",
      journal = {\apjs},
         year = 2023,
        month = dec,
       volume = {269},
       number = {2},
          eid = {60},
        pages = {60},
          doi = {10.3847/1538-4365/ad04d5},
       adsurl = {https://ui.adsabs.harvard.edu/abs/2023ApJS..269...60Y}
}

@MISC{Bradley2022,
       author = {{Bradley}, Larry and {Sip{\H{o}}cz}, Brigitta and {Robitaille}, Thomas and {Tollerud}, Erik and {Vin{\'\i}cius}, Z{\'e} and {Deil}, Christoph and {Barbary}, Kyle and {Wilson}, Tom J and {Busko}, Ivo and {Donath}, Axel and {G{\"u}nther}, Hans Moritz and {Cara}, Mihai and {Lim}, P.~L. and {Me{\ss}linger}, Sebastian and {Conseil}, Simon and {Burnett}, Zach and {Bostroem}, Azalee and {Droettboom}, Michael and {Bray}, E.~M. and {Andersen Bratholm}, Lars and {Jamieson}, William and {Ginsburg}, Adam and {Barentsen}, Geert and {Craig}, Matt and {Morris}, Brett M. and {Perrin}, Marshall and {Rathi}, Shivangee and {Pascual}, Sergio and {Perren}, Gabriel and {Georgiev}, Iskren Y.},
        title = "{astropy/photutils: 1.10.0}",
         year = 2023,
        month = nov,
          eid = {10.5281/zenodo.596036},
          doi = {10.5281/zenodo.596036},
      version = {1.10.0},
    publisher = {Zenodo},
       adsurl = {https://ui.adsabs.harvard.edu/abs/2022zndo....596036B}
}

@MISC{Craig2017,
       author = {{Craig}, Matt and {Crawford}, Steve and {Seifert}, Michael and {Robitaille}, Thomas and {Sipocz}, Brigitta and {Walawender}, Josh and {Vin{\'\i}cius}, Z{\'e} and {Ninan}, Joe Philip and {Droettboom}, Michael and {Youn}, Jiyong and {Tollerud}, Erik and {Bray}, Erik and {Walkerna22} and {Reddy Janga}, VSN and {Stottsco} and {G{\"u}nther}, Hans Moritz and {Rol}, Evert and {Bach}, Yoonsoo P. and {Bradley}, Larry and {Deil}, Christoph and {Price-Whelan}, Adrian and {Barbary}, Kyle and {Horton}, Anthony and {Schoenell}, William and {Nathan} and {Gasdia}, Forrest and {Nelson}, Stefan and {Streicher}, Ole},
        title = "{astropy/ccdproc: v1.3.0.post1}",
         year = 2017,
        month = dec,
          eid = {10.5281/zenodo.1069648},
          doi = {10.5281/zenodo.1069648},
      version = {v1.3.0.post1},
    publisher = {Zenodo},
       adsurl = {https://ui.adsabs.harvard.edu/abs/2017zndo...1069648C}
}

@ARTICLE{Lang2010,
       author = {{Lang}, Dustin and {Hogg}, David W. and {Mierle}, Keir and {Blanton}, Michael and {Roweis}, Sam},
        title = "{Astrometry.net: Blind Astrometric Calibration of Arbitrary Astronomical Images}",
      journal = {\aj},
         year = 2010,
        month = may,
       volume = {139},
       number = {5},
        pages = {1782-1800},
          doi = {10.1088/0004-6256/139/5/1782},
archivePrefix = {arXiv},
       eprint = {0910.2233},
 primaryClass = {astro-ph.IM},
       adsurl = {https://ui.adsabs.harvard.edu/abs/2010AJ....139.1782L}
}

@ARTICLE{Dokkum2001,
       author = {{van Dokkum}, Pieter G.},
        title = "{Cosmic-Ray Rejection by Laplacian Edge Detection}",
      journal = {\pasp},
         year = 2001,
        month = nov,
       volume = {113},
       number = {789},
        pages = {1420-1427},
          doi = {10.1086/323894},
archivePrefix = {arXiv},
       eprint = {astro-ph/0108003},
 primaryClass = {astro-ph},
       adsurl = {https://ui.adsabs.harvard.edu/abs/2001PASP..113.1420V}
}

@ARTICLE{Diego2015,
       author = {{de Diego}, Jos{\'e} A. and {Polednikova}, Jana and {Bongiovanni}, Angel and {P{\'e}rez Garc{\'\i}a}, Ana M. and {De Leo}, Mario A. and {Verdugo}, Tom{\'a}s and {Cepa}, Jordi},
        title = "{Testing Microvariability in Quasar Differential Light Curves Using Several Field Stars}",
      journal = {\aj},
         year = 2015,
        month = aug,
       volume = {150},
       number = {2},
          eid = {44},
        pages = {44},
          doi = {10.1088/0004-6256/150/2/44},
archivePrefix = {arXiv},
       eprint = {1505.02113},
 primaryClass = {astro-ph.IM},
       adsurl = {https://ui.adsabs.harvard.edu/abs/2015AJ....150...44D}
}

@ARTICLE{Diego2014,
       author = {{de Diego}, Jos{\'e} A.},
        title = "{On the Reliability of Microvariability Tests in Quasars}",
      journal = {\aj},
         year = 2014,
        month = nov,
       volume = {148},
       number = {5},
          eid = {93},
        pages = {93},
          doi = {10.1088/0004-6256/148/5/93},
archivePrefix = {arXiv},
       eprint = {1409.0468},
 primaryClass = {astro-ph.GA},
       adsurl = {https://ui.adsabs.harvard.edu/abs/2014AJ....148...93D}
}

@INPROCEEDINGS{Dhalla2010,
       author = {{Dhalla}, Sarah M. and {Webb}, J.~R. and {Bhatta}, G. and {Pollock}, J.~T.},
        title = "{The Nature of Microvariability in Blazar 0716+71}",
    booktitle = {American Astronomical Society Meeting Abstracts \#215},
         year = 2010,
       series = {American Astronomical Society Meeting Abstracts},
       volume = {215},
        month = jan,
          eid = {434.11},
        pages = {434.11},
       adsurl = {https://ui.adsabs.harvard.edu/abs/2010AAS...21543411D}
}

@ARTICLE{Abdo2010,
       author = {{Abdo}, A.~A. and {Ackermann}, M. and {Ajello}, M. and {Antolini}, E. and {Baldini}, L. and {Ballet}, J. and {Barbiellini}, G. and {Bastieri}, D. and {Bechtol}, K. and {Bellazzini}, R. and {Berenji}, B. and {Blandford}, R.~D. and {Bloom}, E.~D. and {Bonamente}, E. and {Borgland}, A.~W. and {Bouvier}, A. and {Bregeon}, J. and {Brez}, A. and {Brigida}, M. and {Bruel}, P. and {Buehler}, R. and {Burnett}, T.~H. and {Buson}, S. and {Caliandro}, G.~A. and {Cameron}, R.~A. and {Caraveo}, P.~A. and {Carrigan}, S. and {Casandjian}, J.~M. and {Cavazzuti}, E. and {Cecchi}, C. and {{\c{C}}elik}, {\"O}. and {Chekhtman}, A. and {Cheung}, C.~C. and {Chiang}, J. and {Ciprini}, S. and {Claus}, R. and {Cohen-Tanugi}, J. and {Cominsky}, L.~R. and {Conrad}, J. and {Costamante}, L. and {Cutini}, S. and {Dermer}, C.~D. and {de Angelis}, A. and {de Palma}, F. and {Silva}, E. do Couto e. and {Drell}, P.~S. and {Dubois}, R. and {Dumora}, D. and {Farnier}, C. and {Favuzzi}, C. and {Fegan}, S.~J. and {Focke}, W.~B. and {Fortin}, P. and {Frailis}, M. and {Fukazawa}, Y. and {Funk}, S. and {Fusco}, P. and {Gargano}, F. and {Gasparrini}, D. and {Gehrels}, N. and {Germani}, S. and {Giebels}, B. and {Giglietto}, N. and {Giommi}, P. and {Giordano}, F. and {Glanzman}, T. and {Godfrey}, G. and {Grenier}, I.~A. and {Grondin}, M. -H. and {Grove}, J.~E. and {Guiriec}, S. and {Hadasch}, D. and {Hayashida}, M. and {Hays}, E. and {Healey}, S.~E. and {Horan}, D. and {Hughes}, R.~E. and {Itoh}, R. and {J{\'o}hannesson}, G. and {Johnson}, A.~S. and {Johnson}, W.~N. and {Kamae}, T. and {Katagiri}, H. and {Kataoka}, J. and {Kawai}, N. and {Kn{\"o}dlseder}, J. and {Kuss}, M. and {Lande}, J. and {Larsson}, S. and {Latronico}, L. and {Lemoine-Goumard}, M. and {Longo}, F. and {Loparco}, F. and {Lott}, B. and {Lovellette}, M.~N. and {Lubrano}, P. and {Madejski}, G.~M. and {Makeev}, A. and {Massaro}, E. and {Mazziotta}, M.~N. and {McEnery}, J.~E. and {Michelson}, P.~F. and {Mitthumsiri}, W. and {Mizuno}, T. and {Moiseev}, A.~A. and {Monte}, C. and {Monzani}, M.~E. and {Morselli}, A. and {Moskalenko}, I.~V. and {Mueller}, M. and {Murgia}, S. and {Nolan}, P.~L. and {Norris}, J.~P. and {Nuss}, E. and {Ohno}, M. and {Ohsugi}, T. and {Omodei}, N. and {Orlando}, E. and {Ormes}, J.~F. and {Ozaki}, M. and {Panetta}, J.~H. and {Parent}, D. and {Pelassa}, V. and {Pepe}, M. and {Pesce-Rollins}, M. and {Piron}, F. and {Porter}, T.~A. and {Rain{\`o}}, S. and {Rando}, R. and {Razzano}, M. and {Reimer}, A. and {Reimer}, O. and {Ritz}, S. and {Rodriguez}, A.~Y. and {Romani}, R.~W. and {Roth}, M. and {Ryde}, F. and {Sadrozinski}, H.~F. -W. and {Sander}, A. and {Scargle}, J.~D. and {Sgr{\`o}}, C. and {Shaw}, M.~S. and {Smith}, P.~D. and {Spandre}, G. and {Spinelli}, P. and {Starck}, J. -L. and {Strickman}, M.~S. and {Suson}, D.~J. and {Takahashi}, H. and {Takahashi}, T. and {Tanaka}, T. and {Thayer}, J.~B. and {Thayer}, J.~G. and {Thompson}, D.~J. and {Tibaldo}, L. and {Torres}, D.~F. and {Tosti}, G. and {Tramacere}, A. and {Uchiyama}, Y. and {Usher}, T.~L. and {Vasileiou}, V. and {Vilchez}, N. and {Vitale}, V. and {Waite}, A.~P. and {Wallace}, E. and {Wang}, P. and {Winer}, B.~L. and {Wood}, K.~S. and {Yang}, Z. and {Ylinen}, T. and {Ziegler}, M.},
        title = "{Gamma-ray Light Curves and Variability of Bright Fermi-detected Blazars}",
      journal = {\apj},
         year = 2010,
        month = oct,
       volume = {722},
       number = {1},
        pages = {520-542},
          doi = {10.1088/0004-637X/722/1/520},
archivePrefix = {arXiv},
       eprint = {1004.0348},
 primaryClass = {astro-ph.HE},
       adsurl = {https://ui.adsabs.harvard.edu/abs/2010ApJ...722..520A}
}

@PHDTHESIS{Nestoras2015,
       author = {{Nestoras}, Ioannis},
        title = "{Broadband radio jet emission and variability of {\ensuremath{\gamma}}-ray blazars}",
       school = {Andreas Eckart University of Cologne, Germany},
         year = 2015,
        month = jul,
       adsurl = {https://ui.adsabs.harvard.edu/abs/2015PhDT........93N}
}

@ARTICLE{Liu2019,
       author = {{Liu}, H.~T. and {Feng}, Hai Cheng and {Xin}, Y.~X. and {Bai}, J.~M. and {Li}, S.~K. and {Wang}, Fang},
        title = "{Search for Intra-day Optical Variability in {\ensuremath{\gamma}}-Ray-loud Blazars S5 0716+714 and 3C 273}",
      journal = {\apj},
         year = 2019,
        month = aug,
       volume = {880},
       number = {2},
          eid = {155},
        pages = {155},
          doi = {10.3847/1538-4357/ab29fc},
archivePrefix = {arXiv},
       eprint = {1906.06568},
 primaryClass = {astro-ph.GA},
       adsurl = {https://ui.adsabs.harvard.edu/abs/2019ApJ...880..155L}
}

@ARTICLE{Pandey2020,
       author = {{Pandey}, Ashwani and {Gupta}, Alok C. and {Kurtanidze}, Sofia O. and {Wiita}, Paul J. and {Damljanovic}, G. and {Bachev}, R. and {Zhang}, Jin and {Kurtanidze}, O.~M. and {Darriba}, A. and {Chigladze}, R.~A. and {Latev}, G. and {Nikolashvili}, M.~G. and {Peneva}, S. and {Semkov}, E. and {Strigachev}, A. and {Tiwari}, S.~N. and {Vince}, O.},
        title = "{Optical Variability of the TeV Blazar 1ES 0806+524 on Diverse Timescales}",
      journal = {\apj},
         year = 2020,
        month = feb,
       volume = {890},
       number = {1},
          eid = {72},
        pages = {72},
          doi = {10.3847/1538-4357/ab698e},
archivePrefix = {arXiv},
       eprint = {2001.02398},
 primaryClass = {astro-ph.HE},
       adsurl = {https://ui.adsabs.harvard.edu/abs/2020ApJ...890...72P}
}

@ARTICLE{Pandian2022,
       author = {{Pandian}, K. Subbu Ulaganatha and {Natarajan}, A. and {Stalin}, C.~S. and {Pandey}, Ashwani and {Muneer}, S. and {Natarajan}, B.},
        title = "{Intra-night optical variability monitoring of {\ensuremath{\gamma}} -ray emitting blazars}",
      journal = {Journal of Astrophysics and Astronomy},
         year = 2022,
        month = dec,
       volume = {43},
       number = {2},
          eid = {48},
        pages = {48},
          doi = {10.1007/s12036-022-09826-7},
archivePrefix = {arXiv},
       eprint = {2202.07353},
 primaryClass = {astro-ph.HE},
       adsurl = {https://ui.adsabs.harvard.edu/abs/2022JApA...43...48P}
}

@ARTICLE{Tripathi2024,
       author = {{Tripathi}, Tushar and {Gupta}, Alok C. and {Takey}, Ali and {Bachev}, Rumen and {Vince}, Oliver and {Strigachev}, Anton and {Kushwaha}, Pankaj and {Elhosseiny}, E.~G. and {Wiita}, Paul J. and {Damljanovic}, G. and {Dhiman}, Vinit and {Fouad}, A. and {Gaur}, Haritma and {Gu}, Minfeng and {Hamed}, G.~E. and {Kishore}, Shubham and {Kurtenkov}, A. and {Rastogi}, Shantanu and {Semkov}, E. and {Zead}, I. and {Zhang}, Zhongli},
        title = "{Optical intraday variability of the blazar S5 0716+714}",
      journal = {\mnras},
         year = 2024,
        month = jan,
       volume = {527},
       number = {3},
        pages = {5220-5237},
          doi = {10.1093/mnras/stad3574},
archivePrefix = {arXiv},
       eprint = {2311.10358},
 primaryClass = {astro-ph.HE},
       adsurl = {https://ui.adsabs.harvard.edu/abs/2024MNRAS.527.5220T}
}

@ARTICLE{Polednikova2016,
       author = {{Polednikova}, J. and {Ederoclite}, A. and {de Diego}, J.~A. and {Cepa}, J. and {Gonz{\'a}lez-Serrano}, J.~I. and {Bongiovanni}, A. and {Oteo}, I. and {Garc{\'\i}a}, A.~M. P{\'e}rez and {P{\'e}rez-Mart{\'\i}nez}, R. and {Pintos-Castro}, I. and {Ram{\'o}n-P{\'e}rez}, M. and {S{\'a}nchez-Portal}, M.},
        title = "{Detecting microvariability in type 2 quasars using enhanced F-test}",
      journal = {\mnras},
         year = 2016,
        month = aug,
       volume = {460},
       number = {4},
        pages = {3950-3959},
          doi = {10.1093/mnras/stw1252},
archivePrefix = {arXiv},
       eprint = {1605.09424},
 primaryClass = {astro-ph.GA},
       adsurl = {https://ui.adsabs.harvard.edu/abs/2016MNRAS.460.3950P}
}

@ARTICLE{Pandey2019,
       author = {{Pandey}, Ashwani and {Gupta}, Alok C. and {Wiita}, Paul J. and {Tiwari}, S.~N.},
        title = "{Optical Flux and Spectral Variability of the TeV Blazar PG 1553+113}",
      journal = {\apj},
         year = 2019,
        month = feb,
       volume = {871},
       number = {2},
          eid = {192},
        pages = {192},
          doi = {10.3847/1538-4357/aaf974},
archivePrefix = {arXiv},
       eprint = {1901.06696},
 primaryClass = {astro-ph.HE},
       adsurl = {https://ui.adsabs.harvard.edu/abs/2019ApJ...871..192P}
}

@ARTICLE{Joshi2011,
       author = {{Joshi}, Ravi and {Chand}, Hum and {Gupta}, Alok C. and {Wiita}, Paul J.},
        title = "{Optical microvariability properties of BALQSOs}",
      journal = {\mnras},
         year = 2011,
        month = apr,
       volume = {412},
       number = {4},
        pages = {2717-2728},
          doi = {10.1111/j.1365-2966.2010.18099.x},
archivePrefix = {arXiv},
       eprint = {1011.5611},
 primaryClass = {astro-ph.CO},
       adsurl = {https://ui.adsabs.harvard.edu/abs/2011MNRAS.412.2717J}
}

@ARTICLE{Heidt1996,
       author = {{Heidt}, J. and {Wagner}, S.~J.},
        title = "{Statistics of optical intraday variability in a complete sample of radio-selected BL Lacertae objects.}",
      journal = {\aap},
         year = 1996,
        month = jan,
       volume = {305},
        pages = {42},
          doi = {10.48550/arXiv.astro-ph/9506032},
archivePrefix = {arXiv},
       eprint = {astro-ph/9506032},
 primaryClass = {astro-ph},
       adsurl = {https://ui.adsabs.harvard.edu/abs/1996A&A...305...42H}
}

@INPROCEEDINGS{Edelson1988,
       author = {{Edelson}, R.~A. and {Krolik}, J.~H.},
        title = "{The discrete correlation function: a new method for analysing unevenly sampled variability data}",
    booktitle = {ESA Special Publication},
         year = 1988,
       editor = {{Longdon}, N. and {Rolfe}, E.~J.},
       series = {ESA Special Publication},
       volume = {2},
        month = jun,
        pages = {387-390},
       adsurl = {https://ui.adsabs.harvard.edu/abs/1988ESASP.281b.387E}
}

@ARTICLE{1998A&A...333..231B,
       author = {{Bessell}, M.~S. and {Castelli}, F. and {Plez}, B.},
        title = "{Model atmospheres broad-band colors, bolometric corrections and temperature calibrations for O - M stars}",
      journal = {\aap},
         year = 1998,
        month = may,
       volume = {333},
        pages = {231-250},
       adsurl = {https://ui.adsabs.harvard.edu/abs/1998A&A...333..231B}
}

@ARTICLE{Cardelli1989,
       author = {{Cardelli}, Jason A. and {Clayton}, Geoffrey C. and {Mathis}, John S.},
        title = "{The Relationship between Infrared, Optical, and Ultraviolet Extinction}",
      journal = {\apj},
         year = 1989,
        month = oct,
       volume = {345},
        pages = {245},
          doi = {10.1086/167900},
       adsurl = {https://ui.adsabs.harvard.edu/abs/1989ApJ...345..245C}
}

@ARTICLE{Schlafly2011,
       author = {{Schlafly}, Edward F. and {Finkbeiner}, Douglas P.},
        title = "{Measuring Reddening with Sloan Digital Sky Survey Stellar Spectra and Recalibrating SFD}",
      journal = {\apj},
         year = 2011,
        month = aug,
       volume = {737},
       number = {2},
          eid = {103},
        pages = {103},
          doi = {10.1088/0004-637X/737/2/103},
archivePrefix = {arXiv},
       eprint = {1012.4804},
 primaryClass = {astro-ph.GA},
       adsurl = {https://ui.adsabs.harvard.edu/abs/2011ApJ...737..103S}
}

@ARTICLE{1999PASJ...51..253M,
       author = {{Matsumoto}, Katsura and {Kato}, Taichi and {Nogami}, Daisaku and {Kawaguchi}, Toshihiro and {Kinnunen}, Timo and {Poyner}, Gary},
        title = "{The 1997 Outburst of BL Lacertae and Detection of a 0.6-mag Rapid Variation}",
      journal = {\pasj},
         year = 1999,
        month = apr,
       volume = {51},
        pages = {253-256},
          doi = {10.1093/pasj/51.2.253},
       adsurl = {https://ui.adsabs.harvard.edu/abs/1999PASJ...51..253M}
}

@ARTICLE{1999ApJ...522..846X,
       author = {{Xie}, G.~Z. and {Li}, K.~H. and {Zhang}, X. and {Bai}, J.~M. and {Liu}, W.~W.},
        title = "{Optical Monitoring Sample of the GEV Gamma-Ray-loud Blazars}",
      journal = {\apj},
         year = 1999,
        month = sep,
       volume = {522},
       number = {2},
        pages = {846-862},
          doi = {10.1086/307673},
       adsurl = {https://ui.adsabs.harvard.edu/abs/1999ApJ...522..846X}
}

@ARTICLE{2003A&A...397..565P,
       author = {{Papadakis}, I.~E. and {Boumis}, P. and {Samaritakis}, V. and {Papamastorakis}, J.},
        title = "{Multi-band optical micro-variability observations of BL Lacertae}",
      journal = {\aap},
         year = 2003,
        month = jan,
       volume = {397},
        pages = {565-573},
          doi = {10.1051/0004-6361:20021581},
archivePrefix = {arXiv},
       eprint = {astro-ph/0211083},
 primaryClass = {astro-ph},
       adsurl = {https://ui.adsabs.harvard.edu/abs/2003A&A...397..565P}
}

@ARTICLE{1995PASP..107..803U,
       author = {{Urry}, C. Megan and {Padovani}, Paolo},
        title = "{Unified Schemes for Radio-Loud Active Galactic Nuclei}",
      journal = {\pasp},
         year = 1995,
        month = sep,
       volume = {107},
        pages = {803},
          doi = {10.1086/133630},
archivePrefix = {arXiv},
       eprint = {astro-ph/9506063},
 primaryClass = {astro-ph},
       adsurl = {https://ui.adsabs.harvard.edu/abs/1995PASP..107..803U}
}

@ARTICLE{1998MNRAS.299..433F,
       author = {{Fossati}, G. and {Maraschi}, L. and {Celotti}, A. and {Comastri}, A. and {Ghisellini}, G.},
        title = "{A unifying view of the spectral energy distributions of blazars}",
      journal = {\mnras},
         year = 1998,
        month = sep,
       volume = {299},
       number = {2},
        pages = {433-448},
          doi = {10.1046/j.1365-8711.1998.01828.x},
archivePrefix = {arXiv},
       eprint = {astro-ph/9804103},
 primaryClass = {astro-ph},
       adsurl = {https://ui.adsabs.harvard.edu/abs/1998MNRAS.299..433F}
}

@ARTICLE{2010ApJ...716...30A,
       author = {{Abdo}, A.~A. and {Ackermann}, M. and {Agudo}, I. and {Ajello}, M. and {Aller}, H.~D. and {Aller}, M.~F. and {Angelakis}, E. and {Arkharov}, A.~A. and {Axelsson}, M. and {Bach}, U. and {Baldini}, L. and {Ballet}, J. and {Barbiellini}, G. and {Bastieri}, D. and {Baughman}, B.~M. and {Bechtol}, K. and {Bellazzini}, R. and {Benitez}, E. and {Berdyugin}, A. and {Berenji}, B. and {Blandford}, R.~D. and {Bloom}, E.~D. and {Boettcher}, M. and {Bonamente}, E. and {Borgland}, A.~W. and {Bregeon}, J. and {Brez}, A. and {Brigida}, M. and {Bruel}, P. and {Burnett}, T.~H. and {Burrows}, D. and {Buson}, S. and {Caliandro}, G.~A. and {Calzoletti}, L. and {Cameron}, R.~A. and {Capalbi}, M. and {Caraveo}, P.~A. and {Carosati}, D. and {Casandjian}, J.~M. and {Cavazzuti}, E. and {Cecchi}, C. and {{\c{C}}elik}, {\"O}. and {Charles}, E. and {Chaty}, S. and {Chekhtman}, A. and {Chen}, W.~P. and {Chiang}, J. and {Chincarini}, G. and {Ciprini}, S. and {Claus}, R. and {Cohen-Tanugi}, J. and {Colafrancesco}, S. and {Cominsky}, L.~R. and {Conrad}, J. and {Costamante}, L. and {Cutini}, S. and {D'ammando}, F. and {Deitrick}, R. and {D'Elia}, V. and {Dermer}, C.~D. and {de Angelis}, A. and {de Palma}, F. and {Digel}, S.~W. and {Donnarumma}, I. and {Silva}, E. do Couto e. and {Drell}, P.~S. and {Dubois}, R. and {Dultzin}, D. and {Dumora}, D. and {Falcone}, A. and {Farnier}, C. and {Favuzzi}, C. and {Fegan}, S.~J. and {Focke}, W.~B. and {Forn{\'e}}, E. and {Fortin}, P. and {Frailis}, M. and {Fuhrmann}, L. and {Fukazawa}, Y. and {Funk}, S. and {Fusco}, P. and {G{\'o}mez}, J.~L. and {Gargano}, F. and {Gasparrini}, D. and {Gehrels}, N. and {Germani}, S. and {Giebels}, B. and {Giglietto}, N. and {Giommi}, P. and {Giordano}, F. and {Giuliani}, A. and {Glanzman}, T. and {Godfrey}, G. and {Grenier}, I.~A. and {Gronwall}, C. and {Grove}, J.~E. and {Guillemot}, L. and {Guiriec}, S. and {Gurwell}, M.~A. and {Hadasch}, D. and {Hanabata}, Y. and {Harding}, A.~K. and {Hayashida}, M. and {Hays}, E. and {Healey}, S.~E. and {Heidt}, J. and {Hiriart}, D. and {Horan}, D. and {Hoversten}, E.~A. and {Hughes}, R.~E. and {Itoh}, R. and {Jackson}, M.~S. and {J{\'o}hannesson}, G. and {Johnson}, A.~S. and {Johnson}, W.~N. and {Jorstad}, S.~G. and {Kadler}, M. and {Kamae}, T. and {Katagiri}, H. and {Kataoka}, J. and {Kawai}, N. and {Kennea}, J. and {Kerr}, M. and {Kimeridze}, G. and {Kn{\"o}dlseder}, J. and {Kocian}, M.~L. and {Kopatskaya}, E.~N. and {Koptelova}, E. and {Konstantinova}, T.~S. and {Kovalev}, Y.~Y. and {Kovalev}, Yu. A. and {Kurtanidze}, O.~M. and {Kuss}, M. and {Lande}, J. and {Larionov}, V.~M. and {Latronico}, L. and {Leto}, P. and {Lindfors}, E. and {Longo}, F. and {Loparco}, F. and {Lott}, B. and {Lovellette}, M.~N. and {Lubrano}, P. and {Madejski}, G.~M. and {Makeev}, A. and {Marchegiani}, P. and {Marscher}, A.~P. and {Marshall}, F. and {Max-Moerbeck}, W. and {Mazziotta}, M.~N. and {McConville}, W. and {McEnery}, J.~E. and {Meurer}, C. and {Michelson}, P.~F. and {Mitthumsiri}, W. and {Mizuno}, T. and {Moiseev}, A.~A. and {Monte}, C. and {Monzani}, M.~E. and {Morselli}, A. and {Moskalenko}, I.~V. and {Murgia}, S. and {Nestoras}, I. and {Nilsson}, K. and {Nizhelsky}, N.~A. and {Nolan}, P.~L. and {Norris}, J.~P. and {Nuss}, E. and {Ohsugi}, T. and {Ojha}, R. and {Omodei}, N. and {Orlando}, E. and {Ormes}, J.~F. and {Osborne}, J. and {Ozaki}, M. and {Pacciani}, L. and {Padovani}, P. and {Pagani}, C. and {Page}, K. and {Paneque}, D. and {Panetta}, J.~H. and {Parent}, D. and {Pasanen}, M. and {Pavlidou}, V. and {Pelassa}, V. and {Pepe}, M. and {Perri}, M. and {Pesce-Rollins}, M. and {Piranomonte}, S. and {Piron}, F. and {Pittori}, C. and {Porter}, T.~A. and {Puccetti}, S. and {Rahoui}, F. and {Rain{\`o}}, S. and {Raiteri}, C. and {Rando}, R. and {Razzano}, M. and {Reimer}, A. and {Reimer}, O.},
        title = "{The Spectral Energy Distribution of Fermi Bright Blazars}",
      journal = {\apj},
         year = 2010,
        month = jun,
       volume = {716},
       number = {1},
        pages = {30-70},
          doi = {10.1088/0004-637X/716/1/30},
archivePrefix = {arXiv},
       eprint = {0912.2040},
 primaryClass = {astro-ph.CO},
       adsurl = {https://ui.adsabs.harvard.edu/abs/2010ApJ...716...30A}
}

@ARTICLE{1995ARA&A..33..163W,
       author = {{Wagner}, S.~J. and {Witzel}, A.},
        title = "{Intraday Variability In Quasars and BL Lac Objects}",
      journal = {\araa},
         year = 1995,
        month = jan,
       volume = {33},
        pages = {163-198},
          doi = {10.1146/annurev.aa.33.090195.001115},
       adsurl = {https://ui.adsabs.harvard.edu/abs/1995ARA&A..33..163W}
}

@ARTICLE{2004A&A...422..505G,
       author = {{Gupta}, A.~C. and {Banerjee}, D.~P.~K. and {Ashok}, N.~M. and {Joshi}, U.~C.},
        title = "{Near infrared intraday variability of Mrk 421}",
      journal = {\aap},
         year = 2004,
        month = aug,
       volume = {422},
        pages = {505-508},
          doi = {10.1051/0004-6361:20040306},
archivePrefix = {arXiv},
       eprint = {astro-ph/0405186},
 primaryClass = {astro-ph},
       adsurl = {https://ui.adsabs.harvard.edu/abs/2004A&A...422..505G}
}

@ARTICLE{2003A&A...400..487C,
       author = {{Ciprini}, S. and {Tosti}, G. and {Raiteri}, C.~M. and {Villata}, M. and {Ibrahimov}, M.~A. and {Nucciarelli}, G. and {Lanteri}, L.},
        title = "{Optical variability of the BL Lacertae object GC 0109+224. Multiband behaviour and time scales from a 7-years monitoring campaign}",
      journal = {\aap},
         year = 2003,
        month = mar,
       volume = {400},
        pages = {487-498},
          doi = {10.1051/0004-6361:20030045},
archivePrefix = {arXiv},
       eprint = {astro-ph/0301325},
 primaryClass = {astro-ph},
       adsurl = {https://ui.adsabs.harvard.edu/abs/2003A&A...400..487C}
}

@ARTICLE{1989Natur.337..627M,
       author = {{Miller}, H.~R. and {Carini}, M.~T. and {Goodrich}, B.~D.},
        title = "{Detection of microvariability for BL Lacertae objects}",
      journal = {\nat},
         year = 1989,
        month = feb,
       volume = {337},
       number = {6208},
        pages = {627-629},
          doi = {10.1038/337627a0},
       adsurl = {https://ui.adsabs.harvard.edu/abs/1989Natur.337..627M}
}

@ARTICLE{2016MNRAS.458.1127G,
       author = {{Gupta}, Alok C. and {Agarwal}, A. and {Bhagwan}, J. and {Strigachev}, A. and {Bachev}, R. and {Semkov}, E. and {Gaur}, H. and {Damljanovic}, G. and {Vince}, O. and {Wiita}, Paul J.},
        title = "{Multiband optical variability of three TeV blazars on diverse time-scales}",
      journal = {\mnras},
         year = 2016,
        month = may,
       volume = {458},
       number = {1},
        pages = {1127-1137},
          doi = {10.1093/mnras/stw377},
archivePrefix = {arXiv},
       eprint = {1602.04200},
 primaryClass = {astro-ph.HE},
       adsurl = {https://ui.adsabs.harvard.edu/abs/2016MNRAS.458.1127G}
}

@ARTICLE{2019MNRAS.488.4093A,
       author = {{Agarwal}, Aditi and {Cellone}, Sergio A. and {Andruchow}, Ileana and {Mammana}, Luis and {Singh}, Mridweeka and {Anupama}, G.~C. and {Mihov}, B. and {Raj}, Ashish and {Slavcheva-Mihova}, L. and {{\"O}zd{\"o}nmez}, Aykut and {Ege}, Erg{\"u}n},
        title = "{Multiband optical variability of 3C 279 on diverse time-scales}",
      journal = {\mnras},
         year = 2019,
        month = sep,
       volume = {488},
       number = {3},
        pages = {4093-4105},
          doi = {10.1093/mnras/stz1981},
archivePrefix = {arXiv},
       eprint = {1908.01465},
 primaryClass = {astro-ph.HE},
       adsurl = {https://ui.adsabs.harvard.edu/abs/2019MNRAS.488.4093A}
}

@ARTICLE{2022MNRAS.513.4645S,
       author = {{Sahakyan}, N. and {Giommi}, P.},
        title = "{A 13-yr-long broad-band view of BL Lac}",
      journal = {\mnras},
         year = 2022,
        month = jul,
       volume = {513},
       number = {3},
        pages = {4645-4656},
          doi = {10.1093/mnras/stac1011},
archivePrefix = {arXiv},
       eprint = {2108.12232},
 primaryClass = {astro-ph.HE},
       adsurl = {https://ui.adsabs.harvard.edu/abs/2022MNRAS.513.4645S}
}

@ARTICLE{2020ApJ...892..105A,
       author = {{Ajello}, M. and {Angioni}, R. and {Axelsson}, M. and {Ballet}, J. and {Barbiellini}, G. and {Bastieri}, D. and {Becerra Gonzalez}, J. and {Bellazzini}, R. and {Bissaldi}, E. and {Bloom}, E.~D. and {Bonino}, R. and {Bottacini}, E. and {Bruel}, P. and {Buson}, S. and {Cafardo}, F. and {Cameron}, R.~A. and {Cavazzuti}, E. and {Chen}, S. and {Cheung}, C.~C. and {Ciprini}, S. and {Costantin}, D. and {Cutini}, S. and {D'Ammando}, F. and {de la Torre Luque}, P. and {de Menezes}, R. and {de Palma}, F. and {Desai}, A. and {Di Lalla}, N. and {Di Venere}, L. and {Dom{\'\i}nguez}, A. and {Dirirsa}, F. Fana and {Ferrara}, E.~C. and {Finke}, J. and {Franckowiak}, A. and {Fukazawa}, Y. and {Funk}, S. and {Fusco}, P. and {Gargano}, F. and {Garrappa}, S. and {Gasparrini}, D. and {Giglietto}, N. and {Giordano}, F. and {Giroletti}, M. and {Green}, D. and {Grenier}, I.~A. and {Guiriec}, S. and {Harita}, S. and {Hays}, E. and {Horan}, D. and {Itoh}, R. and {J{\'o}hannesson}, G. and {Kovac'evic'}, M. and {Krauss}, F. and {Kreter}, M. and {Kuss}, M. and {Larsson}, S. and {Leto}, C. and {Li}, J. and {Liodakis}, I. and {Longo}, F. and {Loparco}, F. and {Lott}, B. and {Lovellette}, M.~N. and {Lubrano}, P. and {Madejski}, G.~M. and {Maldera}, S. and {Manfreda}, A. and {Mart{\'\i}-Devesa}, G. and {Massaro}, F. and {Mazziotta}, M.~N. and {Mereu}, I. and {Meyer}, M. and {Migliori}, G. and {Mirabal}, N. and {Mizuno}, T. and {Monzani}, M.~E. and {Morselli}, A. and {Moskalenko}, I.~V. and {Negro}, M. and {Nemmen}, R. and {Nuss}, E. and {Ojha}, L.~S. and {Ojha}, R. and {Omodei}, N. and {Orienti}, M. and {Orlando}, E. and {Ormes}, J.~F. and {Paliya}, V.~S. and {Pei}, Z. and {Pe{\~n}a-Herazo}, H. and {Persic}, M. and {Pesce-Rollins}, M. and {Petrov}, L. and {Piron}, F. and {Poon}, H. and {Principe}, G. and {Rain{\`o}}, S. and {Rando}, R. and {Rani}, B. and {Razzano}, M. and {Razzaque}, S. and {Reimer}, A. and {Reimer}, O. and {Schinzel}, F.~K. and {Serini}, D. and {Sgr{\`o}}, C. and {Siskind}, E.~J. and {Spandre}, G. and {Spinelli}, P. and {Suson}, D.~J. and {Tachibana}, Y. and {Thompson}, D.~J. and {Torres}, D.~F. and {Torresi}, E. and {Troja}, E. and {Valverde}, J. and {van Zyl}, P. and {Yassine}, M.},
        title = "{The Fourth Catalog of Active Galactic Nuclei Detected by the Fermi Large Area Telescope}",
      journal = {\apj},
         year = 2020,
        month = apr,
       volume = {892},
       number = {2},
          eid = {105},
        pages = {105},
          doi = {10.3847/1538-4357/ab791e},
archivePrefix = {arXiv},
       eprint = {1905.10771},
 primaryClass = {astro-ph.HE},
       adsurl = {https://ui.adsabs.harvard.edu/abs/2020ApJ...892..105A}
}

@ARTICLE{2011ApJ...743..171A,
       author = {{Ackermann}, M. and {Ajello}, M. and {Allafort}, A. and {Antolini}, E. and {Atwood}, W.~B. and {Axelsson}, M. and {Baldini}, L. and {Ballet}, J. and {Barbiellini}, G. and {Bastieri}, D. and {Bechtol}, K. and {Bellazzini}, R. and {Berenji}, B. and {Blandford}, R.~D. and {Bloom}, E.~D. and {Bonamente}, E. and {Borgland}, A.~W. and {Bottacini}, E. and {Bouvier}, A. and {Bregeon}, J. and {Brigida}, M. and {Bruel}, P. and {Buehler}, R. and {Burnett}, T.~H. and {Buson}, S. and {Caliandro}, G.~A. and {Cameron}, R.~A. and {Caraveo}, P.~A. and {Casandjian}, J.~M. and {Cavazzuti}, E. and {Cecchi}, C. and {Charles}, E. and {Cheung}, C.~C. and {Chiang}, J. and {Ciprini}, S. and {Claus}, R. and {Cohen-Tanugi}, J. and {Conrad}, J. and {Costamante}, L. and {Cutini}, S. and {de Angelis}, A. and {de Palma}, F. and {Dermer}, C.~D. and {Digel}, S.~W. and {Silva}, E. do Couto e. and {Drell}, P.~S. and {Dubois}, R. and {Escande}, L. and {Favuzzi}, C. and {Fegan}, S.~J. and {Ferrara}, E.~C. and {Finke}, J. and {Focke}, W.~B. and {Fortin}, P. and {Frailis}, M. and {Fukazawa}, Y. and {Funk}, S. and {Fusco}, P. and {Gargano}, F. and {Gasparrini}, D. and {Gehrels}, N. and {Germani}, S. and {Giebels}, B. and {Giglietto}, N. and {Giommi}, P. and {Giordano}, F. and {Giroletti}, M. and {Glanzman}, T. and {Godfrey}, G. and {Grenier}, I.~A. and {Grove}, J.~E. and {Guiriec}, S. and {Gustafsson}, M. and {Hadasch}, D. and {Hayashida}, M. and {Hays}, E. and {Healey}, S.~E. and {Horan}, D. and {Hou}, X. and {Hughes}, R.~E. and {Iafrate}, G. and {J{\'o}hannesson}, G. and {Johnson}, A.~S. and {Johnson}, W.~N. and {Kamae}, T. and {Katagiri}, H. and {Kataoka}, J. and {Kn{\"o}dlseder}, J. and {Kuss}, M. and {Lande}, J. and {Larsson}, S. and {Latronico}, L. and {Longo}, F. and {Loparco}, F. and {Lott}, B. and {Lovellette}, M.~N. and {Lubrano}, P. and {Madejski}, G.~M. and {Mazziotta}, M.~N. and {McConville}, W. and {McEnery}, J.~E. and {Michelson}, P.~F. and {Mitthumsiri}, W. and {Mizuno}, T. and {Moiseev}, A.~A. and {Monte}, C. and {Monzani}, M.~E. and {Moretti}, E. and {Morselli}, A. and {Moskalenko}, I.~V. and {Murgia}, S. and {Nakamori}, T. and {Naumann-Godo}, M. and {Nolan}, P.~L. and {Norris}, J.~P. and {Nuss}, E. and {Ohno}, M. and {Ohsugi}, T. and {Okumura}, A. and {Omodei}, N. and {Orienti}, M. and {Orlando}, E. and {Ormes}, J.~F. and {Ozaki}, M. and {Paneque}, D. and {Parent}, D. and {Pesce-Rollins}, M. and {Pierbattista}, M. and {Piranomonte}, S. and {Piron}, F. and {Pivato}, G. and {Porter}, T.~A. and {Rain{\`o}}, S. and {Rando}, R. and {Razzano}, M. and {Razzaque}, S. and {Reimer}, A. and {Reimer}, O. and {Ritz}, S. and {Rochester}, L.~S. and {Romani}, R.~W. and {Roth}, M. and {Sanchez}, D.~A. and {Sbarra}, C. and {Scargle}, J.~D. and {Schalk}, T.~L. and {Sgr{\`o}}, C. and {Shaw}, M.~S. and {Siskind}, E.~J. and {Spandre}, G. and {Spinelli}, P. and {Strong}, A.~W. and {Suson}, D.~J. and {Tajima}, H. and {Takahashi}, H. and {Takahashi}, T. and {Tanaka}, T. and {Thayer}, J.~G. and {Thayer}, J.~B. and {Thompson}, D.~J. and {Tibaldo}, L. and {Tinivella}, M. and {Torres}, D.~F. and {Tosti}, G. and {Troja}, E. and {Uchiyama}, Y. and {Vandenbroucke}, J. and {Vasileiou}, V. and {Vianello}, G. and {Vitale}, V. and {Waite}, A.~P. and {Wallace}, E. and {Wang}, P. and {Winer}, B.~L. and {Wood}, D.~L. and {Wood}, K.~S. and {Zimmer}, S.},
        title = "{The Second Catalog of Active Galactic Nuclei Detected by the Fermi Large Area Telescope}",
      journal = {\apj},
         year = 2011,
        month = dec,
       volume = {743},
       number = {2},
          eid = {171},
        pages = {171},
          doi = {10.1088/0004-637X/743/2/171},
archivePrefix = {arXiv},
       eprint = {1108.1420},
 primaryClass = {astro-ph.HE},
       adsurl = {https://ui.adsabs.harvard.edu/abs/2011ApJ...743..171A}
}

@ARTICLE{2024MNRAS.528.6823L,
       author = {{Li}, Huai-Zhen and {Guo}, Di-Fu and {Qin}, Long-Hua and {Yi}, Ting-Feng and {Liu}, Fen and {Gao}, Quan-Gui and {Chang}, Xin},
        title = "{The optical intra-day variability of BL laceratae object 2200 + 420}",
      journal = {\mnras},
         year = 2024,
        month = mar,
       volume = {528},
       number = {4},
        pages = {6823-6835},
          doi = {10.1093/mnras/stae422},
       adsurl = {https://ui.adsabs.harvard.edu/abs/2024MNRAS.528.6823L}
}

@ARTICLE{Raiteri2009,
       author = {{Raiteri}, C.~M. and {Villata}, M. and {Capetti}, A. and {Aller}, M.~F. and {Bach}, U. and {Calcidese}, P. and {Gurwell}, M.~A. and {Larionov}, V.~M. and {Ohlert}, J. and {Nilsson}, K. and {Strigachev}, A. and {Agudo}, I. and {Aller}, H.~D. and {Bachev}, R. and {Ben{\'\i}tez}, E. and {Berdyugin}, A. and {B{\"o}ttcher}, M. and {Buemi}, C.~S. and {Buttiglione}, S. and {Carosati}, D. and {Charlot}, P. and {Chen}, W.~P. and {Dultzin}, D. and {Forn{\'e}}, E. and {Fuhrmann}, L. and {G{\'o}mez}, J.~L. and {Gupta}, A.~C. and {Heidt}, J. and {Hiriart}, D. and {Hsiao}, W. -S. and {Jel{\'\i}nek}, M. and {Jorstad}, S.~G. and {Kimeridze}, G.~N. and {Konstantinova}, T.~S. and {Kopatskaya}, E.~N. and {Kostov}, A. and {Kurtanidze}, O.~M. and {L{\"a}hteenm{\"a}ki}, A. and {Lanteri}, L. and {Larionova}, L.~V. and {Leto}, P. and {Latev}, G. and {Le Campion}, J. -F. and {Lee}, C. -U. and {Ligustri}, R. and {Lindfors}, E. and {Marscher}, A.~P. and {Mihov}, B. and {Nikolashvili}, M.~G. and {Nikolov}, Y. and {Ovcharov}, E. and {Principe}, D. and {Pursimo}, T. and {Ragozzine}, B. and {Robb}, R.~M. and {Ros}, J.~A. and {Sadun}, A.~C. and {Sagar}, R. and {Semkov}, E. and {Sigua}, L.~A. and {Smart}, R.~L. and {Sorcia}, M. and {Takalo}, L.~O. and {Tornikoski}, M. and {Trigilio}, C. and {Uckert}, K. and {Umana}, G. and {Valcheva}, A. and {Volvach}, A.},
        title = "{WEBT multiwavelength monitoring and XMM-Newton observations of <ASTROBJ>BL Lacertae</ASTROBJ> in 2007-2008. Unveiling different emission components}",
      journal = {\aap},
         year = 2009,
        month = nov,
       volume = {507},
       number = {2},
        pages = {769-779},
          doi = {10.1051/0004-6361/200912953},
archivePrefix = {arXiv},
       eprint = {0909.1701},
 primaryClass = {astro-ph.HE},
       adsurl = {https://ui.adsabs.harvard.edu/abs/2009A&A...507..769R}
}

@ARTICLE{1997A&A...327...61G,
       author = {{Ghisellini}, G. and {Villata}, M. and {Raiteri}, C.~M. and {Bosio}, S. and {de Francesco}, G. and {Latini}, G. and {Maesano}, M. and {Massaro}, E. and {Montagni}, F. and {Nesci}, R. and {Tosti}, G. and {Fiorucci}, M. and {Pian}, E. and {Maraschi}, L. and {Treves}, A. and {Comastri}, A. and {Mignoli}, M.},
        title = "{Optical-IUE observations of the gamma-ray loud BL Lacertae object S5 0716+714: data and interpretation.}",
      journal = {\aap},
         year = 1997,
        month = nov,
       volume = {327},
        pages = {61-71},
          doi = {10.48550/arXiv.astro-ph/9706254},
archivePrefix = {arXiv},
       eprint = {astro-ph/9706254},
 primaryClass = {astro-ph},
       adsurl = {https://ui.adsabs.harvard.edu/abs/1997A&A...327...61G}
}

@ARTICLE{2003ApJ...586L..25G,
       author = {{Gopal-Krishna} and {Stalin}, C.~S. and {Sagar}, Ram and {Wiita}, Paul J.},
        title = "{Clear Evidence for Intranight Optical Variability in Radio-quiet Quasars}",
      journal = {\apjl},
         year = 2003,
        month = mar,
       volume = {586},
       number = {1},
        pages = {L25-L28},
          doi = {10.1086/374655},
archivePrefix = {arXiv},
       eprint = {astro-ph/0302188},
 primaryClass = {astro-ph},
       adsurl = {https://ui.adsabs.harvard.edu/abs/2003ApJ...586L..25G}
}

@ARTICLE{2017MNRAS.469.3588M,
       author = {{Meng}, Nankun and {Wu}, Jianghua and {Webb}, James R. and {Zhang}, Xiaoyuan and {Dai}, Yan},
        title = "{Intraday optical variability of BL Lacertae}",
      journal = {\mnras},
         year = 2017,
        month = aug,
       volume = {469},
       number = {3},
        pages = {3588-3596},
          doi = {10.1093/mnras/stx1055},
archivePrefix = {arXiv},
       eprint = {1706.07183},
 primaryClass = {astro-ph.GA},
       adsurl = {https://ui.adsabs.harvard.edu/abs/2017MNRAS.469.3588M}
}

@ARTICLE{2021RAA....21..259L,
       author = {{Li}, Tian and {Wu}, Jiang-Hua and {Meng}, Nan-Kun and {Dai}, Yan and {Zhang}, Xiao-Yuan},
        title = "{Intra-day variability of BL Lacertae from 2016 to 2018}",
      journal = {Research in Astronomy and Astrophysics},
         year = 2021,
        month = nov,
       volume = {21},
       number = {10},
          eid = {259},
        pages = {259},
          doi = {10.1088/1674-4527/21/10/259},
archivePrefix = {arXiv},
       eprint = {2108.04699},
 primaryClass = {astro-ph.GA},
       adsurl = {https://ui.adsabs.harvard.edu/abs/2021RAA....21..259L}
}

@ARTICLE{Agarwal2025,
       author = {{Agarwal}, Aditi and {Agrawal}, V. and {Zola}, S. and {Jana}, Swarnendu and {Bisht}, M.~S. and {Raj}, A. and {Kouprianov}, V. and {Reichart}, Daniel E. and {Caton}, D.~B. and {Dawidson}, James W.},
        title = "{Multiband flux and spectral variability study of the flaring activity in BL Lacertae during its 2020 outburst}",
      journal = {\mnras},
         year = 2025,
        month = mar,
       volume = {537},
       number = {3},
        pages = {2586-2601},
          doi = {10.1093/mnras/staf183},
       adsurl = {https://ui.adsabs.harvard.edu/abs/2025MNRAS.537.2586A}
}

@ARTICLE{1987ApJ...322..650B,
       author = {{Begelman}, Mitchell C. and {Sikora}, Marek},
        title = "{Inverse Compton Scattering of Ambient Radiation by a Cold Relativistic Jet: A Source of Beamed, Polarized X-Ray and Optical Observations of X-Ray--selected BL Lacertae Objects}",
      journal = {\apj},
         year = 1987,
        month = nov,
       volume = {322},
        pages = {650},
          doi = {10.1086/165760},
       adsurl = {https://ui.adsabs.harvard.edu/abs/1987ApJ...322..650B}
}

@ARTICLE{1989ApJ...340..181G,
       author = {{Ghisellini}, Gabriele and {Maraschi}, Laura},
        title = "{Bulk Acceleration in Relativistic Jets and the Spectral Properties of Blazars}",
      journal = {\apj},
         year = 1989,
        month = may,
       volume = {340},
        pages = {181},
          doi = {10.1086/167383},
       adsurl = {https://ui.adsabs.harvard.edu/abs/1989ApJ...340..181G}
}

@ARTICLE{2002A&A...390..407V,
       author = {{Villata}, M. and {Raiteri}, C.~M. and {Kurtanidze}, O.~M. and {Nikolashvili}, M.~G. and {Ibrahimov}, M.~A. and {Papadakis}, I.~E. and {Tsinganos}, K. and {Sadakane}, K. and {Okada}, N. and {Takalo}, L.~O. and {Sillanp{\"a}{\"a}}, A. and {Tosti}, G. and {Ciprini}, S. and {Frasca}, A. and {Marilli}, E. and {Robb}, R.~M. and {Noble}, J.~C. and {Jorstad}, S.~G. and {Hagen-Thorn}, V.~A. and {Larionov}, V.~M. and {Nesci}, R. and {Maesano}, M. and {Schwartz}, R.~D. and {Basler}, J. and {Gorham}, P.~W. and {Iwamatsu}, H. and {Kato}, T. and {Pullen}, C. and {Ben{\'\i}tez}, E. and {de Diego}, J.~A. and {Moilanen}, M. and {Oksanen}, A. and {Rodriguez}, D. and {Sadun}, A.~C. and {Kelly}, M. and {Carini}, M.~T. and {Miller}, H.~R. and {Catalano}, S. and {Dultzin-Hacyan}, D. and {Fan}, J.~H. and {Ishioka}, R. and {Karttunen}, H. and {Kein{\"a}nen}, P. and {Kudryavtseva}, N.~A. and {Lainela}, M. and {Lanteri}, L. and {Larionova}, E.~G. and {Matsumoto}, K. and {Mattox}, J.~R. and {Montagni}, F. and {Nucciarelli}, G. and {Ostorero}, L. and {Papamastorakis}, J. and {Pasanen}, M. and {Sobrito}, G. and {Uemura}, M.},
        title = "{The WEBT <ASTROBJ>BL Lacertae</ASTROBJ> Campaign 2000}",
      journal = {\aap},
         year = 2002,
        month = aug,
       volume = {390},
        pages = {407-421},
          doi = {10.1051/0004-6361:20020662},
archivePrefix = {arXiv},
       eprint = {astro-ph/0205479},
 primaryClass = {astro-ph},
       adsurl = {https://ui.adsabs.harvard.edu/abs/2002A&A...390..407V}
}

@ARTICLE{2023MNRAS.522..102R,
       author = {{Raiteri}, C.~M. and {Villata}, M. and {Jorstad}, S.~G. and {Marscher}, A.~P. and {Acosta Pulido}, J.~A. and {Carosati}, D. and {Chen}, W.~P. and {Joner}, M.~D. and {Kurtanidze}, S.~O. and {Lorey}, C. and {Marchini}, A. and {Matsumoto}, K. and {Mirzaqulov}, D.~O. and {Savchenko}, S.~S. and {Strigachev}, A. and {Vince}, O. and {Aceti}, P. and {Apolonio}, G. and {Arena}, C. and {Arkharov}, A. and {Bachev}, R. and {Bader}, N. and {Banfi}, M. and {Bonnoli}, G. and {Borman}, G.~A. and {Bozhilov}, V. and {Brown}, L.~F. and {Carbonell}, W. and {Carnerero}, M.~I. and {Damljanovic}, G. and {Dhiman}, V. and {Ehgamberdiev}, S.~A. and {Elsaesser}, D. and {Feige}, M. and {Gabellini}, D. and {Gal{\'a}n}, D. and {Galli}, G. and {Gaur}, H. and {Gazeas}, K. and {Grishina}, T.~S. and {Gupta}, A.~C. and {Hagen-Thorn}, V.~A. and {Hallum}, M.~K. and {Hart}, M. and {Hasuda}, K. and {Heidemann}, K. and {Horst}, B. and {Hou}, W. -J. and {Ibryamov}, S. and {Ivanidze}, R.~Z. and {Jovanovic}, M.~D. and {Kimeridze}, G.~N. and {Kishore}, S. and {Klimanov}, S. and {Kopatskaya}, E.~N. and {Kurtanidze}, O.~M. and {Kushwaha}, P. and {Lane}, D.~J. and {Larionova}, E.~G. and {Leonini}, S. and {Lin}, H.~C. and {Mannheim}, K. and {Marino}, G. and {Minev}, M. and {Modaressi}, A. and {Morozova}, D.~A. and {Mortari}, F. and {Nazarov}, S.~V. and {Nikolashvili}, M.~G. and {Otero Santos}, J. and {Ovcharov}, E. and {Papini}, R. and {Pinter}, V. and {Privitera}, C.~A. and {Pursimo}, T. and {Reinhart}, D. and {Roberts}, J. and {Romanov}, F.~D. and {Rosenlehner}, K. and {Sakamoto}, T. and {Salvaggio}, F. and {Schoch}, K. and {Semkov}, E. and {Seufert}, J. and {Shakhovskoy}, D. and {Sigua}, L.~A. and {Singh}, C. and {Steineke}, R. and {Stojanovic}, M. and {Tripathi}, T. and {Troitskaya}, Y.~V. and {Troitskiy}, I.~S. and {Tsai}, A. and {Valcheva}, A. and {Vasilyev}, A.~A. and {Vrontaki}, K. and {Weaver}, Z.~R. and {Wooley}, J.~H.~F. and {Zaharieva}, E. and {Zhovtan}, A.~V.},
        title = "{The optical behaviour of BL Lacertae at its maximum brightness levels: a blend of geometry and energetics}",
      journal = {\mnras},
         year = 2023,
        month = jun,
       volume = {522},
       number = {1},
        pages = {102-116},
          doi = {10.1093/mnras/stad942},
archivePrefix = {arXiv},
       eprint = {2302.10555},
 primaryClass = {astro-ph.HE},
       adsurl = {https://ui.adsabs.harvard.edu/abs/2023MNRAS.522..102R}
}

@ARTICLE{2020ApJ...900..137W,
       author = {{Weaver}, Z.~R. and {Williamson}, K.~E. and {Jorstad}, S.~G. and {Marscher}, A.~P. and {Larionov}, V.~M. and {Raiteri}, C.~M. and {Villata}, M. and {Acosta-Pulido}, J.~A. and {Bachev}, R. and {Baida}, G.~V. and {Balonek}, T.~J. and {Ben{\'\i}tez}, E. and {Borman}, G.~A. and {Bozhilov}, V. and {Carnerero}, M.~I. and {Carosati}, D. and {Chen}, W.~P. and {Damljanovic}, G. and {Dhiman}, V. and {Dougherty}, D.~J. and {Ehgamberdiev}, S.~A. and {Grishina}, T.~S. and {Gupta}, A.~C. and {Hart}, M. and {Hiriart}, D. and {Hsiao}, H.~Y. and {Ibryamov}, S. and {Joner}, M. and {Kimeridze}, G.~N. and {Kopatskaya}, E.~N. and {Kurtanidze}, O.~M. and {Kurtanidze}, S.~O. and {Larionova}, E.~G. and {Matsumoto}, K. and {Matsumura}, R. and {Minev}, M. and {Mirzaqulov}, D.~O. and {Morozova}, D.~A. and {Nikiforova}, A.~A. and {Nikolashvili}, M.~G. and {Ovcharov}, E. and {Rizzi}, N. and {Sadun}, A. and {Savchenko}, S.~S. and {Semkov}, E. and {Slater}, J.~J. and {Smith}, K.~L. and {Stojanovic}, M. and {Strigachev}, A. and {Troitskaya}, Yu. V. and {Troitsky}, I.~S. and {Tsai}, A.~L. and {Vince}, O. and {Valcheva}, A. and {Vasilyev}, A.~A. and {Zaharieva}, E. and {Zhovtan}, A.~V.},
        title = "{Multiwavelength Variability of BL Lacertae Measured with High Time Resolution}",
      journal = {\apj},
         year = 2020,
        month = sep,
       volume = {900},
       number = {2},
          eid = {137},
        pages = {137},
          doi = {10.3847/1538-4357/aba693},
archivePrefix = {arXiv},
       eprint = {2007.07999},
 primaryClass = {astro-ph.GA},
       adsurl = {https://ui.adsabs.harvard.edu/abs/2020ApJ...900..137W}
}

@ARTICLE{2006MNRAS.373..209H,
       author = {{Hu}, Shao Ming and {Wu}, J.~H. and {Zhao}, G. and {Zhou}, X.},
        title = "{Optical multiband observations of BL Lacertae during the outburst of 2005}",
      journal = {\mnras},
         year = 2006,
        month = nov,
       volume = {373},
       number = {1},
        pages = {209-216},
          doi = {10.1111/j.1365-2966.2006.10995.x},
archivePrefix = {arXiv},
       eprint = {astro-ph/0701129},
 primaryClass = {astro-ph},
       adsurl = {https://ui.adsabs.harvard.edu/abs/2006MNRAS.373..209H}
}

@ARTICLE{2022ApJ...926...91F,
       author = {{Fang}, Yue and {Zhang}, Yan and {Chen}, Qihang and {Wu}, Jianghua},
        title = "{Intraday Optical Multiband Observation of BL Lacertae}",
      journal = {\apj},
         year = 2022,
        month = feb,
       volume = {926},
       number = {1},
          eid = {91},
        pages = {91},
          doi = {10.3847/1538-4357/ac4490},
       adsurl = {https://ui.adsabs.harvard.edu/abs/2022ApJ...926...91F}
}

@ARTICLE{1985ApJ...296...46S,
       author = {{Simonetti}, J.~H. and {Cordes}, J.~M. and {Heeschen}, D.~S.},
        title = "{Flicker of extragalactic radio sources at two frequencies.}",
      journal = {\apj},
         year = 1985,
        month = sep,
       volume = {296},
        pages = {46-59},
          doi = {10.1086/163418},
       adsurl = {https://ui.adsabs.harvard.edu/abs/1985ApJ...296...46S}
}

@ARTICLE{2005A&A...443..451F,
       author = {{Favre}, P. and {Courvoisier}, T.~J. -L. and {Paltani}, S.},
        title = "{AGN variability time scales and the discrete-event model}",
      journal = {\aap},
         year = 2005,
        month = nov,
       volume = {443},
       number = {2},
        pages = {451-463},
          doi = {10.1051/0004-6361:20041903},
archivePrefix = {arXiv},
       eprint = {astro-ph/0508620},
 primaryClass = {astro-ph},
       adsurl = {https://ui.adsabs.harvard.edu/abs/2005A&A...443..451F}
}

@ARTICLE{2015MNRAS.451.3882A,
       author = {{Agarwal}, Aditi and {Gupta}, Alok C. and {Bachev}, R. and {Strigachev}, A. and {Semkov}, E. and {Wiita}, Paul J. and {B{\"o}ttcher}, M. and {Boeva}, S. and {Gaur}, H. and {Gu}, M.~F. and {Peneva}, S. and {Ibryamov}, S. and {Pandey}, U.~S.},
        title = "{Multiband optical-NIR variability of blazars on diverse time-scales}",
      journal = {\mnras},
         year = 2015,
        month = aug,
       volume = {451},
       number = {4},
        pages = {3882-3897},
          doi = {10.1093/mnras/stv1208},
archivePrefix = {arXiv},
       eprint = {1506.00601},
 primaryClass = {astro-ph.GA},
       adsurl = {https://ui.adsabs.harvard.edu/abs/2015MNRAS.451.3882A}
}

@ARTICLE{2001ApJ...555..775C,
       author = {{Collier}, Stefan and {Peterson}, Bradley M.},
        title = "{Characteristic Ultraviolet/Optical Timescales in Active Galactic Nuclei}",
      journal = {\apj},
         year = 2001,
        month = jul,
       volume = {555},
       number = {2},
        pages = {775-785},
          doi = {10.1086/321517},
       adsurl = {https://ui.adsabs.harvard.edu/abs/2001ApJ...555..775C}
}

@ARTICLE{2018MNRAS.478.2557G,
       author = {{Gallo}, L.~C. and {Blue}, D.~M. and {Grupe}, D. and {Komossa}, S. and {Wilkins}, D.~R.},
        title = "{Eleven years of monitoring the Seyfert 1 Mrk 335 with Swift: Characterizing the X-ray and UV/optical variability}",
      journal = {\mnras},
         year = 2018,
        month = aug,
       volume = {478},
       number = {2},
        pages = {2557-2568},
          doi = {10.1093/mnras/sty1134},
archivePrefix = {arXiv},
       eprint = {1805.00300},
 primaryClass = {astro-ph.HE},
       adsurl = {https://ui.adsabs.harvard.edu/abs/2018MNRAS.478.2557G}
}

@ARTICLE{2024MNRAS.533..120C,
       author = {{Chang}, X. and {Xiong}, D.~R. and {Yi}, T.~F. and {Liu}, C.~X. and {Bhatta}, G. and {Xu}, J.~R. and {Gong}, Y.~L.},
        title = "{Optical intraday variability analysis for the BL Lacertae object 1ES 1426+42.8}",
      journal = {\mnras},
         year = 2024,
        month = sep,
       volume = {533},
       number = {1},
        pages = {120-130},
          doi = {10.1093/mnras/stae1839},
archivePrefix = {arXiv},
       eprint = {2409.06983},
 primaryClass = {astro-ph.GA},
       adsurl = {https://ui.adsabs.harvard.edu/abs/2024MNRAS.533..120C}
}

@ARTICLE{2023ApJS..269...60Y,
       author = {{Yuan}, Y.~H. and {Du}, G.~J. and {Fan}, J.~H. and {Liu}, Y. and {Yang}, J.~H. and {Ding}, G.~Z. and {Pei}, Z.~Y.},
        title = "{Optical Monitoring and Intraday Variabilities of BL Lacertae}",
      journal = {\apjs},
         year = 2023,
        month = dec,
       volume = {269},
       number = {2},
          eid = {60},
        pages = {60},
          doi = {10.3847/1538-4365/ad04d5},
       adsurl = {https://ui.adsabs.harvard.edu/abs/2023ApJS..269...60Y}
}

@ARTICLE{2021RAA....21..138Y,
       author = {{Yuan}, Yu-Hai and {Fan}, Jun-Hui and {Wu}, Hong and {Hao}, Jing-Meng and {Huang}, Wei-Rong and {Liu}, Xiao-Lan and {Huang}, Hong-Ren},
        title = "{Optical monitoring and intra-day variabilities of BL Lac Objects OJ 287}",
      journal = {Research in Astronomy and Astrophysics},
         year = 2021,
        month = aug,
       volume = {21},
       number = {6},
          eid = {138},
        pages = {138},
          doi = {10.1088/1674-4527/21/6/138},
       adsurl = {https://ui.adsabs.harvard.edu/abs/2021RAA....21..138Y}
}

@ARTICLE{2021PASP..133g4101Y,
       author = {{Yuan}, Y.~H. and {Fan}, J.~H.},
        title = "{Optical Monitoring and Intraday Variabilities of the BL Lac Object PKS 0735+178}",
      journal = {\pasp},
         year = 2021,
        month = jul,
       volume = {133},
       number = {1025},
          eid = {074101},
        pages = {074101},
          doi = {10.1088/1538-3873/ac015f},
       adsurl = {https://ui.adsabs.harvard.edu/abs/2021PASP..133g4101Y}
}

@software{2022zndo...7253034J,
       author = {{Jankov}, Isidora and {Kova{\v{c}}evi{\'c}}, Andjelka B. and {Ili{\'c}}, Dragana and {S{\'a}nchez-S{\'a}ez}, Paula and {Nikutta}, Robert},
        title = "{pyZDCF: Initial Release}",
         year = 2022,
        month = oct,
          eid = {10.5281/zenodo.7253034},
          doi = {10.5281/zenodo.7253034},
      version = {v1.0.0},
    publisher = {Zenodo},
       adsurl = {https://ui.adsabs.harvard.edu/abs/2022zndo...7253034J}
}

@ARTICLE{2010MNRAS.404..931E,
       author = {{Emmanoulopoulos}, D. and {McHardy}, I.~M. and {Uttley}, P.},
        title = "{On the use of structure functions to study blazar variability: caveats and problems}",
      journal = {\mnras},
         year = 2010,
        month = may,
       volume = {404},
       number = {2},
        pages = {931-946},
          doi = {10.1111/j.1365-2966.2010.16328.x},
archivePrefix = {arXiv},
       eprint = {1001.2045},
 primaryClass = {astro-ph.CO},
       adsurl = {https://ui.adsabs.harvard.edu/abs/2010MNRAS.404..931E}
}

@ARTICLE{1992ApJ...396..469H,
       author = {{Hughes}, P.~A. and {Aller}, H.~D. and {Aller}, M.~F.},
        title = "{The University of Michigan Radio Astronomy Data Base. I. Structure Function Analysis and the Relation between BL Lacertae Objects and Quasi-stellar Objects}",
      journal = {\apj},
         year = 1992,
        month = sep,
       volume = {396},
        pages = {469},
          doi = {10.1086/171734},
       adsurl = {https://ui.adsabs.harvard.edu/abs/1992ApJ...396..469H}
}

@ARTICLE{2023MNRAS.524.4333D,
       author = {{Diwan}, Rishank and {Prince}, Raj and {Agarwal}, Aditi and {Bose}, Debanjan and {Majumdar}, Pratik and {{\"O}zd{\"o}nmez}, Aykut and {Chandra}, Sunil and {Khatoon}, Rukaiya and {Ege}, Erg{\"u}n},
        title = "{Multiwavelength study of TeV blazar 1ES 1218+304 using gamma-ray, X-ray and optical observations}",
      journal = {\mnras},
         year = 2023,
        month = sep,
       volume = {524},
       number = {3},
        pages = {4333-4345},
          doi = {10.1093/mnras/stad2088},
archivePrefix = {arXiv},
       eprint = {2301.00991},
 primaryClass = {astro-ph.HE},
       adsurl = {https://ui.adsabs.harvard.edu/abs/2023MNRAS.524.4333D}
}

@ARTICLE{2022ApJ...933...42A,
       author = {{Agarwal}, A. and {Pandey}, Ashwani and {{\"O}zd{\"o}nmez}, Aykut and {Ege}, Erg{\"u}n and {Kumar Das}, Avik and {Karakulak}, Volkan},
        title = "{Characterizing the Optical Nature of the Blazar S5 1803+784 during Its 2020 Flare}",
      journal = {\apj},
         year = 2022,
        month = jul,
       volume = {933},
       number = {1},
          eid = {42},
        pages = {42},
          doi = {10.3847/1538-4357/ac6cef},
archivePrefix = {arXiv},
       eprint = {2205.02171},
 primaryClass = {astro-ph.HE},
       adsurl = {https://ui.adsabs.harvard.edu/abs/2022ApJ...933...42A}
}

@ARTICLE{2025ApJ...979...56O,
       author = {{Ozdonmez}, A. and {Er}, H. and {Tekkesinoglu}, M. and {Ege}, E. and {Kenger}, M.~E. and {Ozkesen}, I.~C. and {Polato{\u{g}}lu}, A.},
        title = "{Characterizing Multiband Optical Emission and Variability of 1ES 1215+303 on Diverse Timescales}",
      journal = {\apj},
         year = 2025,
        month = jan,
       volume = {979},
       number = {1},
          eid = {56},
        pages = {56},
          doi = {10.3847/1538-4357/ad9d45},
       adsurl = {https://ui.adsabs.harvard.edu/abs/2025ApJ...979...56O}
}

@ARTICLE{2024PASA...41...52O,
       author = {{{\"O}zd{\"o}nmez}, Aykut and {Tekke{\textcommabelow s}ino{\v{g}}lu}, Murat},
        title = "{Multi-band optical variability on diverse timescales of blazar 1E 1458.8+2249}",
      journal = {\pasa},
         year = 2024,
        month = sep,
       volume = {41},
          eid = {e052},
        pages = {e052},
          doi = {10.1017/pasa.2024.59},
archivePrefix = {arXiv},
       eprint = {2410.18675},
 primaryClass = {astro-ph.HE},
       adsurl = {https://ui.adsabs.harvard.edu/abs/2024PASA...41...52O}
}

@ARTICLE{2024ApJ...971...74E,
       author = {{Ege}, Erg{\"u}n and {{\"O}zd{\"o}nmez}, Aykut and {Agarwal}, Aditi and {Ak}, Tansel},
        title = "{Investigating Optical Variability of the Blazar S5 0716+714 on Diverse Timescales}",
      journal = {\apj},
         year = 2024,
        month = aug,
       volume = {971},
       number = {1},
          eid = {74},
        pages = {74},
          doi = {10.3847/1538-4357/ad5cef},
archivePrefix = {arXiv},
       eprint = {2407.09419},
 primaryClass = {astro-ph.HE},
       adsurl = {https://ui.adsabs.harvard.edu/abs/2024ApJ...971...74E}
}

@ARTICLE{2022MNRAS.513.2239P,
       author = {{Priya}, Shruti and {Prince}, Raj and {Agarwal}, Aditi and {Bose}, Debanjan and {{\"O}zd{\"o}nmez}, Aykut and {Ege}, Erg{\"u}n},
        title = "{Multiwavelength temporal and spectral analysis of Blazar S5 1803+78}",
      journal = {\mnras},
         year = 2022,
        month = jun,
       volume = {513},
       number = {2},
        pages = {2239-2251},
          doi = {10.1093/mnras/stac1009},
archivePrefix = {arXiv},
       eprint = {2204.04739},
 primaryClass = {astro-ph.HE},
       adsurl = {https://ui.adsabs.harvard.edu/abs/2022MNRAS.513.2239P}
}

@ARTICLE{2007A&A...469..899H,
       author = {{Hovatta}, T. and {Tornikoski}, M. and {Lainela}, M. and {Lehto}, H.~J. and {Valtaoja}, E. and {Torniainen}, I. and {Aller}, M.~F. and {Aller}, H.~D.},
        title = "{Statistical analyses of long-term variability of AGN at high radio frequencies}",
      journal = {\aap},
         year = 2007,
        month = jul,
       volume = {469},
       number = {3},
        pages = {899-912},
          doi = {10.1051/0004-6361:20077529},
archivePrefix = {arXiv},
       eprint = {0705.3293},
 primaryClass = {astro-ph},
       adsurl = {https://ui.adsabs.harvard.edu/abs/2007A&A...469..899H}
}

@ARTICLE{2015MNRAS.450..541A,
       author = {{Agarwal}, Aditi and {Gupta}, Alok C.},
        title = "{Multiband optical variability studies of BL Lacertae}",
      journal = {\mnras},
         year = 2015,
        month = jun,
       volume = {450},
       number = {1},
        pages = {541-551},
          doi = {10.1093/mnras/stv625},
archivePrefix = {arXiv},
       eprint = {1506.00574},
 primaryClass = {astro-ph.GA},
       adsurl = {https://ui.adsabs.harvard.edu/abs/2015MNRAS.450..541A}
}

%%%%%%%%%%%%%%%%%%%%%%%%%%%%%%%%%%%%%%%%%%%%%%%%%%%%%%%%%%%%%%%%%%%%%%%%%%%%%%%%%%%%%%%%
% APPENDICES 
%%%%%%%%%%%%%%%%%%%%%%%%%%%%%%%%%%%%%%%%%%%%%%%%%%%%%%%%%%%%%%%%%%%%%%%%%%%%%%%%%%%%%%%%

\appendix

\section{Results of Intraday Flux Variability}
\label{app:a}
The LCs for all 62 observing nights are presented in Fig.~\ref{fig:all_IDV_LC}. Nights with observations in two or more photometric bands are shown on the first two pages, while nights with only a single band are displayed on the final page. The plots are arranged chronologically by observing date. To facilitate comparison, different telescope identifiers are represented by distinct markers (A: circle, B: diamond, C: pentagon, D: asterisk), and different colors are used for the optical bands.
The results of the IDV statistical analyses described in Section~\ref{sec:stat_analysis} are summarized in Table~\ref{tab:var_results}, with additional details provided in the table caption.

%%%%%%%%%%%%%%%%%%%%%%%%%%%%%%%%       IDV LCs      %%%%%%%%%%%%%%%%%%%%%%%%%%%%%%%%
% 2 pages of figures for multi-band LCs
% 1 page of figures for single band LCs
% 2 pages of tables for statistical results

\begin{figure*}
\centering
\includegraphics[width=\textwidth,height=\textheight,keepaspectratio]{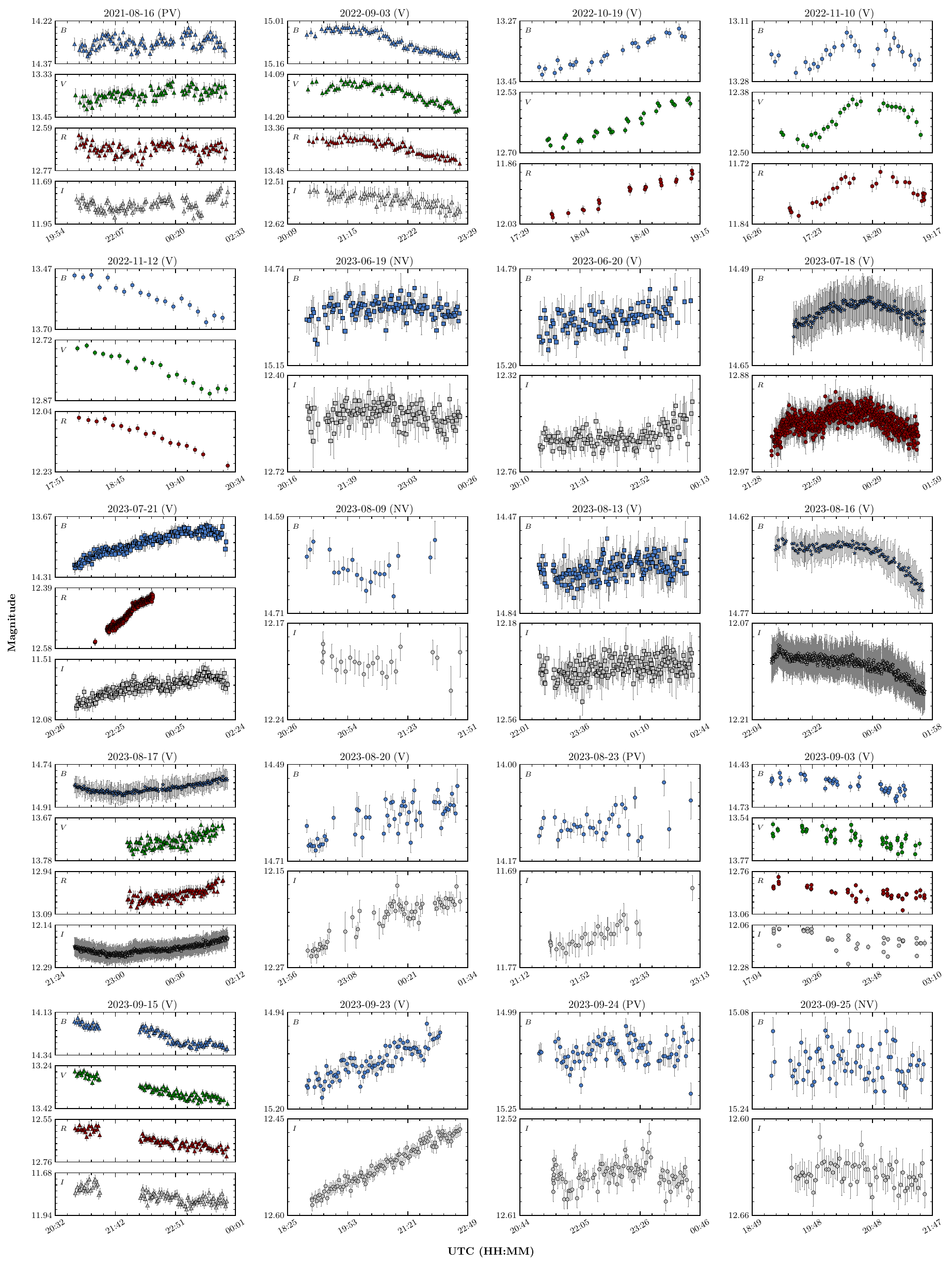}
 \caption{Optical IDV LCs of BL Lacertae for 39 nights with observations in two, three, or four photometric bands. The $B$-, $V$-, $R$-, and $I$-band LCs are plotted in blue, green, red, and gray, respectively. Each panel shows the observation date and the corresponding variability status.}
\label{fig:all_IDV_LC}
\end{figure*}

\begin{figure*}
\ContinuedFloat
\centering
\includegraphics[width=\textwidth,height=\textheight,keepaspectratio]{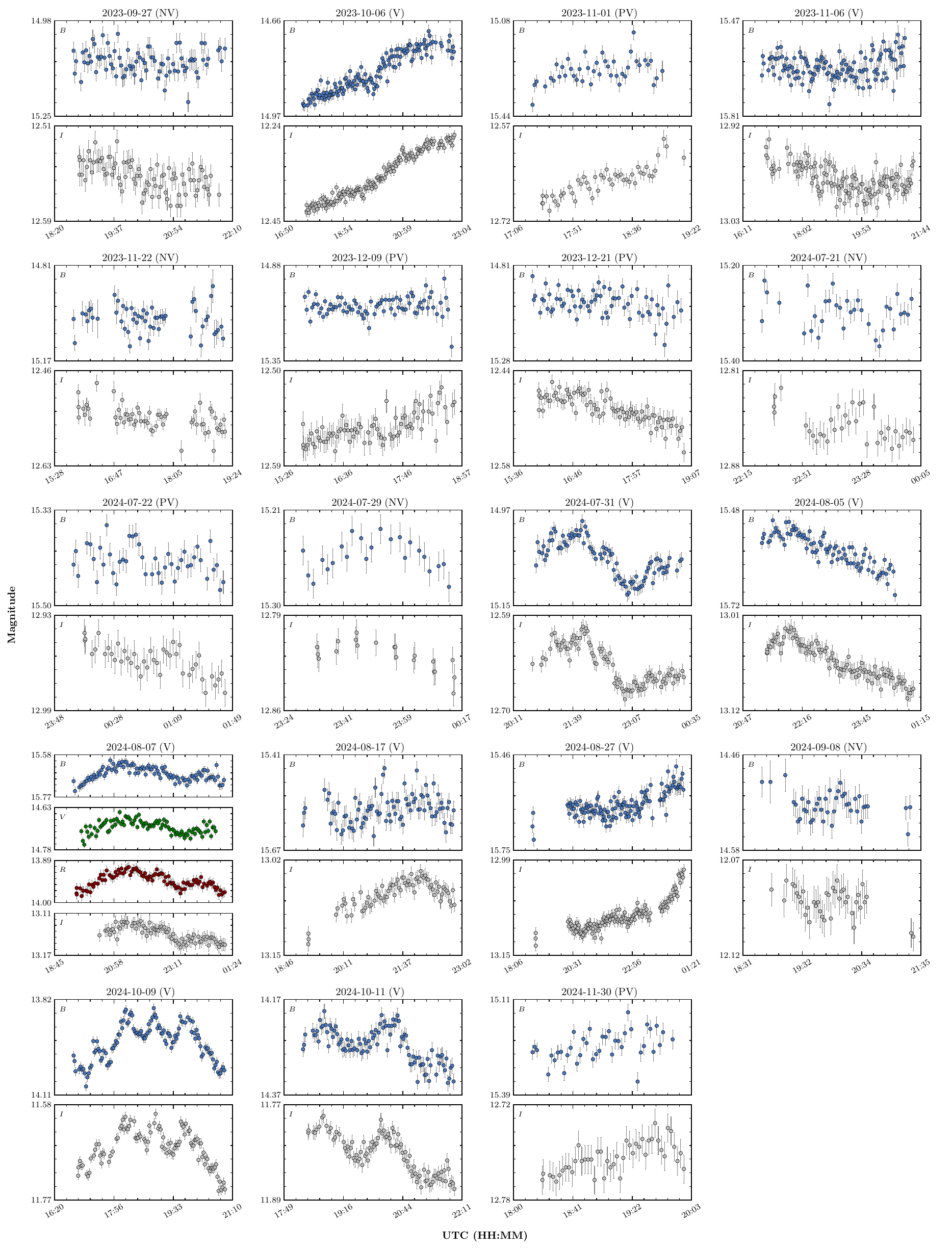}
\caption{Continued.}
\end{figure*}

\begin{figure*}
\ContinuedFloat
\centering
\includegraphics[width=\textwidth,height=\textheight,keepaspectratio]{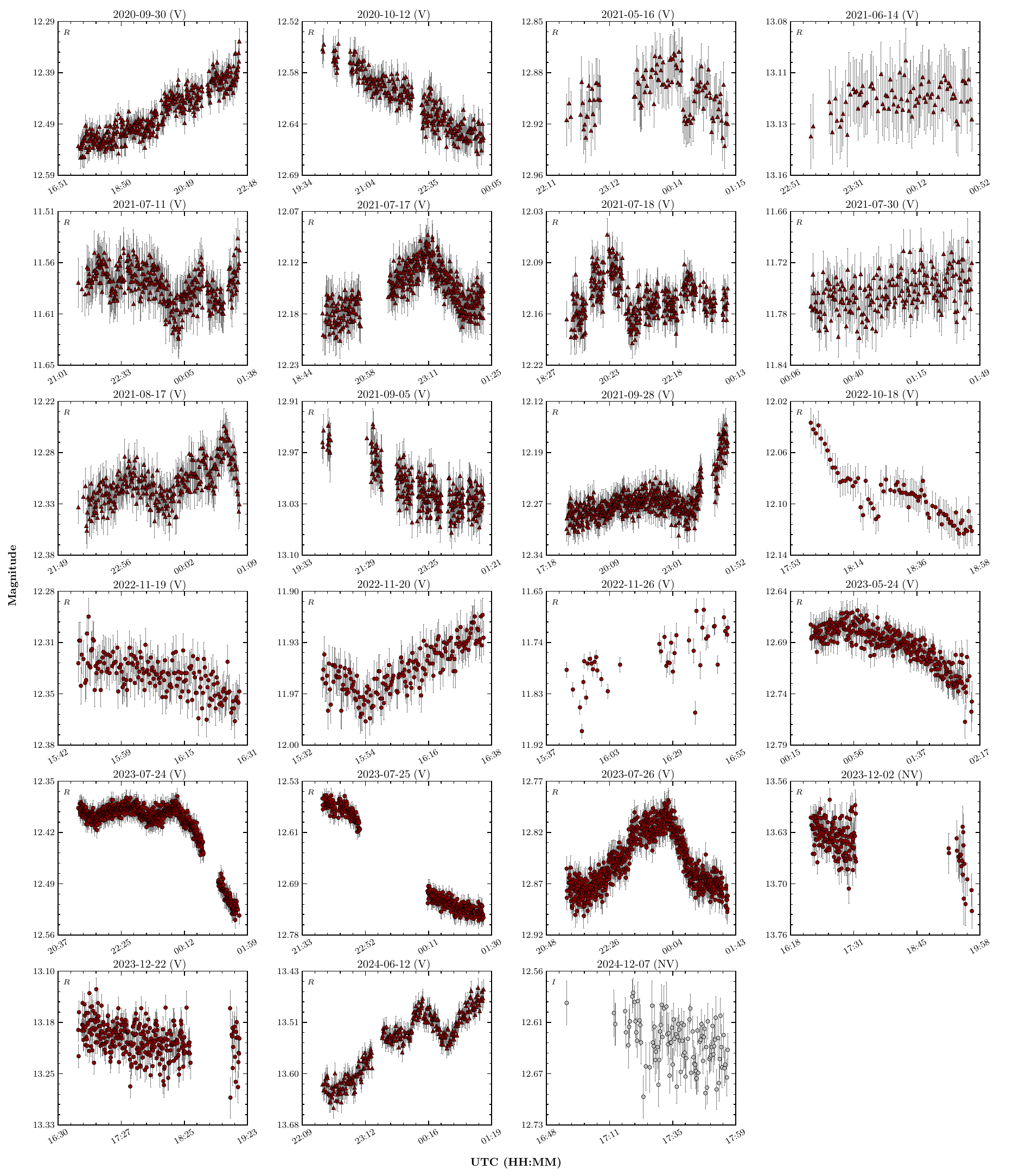}
\caption{Same as the previous plots, but for 23 nights with observations in a single photometric band. Each panel displays the observation date and the corresponding variability status.}
\end{figure*}

%%%%%%%%%%%%%%%%%%%%%%%%%%%%%%%%       IDV table   [final form]    %%%%%%%%%%%%%%%%%%%%%%%%%%%%%%%%
% 2 pages of tables for statistical results

%\begin{landscape}
\FloatBarrier

\onecolumn
% \centering
\scriptsize
% \tiny
% \small
% \normalsize	
% \footnotesize >> could be used with only 13 col
\begin{longtable}{cccccccccccccc}
\caption{\small The variability results obtained from the power-enhanced F-test and the nested ANOVA test are presented below, along with the corresponding variability amplitudes for each variable LC. Individual `Band status' are combined to determine the `Night status', which reflects the observing night's overall variability.}
%\textcolor{red}{Make dates compilant with MNRAS style, bands should be italic} 
\label{tab:var_results}\\
\toprule
% \begin{tabular}

Obs. Date & Telescope & Band & Avg.  &  $t_\mathrm{obs}$ & \multicolumn{3}{c}{Power-enhanced $F$-test} & \multicolumn{3}{c}{Nested ANOVA Test} & Band & Night & Amplitude \\
\cmidrule(lr){6-8} \cmidrule(lr){9-11}
   & code &  & mag. & [h] & DOF($\nu_1$,$\nu_2$) & $F_\mathrm{enh}$ & $F_{\rm c}$ & DOF($\nu_1$,$\nu_2$) & $F$ & $F_{\rm c}$ & status  & status & [per cent] \\
\midrule
\endfirsthead

\multicolumn{14}{c}%
% {{\bfseries \tablename\ \thetable{} -- continued from previous page}} \\
{{\tablename\ \thetable{} -- Variability results (continued.)}} \\
\toprule
Obs. Date & Telescope & Band & Avg.  &  $t_\mathrm{obs}$ & \multicolumn{3}{c}{Power-enhanced F-test} & \multicolumn{3}{c}{Nested ANOVA Test} & Band & Night & Amplitude \\
\cmidrule(lr){6-8} \cmidrule(lr){9-11}
YYYY-MM-DD & code &  & Mag. & [h] & DOF($\nu_1$,$\nu_2$) & $F_\mathrm{enh}$ & $F_{\rm c}$ & DOF($\nu_1$,$\nu_2$) & $F$ & $F_{\rm c}$ &  status & status  & [per cent] \\
\midrule
\endhead

% \bottomrule
% \multicolumn{13}{r}{{Continued on next page}} \\
% \multicolumn{13}{r}{{To be continued}} \\
% \endfoot

% \bottomrule
\endlastfoot

2020 Sep 30 & B & $R$ & 12.47 & 5.07 & (386, 772) & 19.91 & 1.22 & (76, 308) & 95.35 & 1.49 & V & V & 22.21 \\[1ex] 

2020 Oct 12 & B & $R$ & 12.61 & 3.83 & (271, 542) & 21.32 & 1.27 & (53, 216) & 97.04 & 1.61 & V & V & 12.08 \\ [1ex]

2021 May 16 & B & $R$ & 12.9 & 2.6 & (111, 222) & 2.5 & 1.45 & (21, 88) & 10.55 & 2.07 & V & V & 6.43 \\[1ex] 

2021 Jun 14 & B & $R$ & 13.12 & 1.72 & (86, 172) & 2.18 & 1.53 & (16, 68) & 5.09 & 2.28 & V & V & 3.01 \\[1ex] 

2021 Jul 11 & B & $R$ & 11.58 & 3.93 & (358, 716) & 5.13 & 1.23 & (70, 284) & 17.95 & 1.52 & V & V & 8.68 \\[1ex] 

2021 Jul 17 & B & $R$ & 12.15 & 5.69 & (437, 874) & 10.96 & 1.21 & (86, 348) & 54.01 & 1.46 & V & V & 11.29 \\[1ex] 

2021 Jul 18 & B & $R$ & 12.14 & 4.91 & (372, 744) & 3.7 & 1.23 & (73, 296) & 3.22 & 1.5 & V & V & 13.57 \\[1ex] 

2021 Jul 30 & B & $R$ & 11.76 & 1.47 & (211, 422) & 2.21 & 1.31 & (41, 168) & 3.35 & 1.71 & V & V & 10.65 \\[1ex] 

2021 Aug 16 & B & $B$ & 14.3 & 5.56 & (96, 192) & 2.79 & 1.49 & (18, 76) & 3.49 & 2.18 & V & PV & 10.06 \\ 
2021 Aug 16 & B & $V$ & 13.38 & 5.55 & (94, 188) & 3.5 & 1.5 & (18, 76) & 8.11 & 2.18 & V & PV & 6.88 \\ 
2021 Aug 16 & B & $R$ & 12.68 & 5.53 & (97, 194) & 0.88 & 1.49 & (18, 76) & 2.34 & 2.18 & NV & PV &   \\ 
2021 Aug 16 & B & $I$ & 11.83 & 5.58 & (93, 186) & 2.03 & 1.5 & (17, 72) & 0.83 & 2.23 & NV & PV &   \\[1ex] 

2021 Aug 17 & B & $R$ & 12.31 & 2.84 & (284, 568) & 1.97 & 1.27 & (56, 228) & 11.12 & 1.59 & V & V & 10.4 \\[1ex] 

2021 Sep 05 & B & $R$ & 13.02 & 4.93 & (294, 588) & 5.73 & 1.26 & (58, 236) & 21.87 & 1.57 & V & V & 13.0 \\ [1ex]

2021 Sep 28 & B & $R$ & 12.26 & 7.3 & (546, 1092) & 7.73 & 1.19 & (108, 436) & 40.79 & 1.4 & V & V & 15.71 \\[1ex] 

2022 Sep 03 & B & $B$ & 15.08 & 2.82 & (68, 136) & 20.42 & 1.61 & (12, 52) & 133.6 & 2.55 & V & V & 10.56 \\ 
2022 Sep 03 & B & $V$ & 14.14 & 2.81 & (68, 136) & 14.4 & 1.61 & (12, 52) & 80.38 & 2.55 & V & V & 7.45 \\ 
2022 Sep 03 & B & $R$ & 13.42 & 2.8 & (66, 132) & 18.4 & 1.62 & (12, 52) & 84.31 & 2.55 & V & V & 7.83 \\ 
2022 Sep 03 & B & $I$ & 12.56 & 2.79 & (66, 132) & 7.08 & 1.62 & (12, 52) & 26.47 & 2.55 & V & V & 5.95 \\[1ex] 

2022 Oct 18 & A & $R$ & 12.09 & 0.93 & (77, 154) & 39.71 & 1.56 & (14, 60) & 45.99 & 2.39 & V & V & 8.25 \\[1ex] 

2022 Oct 19 & A & $B$ & 13.37 & 1.42 & (25, 50) & 50.5 & 2.17 & (4, 20) & 79.87 & 4.43 & V & V & 13.97 \\ 
2022 Oct 19 & A & $V$ & 12.62 & 1.39 & (29, 58) & 26.21 & 2.05 & (5, 24) & 127.58 & 3.9 & V & V & 13.65 \\ 
2022 Oct 19 & A & $R$ & 11.95 & 1.37 & (22, 44) & 41.1 & 2.28 & (3, 16) & 48.4 & 5.29 & V & V & 12.9 \\[1ex] 

2022 Nov 10 & A & $B$ & 13.2 & 2.33 & (26, 52) & 13.54 & 2.14 & (4, 20) & 6.58 & 4.43 & V & V & 11.67 \\ 
2022 Nov 10 & A & $V$ & 12.44 & 2.2 & (28, 56) & 23.48 & 2.08 & (4, 20) & 38.31 & 4.43 & V & V & 9.53 \\ 
2022 Nov 10 & A & $R$ & 11.78 & 2.13 & (28, 56) & 21.02 & 2.08 & (4, 20) & 26.13 & 4.43 & V & V & 8.46 \\[1ex] 

2022 Nov 12 & A & $B$ & 13.57 & 2.23 & (18, 36) & 21.82 & 2.48 & (2, 12) & 60.51 & 6.93 & V & V & 17.98 \\ 
2022 Nov 12 & A & $V$ & 12.79 & 2.24 & (18, 36) & 18.54 & 2.48 & (2, 12) & 38.62 & 6.93 & V & V & 12.04 \\ 
2022 Nov 12 & A & $R$ & 12.11 & 2.24 & (16, 32) & 67.41 & 2.62 & (2, 12) & 69.77 & 6.93 & V & V & 14.42 \\[1ex] 

2022 Nov 19 & A & $R$ & 12.33 & 0.69 & (176, 352) & 1.88 & 1.35 & (34, 140) & 7.18 & 1.79 & V & V & 7.2 \\[1ex] 

2022 Nov 20 & A & $R$ & 11.95 & 0.93 & (177, 354) & 6.19 & 1.35 & (34, 140) & 27.55 & 1.79 & V & V & 6.82 \\[1ex] 

2022 Nov 26 & A & $R$ & 11.77 & 1.11 & (42, 84) & 7.76 & 1.82 & (7, 32) & 5.44 & 3.26 & V & V & 20.88 \\[1ex] 

2023 May 24 & A & $R$ & 12.69 & 1.73 & (320, 640) & 5.35 & 1.25 & (63, 256) & 27.05 & 1.55 & V & V & 10.88 \\[1ex] 

2023 Jun 19 & C & $B$ & 14.91 & 3.54 & (116, 232) & 1.61 & 1.44 & (22, 92) & 1.21 & 2.04 & NV & NV &   \\ 
2023 Jun 19 & C & $I$ & 12.53 & 3.54 & (121, 242) & 1.25 & 1.43 & (23, 96) & 2.22 & 2.01 & NV & NV &   \\[1ex] 

2023 Jun 20 & C & $B$ & 15.01 & 3.41 & (103, 206) & 2.34 & 1.47 & (19, 80) & 3.61 & 2.14 & V & V & 24.77 \\ 
2023 Jun 20 & C & $I$ & 12.6 & 3.43 & (120, 240) & 1.75 & 1.43 & (23, 96) & 4.49 & 2.01 & V & V & 22.62 \\[1ex] 

2023 Jul 18 & D & $B$ & 14.56 & 3.29 & (179, 358) & 2.9 & 1.34 & (35, 144) & 14.36 & 1.78 & V & V & 4.54 \\ 
2023 Jul 18 & A & $R$ & 12.92 & 3.68 & (702, 1404) & 2.95 & 1.16 & (139, 560) & 11.81 & 1.35 & V & V & 5.48 \\[1ex] 

2023 Jul 21 & C & $B$ & 13.95 & 5.01 & (194, 388) & 39.37 & 1.33 & (38, 156) & 95.66 & 1.74 & V & V & 44.48 \\ 
2023 Jul 21 & A & $R$ & 12.47 & 1.92 & (295, 590) & 56.39 & 1.26 & (58, 236) & 214.66 & 1.57 & V & V & 15.62 \\ 
2023 Jul 21 & C & $I$ & 11.77 & 5.01 & (201, 402) & 6.83 & 1.32 & (39, 160) & 47.11 & 1.73 & V & V & 36.76 \\[1ex] 

2023 Jul 24 & A & $R$ & 12.41 & 4.56 & (793, 1586) & 49.28 & 1.15 & (157, 632) & 292.95 & 1.33 & V & V & 16.73 \\[1ex] 

2023 Jul 25 & A & $R$ & 12.67 & 3.37 & (362, 724) & 127.56 & 1.23 & (71, 288) & 443.13 & 1.51 & V & V & 20.63 \\[1ex] 

2023 Jul 26 & A & $R$ & 12.85 & 4.18 & (779, 1558) & 12.65 & 1.15 & (155, 624) & 70.82 & 1.33 & V & V & 11.37 \\[1ex] 

2023 Aug 09 & A & $B$ & 14.65 & 1.02 & (20, 40) & 0.66 & 2.37 & (3, 16) & 3.47 & 5.29 & NV & NV & \\ 
2023 Aug 09 & A & $I$ & 12.2 & 1.09 & (22, 44) & 1.31 & 2.28 & (3, 16) & 2.23 & 5.29 & NV & NV &   \\[1ex] 

2023 Aug 13 & C & $B$ & 14.67 & 3.86 & (164, 328) & 1.62 & 1.36 & (32, 132) & 3.88 & 1.82 & V & V & 21.29 \\ 
2023 Aug 13 & C & $I$ & 12.36 & 4.01 & (171, 342) & 1.43 & 1.35 & (33, 136) & 2.75 & 1.81 & V & V & 19.53 \\[1ex] 

2023 Aug 16 & D & $B$ & 14.68 & 3.2 & (134, 268) & 18.63 & 1.41 & (26, 108) & 68.75 & 1.93 & V & V & 7.08 \\ 
2023 Aug 16 & D & $I$ & 12.13 & 3.32 & (584, 1168) & 11.33 & 1.18 & (116, 468) & 95.39 & 1.39 & V & V & 6.13 \\[1ex] 

2023 Aug 17 & D & $B$ & 14.83 & 4.06 & (178, 356) & 7.98 & 1.34 & (34, 140) & 30.05 & 1.79 & V & V & 5.88 \\ 
2023 Aug 17 & B & $V$ & 13.73 & 2.55 & (99, 198) & 4.23 & 1.48 & (19, 80) & 11.72 & 2.14 & V & V & 7.14 \\ 
2023 Aug 17 & B & $R$ & 13.02 & 2.55 & (100, 200) & 1.99 & 1.48 & (19, 80) & 1.28 & 2.14 & NV & V &   \\ 
2023 Aug 17 & D & $I$ & 12.23 & 4.08 & (723, 1446) & 16.94 & 1.16 & (143, 576) & 96.9 & 1.34 & V & V & 5.94 \\[1ex] 

2023 Aug 20 & A & $B$ & 14.62 & 3.05 & (52, 104) & 5.2 & 1.72 & (9, 40) & 10.56 & 2.89 & V & V & 14.71 \\ 
2023 Aug 20 & A & $I$ & 12.21 & 3.08 & (55, 110) & 5.08 & 1.69 & (10, 44) & 53.16 & 2.75 & V & V & 8.78 \\[1ex] 

2023 Aug 23 & A & $B$ & 14.1 & 1.71 & (32, 64) & 2.34 & 1.98 & (5, 24) & 2.26 & 3.9 & NV & PV &   \\ 
2023 Aug 23 & A & $I$ & 11.74 & 1.6 & (28, 56) & 4.75 & 2.08 & (4, 20) & 25.68 & 4.43 & V & PV & 4.93 \\[1ex] 

2023 Sep 03 & A & $B$ & 14.58 & 7.49 & (41, 82) & 3.76 & 1.84 & (7, 32) & 10.14 & 3.26 & V & V & 19.4 \\ 
2023 Sep 03 & A & $V$ & 13.64 & 8.23 & (48, 96) & 3.25 & 1.75 & (8, 36) & 8.66 & 3.05 & V & V & 15.97 \\ 
2023 Sep 03 & A & $R$ & 12.91 & 8.47 & (48, 96) & 5.51 & 1.75 & (8, 36) & 10.26 & 3.05 & V & V & 23.0 \\ 
2023 Sep 03 & A & $I$ & 12.14 & 8.46 & (38, 76) & 20.22 & 1.88 & (6, 28) & 6.32 & 3.53 & V & V & 18.35 \\[1ex] 

2023 Sep 15 & B & $B$ & 14.24 & 2.95 & (90, 180) & 23.02 & 1.51 & (17, 72) & 80.68 & 2.23 & V & V & 15.12 \\ 
2023 Sep 15 & B & $V$ & 13.34 & 2.95 & (92, 184) & 0.78 & 1.51 & (17, 72) & 4.66 & 2.23 & NV & V &   \\ 
2023 Sep 15 & B & $R$ & 12.65 & 2.95 & (90, 180) & 3.26 & 1.51 & (17, 72) & 2.43 & 2.23 & V & V & 15.24 \\ 
2023 Sep 15 & B & $I$ & 11.82 & 2.93 & (89, 178) & 1.71 & 1.52 & (17, 72) & 2.5 & 2.23 & V & V & 17.61 \\[1ex] 

2023 Sep 23 & A & $B$ & 15.08 & 3.26 & (84, 168) & 6.53 & 1.53 & (16, 68) & 23.5 & 2.28 & V & V & 19.93 \\ 
2023 Sep 23 & A & $I$ & 12.52 & 3.62 & (100, 200) & 22.31 & 1.48 & (19, 80) & 169.19 & 2.14 & V & V & 11.13 \\[1ex] 

2023 Sep 24 & A & $B$ & 15.09 & 3.43 & (80, 160) & 1.2 & 1.55 & (15, 64) & 1.52 & 2.33 & NV & PV &   \\ 
2023 Sep 24 & A & $I$ & 12.58 & 3.1 & (82, 164) & 2.73 & 1.54 & (15, 64) & 2.55 & 2.33 & V & PV & 6.27 \\[1ex] 

2023 Sep 25 & A & $B$ & 15.16 & 2.52 & (62, 124) & 1.02 & 1.64 & (11, 48) & 0.79 & 2.64 & NV & NV &   \\ 
2023 Sep 25 & A & $I$ & 12.64 & 2.21 & (59, 118) & 1.22 & 1.66 & (11, 48) & 3.75 & 2.64 & NV & NV &   \\[1ex] 

2023 Sep 27 & A & $B$ & 15.1 & 3.27 & (80, 160) & 1.32 & 1.55 & (15, 64) & 0.65 & 2.33 & NV & NV &   \\ 
2023 Sep 27 & A & $I$ & 12.55 & 3.02 & (80, 160) & 1.54 & 1.55 & (15, 64) & 4.45 & 2.33 & NV & NV &   \\[1ex] 

2023 Oct 06 & A & $B$ & 14.83 & 5.29 & (134, 268) & 15.94 & 1.41 & (26, 108) & 67.63 & 1.93 & V & V & 23.76 \\ 
2023 Oct 06 & A & $I$ & 12.35 & 5.18 & (141, 282) & 45.84 & 1.39 & (27, 112) & 295.62 & 1.91 & V & V & 16.25 \\[1ex] 

2023 Nov 01 & A & $B$ & 15.27 & 1.65 & (41, 82) & 1.85 & 1.84 & (7, 32) & 3.24 & 3.26 & NV & PV &   \\ 
2023 Nov 01 & A & $I$ & 12.66 & 1.8 & (44, 88) & 5.84 & 1.8 & (8, 36) & 28.1 & 3.05 & V & PV & 11.41 \\[1ex] 

2023 Nov 06 & A & $B$ & 15.63 & 4.45 & (121, 242) & 1.46 & 1.43 & (23, 96) & 2.11 & 2.01 & V & V & 24.1 \\ 
2023 Nov 06 & A & $I$ & 12.98 & 4.59 & (113, 226) & 1.94 & 1.45 & (21, 88) & 9.15 & 2.07 & V & V & 7.95 \\[1ex] 

2023 Nov 22 & A & $B$ & 15.02 & 3.3 & (53, 106) & 0.96 & 1.71 & (9, 40) & 0.54 & 2.89 & NV & NV &   \\ 
2023 Nov 22 & A & $I$ & 12.55 & 3.23 & (58, 116) & 1.33 & 1.67 & (10, 44) & 4.22 & 2.75 & NV & NV &   \\[1ex] 

2023 Dec 02 & A & $R$ & 13.64 & 3.13 & (175, 350) & 0.9 & 1.35 & (34, 140) & 2.82 & 1.79 & NV & NV &   \\[1ex] 

2023 Dec 09 & A & $B$ & 15.08 & 2.9 & (75, 150) & 1.15 & 1.57 & (14, 60) & 0.97 & 2.39 & NV & PV &   \\ 
2023 Dec 09 & A & $I$ & 12.56 & 2.99 & (75, 150) & 2.22 & 1.57 & (14, 60) & 7.2 & 2.39 & V & PV & 5.92 \\[1ex] 

2023 Dec 21 & A & $B$ & 14.99 & 2.95 & (64, 128) & 1.27 & 1.63 & (12, 52) & 1.89 & 2.55 & NV & PV &   \\ 
2023 Dec 21 & A & $I$ & 12.5 & 2.88 & (79, 158) & 3.32 & 1.55 & (15, 64) & 14.08 & 2.33 & V & PV & 9.67 \\[1ex] 

2023 Dec 22 & A & $R$ & 13.2 & 2.45 & (319, 638) & 1.86 & 1.25 & (63, 256) & 2.56 & 1.55 & V & V & 16.3 \\[1ex] 

2024 Jun 12 & B & $R$ & 13.55 & 2.7 & (301, 602) & 1.77 & 1.26 & (59, 240) & 14.43 & 1.57 & V & V & 20.07 \\[1ex] 

2024 Jul 21 & A & $B$ & 15.3 & 1.55 & (32, 64) & 1.03 & 1.98 & (5, 24) & 2.38 & 3.9 & NV & NV &   \\ 
2024 Jul 21 & A & $I$ & 12.85 & 1.44 & (33, 66) & 1.63 & 1.96 & (5, 24) & 6.02 & 3.9 & NV & NV &   \\[1ex] 

2024 Jul 22 & A & $B$ & 15.42 & 1.71 & (43, 86) & 0.97 & 1.81 & (7, 32) & 2.46 & 3.26 & NV & PV &   \\ 
2024 Jul 22 & A & $I$ & 12.96 & 1.6 & (41, 82) & 2.99 & 1.84 & (7, 32) & 11.38 & 3.26 & V & PV & 3.79 \\[1ex] 

2024 Jul 29 & A & $B$ & 15.25 & 0.73 & (22, 44) & 2.65 & 2.28 & (3, 16) & 4.76 & 5.29 & NV & NV &   \\ 
2024 Jul 29 & A & $I$ & 12.82 & 0.69 & (19, 38) & 1.57 & 2.42 & (3, 16) & 14.96 & 5.29 & NV & NV &   \\[1ex] 

2024 Jul 31 & A & $B$ & 15.06 & 3.6 & (88, 176) & 8.26 & 1.52 & (16, 68) & 32.44 & 2.28 & V & V & 14.48 \\ 
2024 Jul 31 & A & $I$ & 12.64 & 3.74 & (88, 176) & 10.88 & 1.52 & (16, 68) & 61.16 & 2.28 & V & V & 8.12 \\[1ex] 

2024 Aug 05 & A & $B$ & 15.58 & 3.33 & (83, 166) & 7.27 & 1.54 & (15, 64) & 22.38 & 2.33 & V & V & 18.98 \\ 
2024 Aug 05 & A & $I$ & 13.06 & 3.67 & (98, 196) & 10.59 & 1.49 & (18, 76) & 45.37 & 2.18 & V & V & 7.9 \\[1ex] 

2024 Aug 07 & A & $B$ & 15.67 & 5.62 & (99, 198) & 4.07 & 1.48 & (19, 80) & 16.97 & 2.14 & V & V & 13.42 \\ 
2024 Aug 07 & A & $V$ & 14.7 & 5.01 & (90, 180) & 2.92 & 1.51 & (17, 72) & 9.45 & 2.23 & V & V & 11.02 \\ 
2024 Aug 07 & A & $R$ & 13.94 & 5.54 & (97, 194) & 6.65 & 1.49 & (18, 76) & 29.22 & 2.18 & V & V & 7.31 \\ 
2024 Aug 07 & A & $I$ & 13.14 & 4.69 & (79, 158) & 5.81 & 1.55 & (15, 64) & 21.02 & 2.33 & V & V & 3.69 \\[1ex] 

2024 Aug 17 & A & $B$ & 15.56 & 3.61 & (84, 168) & 1.82 & 1.53 & (16, 68) & 2.36 & 2.28 & V & V & 17.53 \\ 
2024 Aug 17 & A & $I$ & 13.06 & 3.51 & (72, 144) & 5.46 & 1.59 & (13, 56) & 23.79 & 2.47 & V & V & 10.3 \\[1ex] 

2024 Aug 27 & A & $B$ & 15.61 & 6.15 & (119, 238) & 3.36 & 1.43 & (23, 96) & 10.31 & 2.01 & V & V & 21.95 \\ 
2024 Aug 27 & A & $I$ & 13.08 & 6.05 & (119, 238) & 10.41 & 1.43 & (23, 96) & 49.13 & 2.01 & V & V & 12.53 \\[1ex] 

2024 Sep 08 & A & $B$ & 14.52 & 2.56 & (40, 80) & 1.19 & 1.85 & (7, 32) & 2.18 & 3.26 & NV & NV &   \\ 
2024 Sep 08 & A & $I$ & 12.09 & 2.45 & (38, 76) & 2.09 & 1.88 & (6, 28) & 0.93 & 3.53 & NV & NV &   \\[1ex] 

2024 Oct 09 & A & $B$ & 13.95 & 4.1 & (112, 224) & 43.26 & 1.45 & (21, 88) & 38.16 & 2.07 & V & V & 23.73 \\ 
2024 Oct 09 & A & $I$ & 11.67 & 3.99 & (96, 192) & 18.15 & 1.49 & (18, 76) & 34.23 & 2.18 & V & V & 15.64 \\[1ex] 

2024 Oct 11 & A & $B$ & 14.26 & 3.71 & (98, 196) & 10.26 & 1.49 & (18, 76) & 23.98 & 2.18 & V & V & 14.54 \\ 
2024 Oct 11 & A & $I$ & 11.83 & 3.6 & (91, 182) & 17.08 & 1.51 & (17, 72) & 54.98 & 2.23 & V & V & 9.41 \\[1ex] 

2024 Nov 30 & A & $B$ & 15.25 & 1.62 & (39, 78) & 1.96 & 1.86 & (7, 32) & 2.17 & 3.26 & NV & PV &   \\ 
2024 Nov 30 & A & $I$ & 12.75 & 1.63 & (44, 88) & 2.65 & 1.8 & (8, 36) & 11.11 & 3.05 & V & PV & 3.47 \\[1ex] 

2024 Dec 07 & A & $I$ & 12.64 & 1.01 & (103, 206) & 1.44 & 1.47 & (19, 80) & 3.04 & 2.14 & NV &  NV &   \\ 

\bottomrule
% \end{tabular}
\end{longtable}

%%%%%%%%%%%%%%%%%%%%%%%%%%%%%%%%%%%%%%%%%%%%%%%%%%%%%%%%%%%%%%%%%%%%%%%%%%%%%%%%%%%%%%%%%%%%%%%%%%%%%%%%%%%%%%%%%
%  Results of SI vs log(Fluxes) 
%%%%%%%%%%%%%%%%%%%%%%%%%%%%%%%%%%%%%%%%%%%%%%%%%%%%%%%%%%%%%%%%%%%%%%%%%%%%%%%%%%%%%%%%%%

\section{Results of SEDs}
\label{app:b}

\begin{figure*}
\centering
\includegraphics[width=\textwidth,height=\textheight,keepaspectratio]{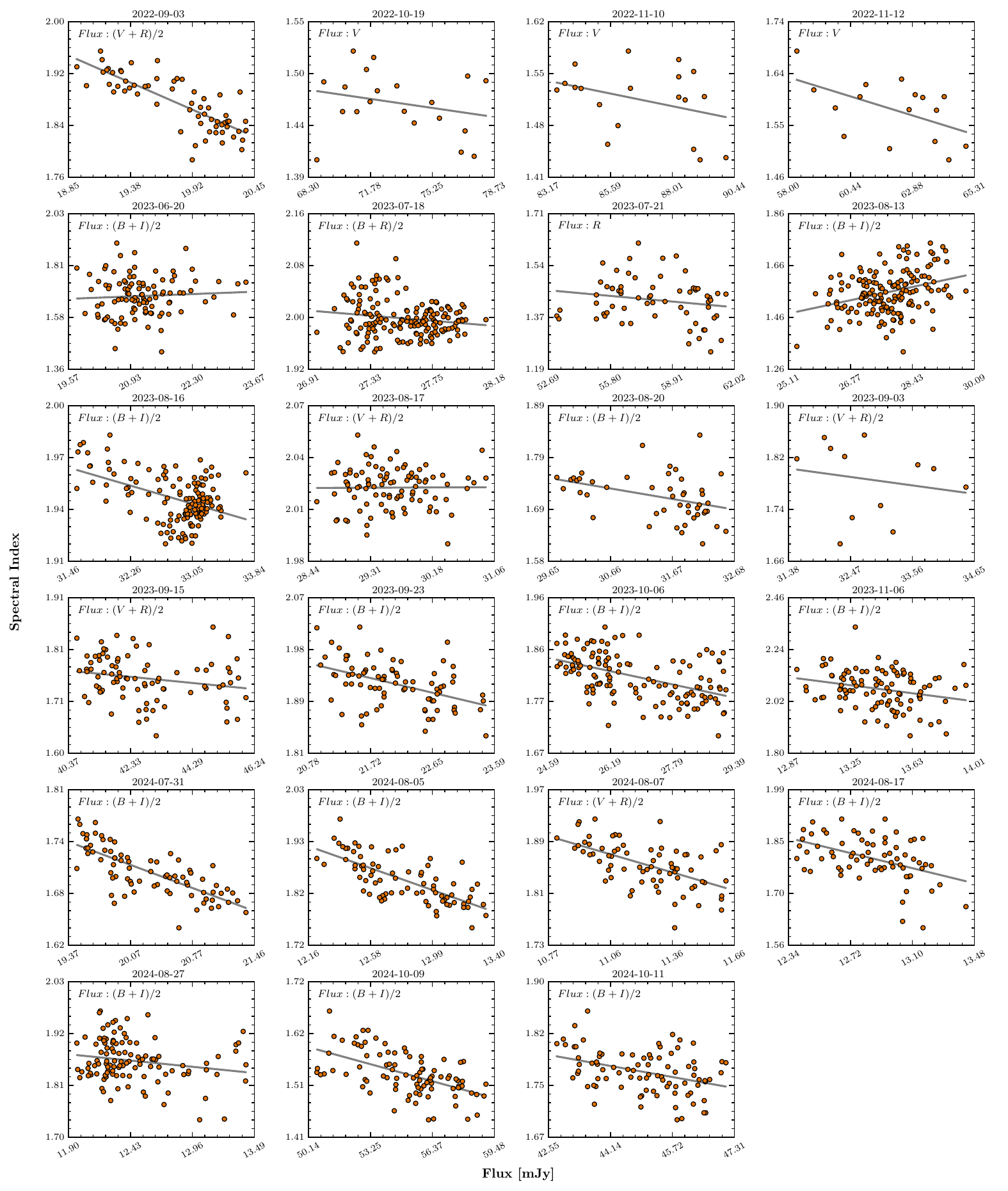}
\caption{Spectral index versus $\log(\mathrm{flux})$ for nights exhibiting multi-band variability of BL Lacertae. Each panel indicates the observation date and the corresponding intermediate flux plotted on the X-axis, while the solid line represents the linear fitting results listed in Table~\ref{tab:Spec_Var}.}
\label{fig:SI_vs_flux}
\end{figure*}

\begin{figure*}
\centering
\includegraphics[width=\textwidth,height=\textheight,keepaspectratio]{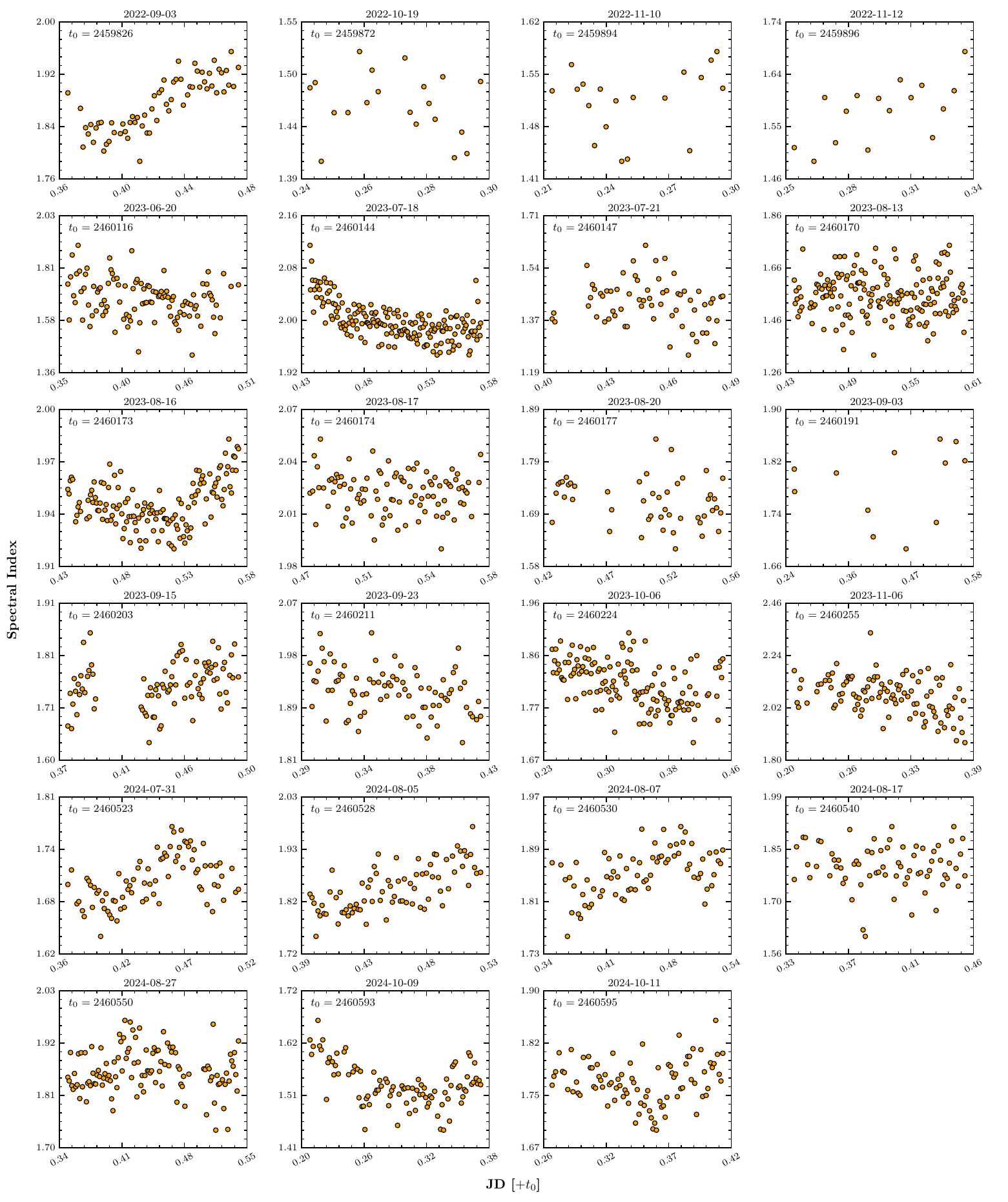}
\caption{Spectral index as a function of time for multi-band variable nights of BL Lacertae. Observing dates and starting time of observation are listed in each panel.}
\label{fig:SI_vs_time}
\end{figure*}

%\twocolumn

% Don't change these lines
\bsp	% typesetting comment
\label{lastpage}

%%%%%%%%%%%%%%%%%%%%%%%%%%%%%%%%%%%%%%%%%%%%%%%%%%%%%%%%%%%%%%%%%%%%%%%%%%%

\end{document}